\documentclass[usenatbib]{mn2e}
\usepackage{apjfonts,amsfonts,amsmath,amssymb,bm,ctable,verbatim}

\newcommand{\Alf}{{Alfv\'en}}
\newcommand{\CRegy}{\gamma_{\rm L}}

\newcommand{\changedtext}[1]{#1}

\newcommand{\gizmourl}{\href{http://www.tapir.caltech.edu/~phopkins/Site/GIZMO.html}{\url{http://www.tapir.caltech.edu/~phopkins/Site/GIZMO.html}}}
\newcommand{\FIREurl}{\href{http://fire.northwestern.edu}{\url{http://fire.northwestern.edu}}}
\newcommand{\msun}{M_{\sun}}

\newcommand{\paperone}{Paper {\small I}}
\newcommand{\papertwo}{Paper {\small II}}
\newcommand{\paperonetwo}{Papers {\small I} \&\ {\small II}}
\newcommand{\etal}{et al.}
\newcommand\plotone[2]{\centering \leavevmode \includegraphics[width={#2\columnwidth}]{#1}}
\newcommand{\plotside}[2]{\centering \leavevmode \includegraphics[width={#2\textwidth}]{#1}}
\newcommand{\datastatement}[1]{\begin{small}\section*{Data Availability Statement}\end{small}{\noindent #1}\vspace{5pt}}
\newcommand{\acknowledgments}[1]{\begin{small}\section*{Acknowledgments}\end{small}{\noindent #1}\vspace{5pt}}
\newcommand{\bhat}{\hat{\bf b}}
\newcommand{\dBprl}{\delta{\bf B}[r_{\rm L}]}
\makeatletter
\newcommand*{\@rowstyle}{}
\newcommand*{\rowstyle}[1]{\gdef\@rowstyle{#1}\@rowstyle\ignorespaces}
\newcolumntype{=}{>{\gdef\@rowstyle{}}}
\newcolumntype{+}{>{\@rowstyle}}
\makeatother
\newcommand{\toohigh}{\rowstyle{\color{red}}}
\newcommand{\toolow}{\rowstyle{\color{cyan}}}
\newcommand{\justright}{\rowstyle{\color{black}}}

\title[Contrasting CR Transport in Galaxies]{Testing Physical Models for Cosmic Ray Transport Coefficients on Galactic Scales: Self-Confinement and Extrinsic Turbulence at $\sim$GeV Energies}

\author[Hopkins \etal]{
\parbox[t]{\textwidth}{
Philip F.~Hopkins$^{1}$, 
Jonathan Squire$^{2}$,
T.~K.\ Chan$^{3,4}$, 
Eliot Quataert$^{5}$, \\
Suoqing Ji$^{1}$, 
Du\v{s}an Kere\v{s}$^{2}$, 
Claude-Andr{\'e} Faucher-Gigu{\`e}re$^{6}$
}\vspace*{4pt} \\
$^1$ TAPIR, Mailcode 350-17, California Institute of Technology, Pasadena, CA 91125, USA. E-mail:phopkins@caltech.edu \\
$^2$ Physics Department, University of Otago, 730 Cumberland St., Dunedin 9016, New Zealand \\
$^3$ Department of Physics, Center for Astrophysics and Space Science, University of California at San Diego, 9500 Gilman Drive, La Jolla, CA 92093 \\ 
$^4$ Institute for Computational Cosmology, Durham University, South Road, Durham, DH1 3LE, UK \\
$^5$ Department of Astronomy and Theoretical Astrophysics Center, University of California Berkeley, Berkeley, CA 94720 \\
$^6$ Department of Physics and Astronomy and CIERA, Northwestern University, 2145 Sheridan Road, Evanston, IL 60208, USA \\ 
 }

\date{}
\begin{document}
\maketitle

\begin{abstract}
The microphysics of $\sim$\,GeV cosmic ray (CR) transport on galactic scales remain deeply uncertain, with almost all studies adopting simple prescriptions (e.g.\ constant-diffusivity). We explore different physically-motivated, anisotropic, dynamical CR transport scalings in high-resolution cosmological FIRE simulations of dwarf and $\sim L_{\ast}$ galaxies where scattering rates vary with local plasma properties motivated by extrinsic turbulence (ET) or self-confinement (SC) scenarios, with varying assumptions about e.g.\ turbulent power spectra on un-resolved scales, \Alf-wave damping, etc. We self-consistently predict observables including $\gamma$-rays ($L_{\gamma}$), grammage, residence times, and CR energy densities to constrain the models. We demonstrate many non-linear dynamical effects (not captured in simpler models) tend to enhance confinement. For example, in multi-phase media, even allowing arbitrary fast transport in neutral gas does not substantially reduce CR residence times (or $L_{\gamma}$), as transport is rate-limited by the ionized WIM and ``inner CGM'' gaseous halo ($10^{4}-10^{6}$\,K gas within $\lesssim 10-30\,$kpc), and $L_{\gamma}$ can be dominated by trapping in small ``patches.'' Most physical ET models contribute negligible scattering of $\sim1-10$\,GeV CRs, but it is crucial to account for anisotropy and damping (especially of fast modes) or else scattering rates would violate observations. We show that the most widely-assumed scalings for SC models produce excessive confinement by factors $\gtrsim 100$ in the WIM and inner CGM, where turbulent and Landau damping dominate. This suggests either a breakdown of quasi-linear theory used to derive the CR transport parameters in SC, or that other novel damping mechanisms dominate in intermediate-density ionized gas.
\end{abstract}

\begin{keywords}
cosmic rays --- plasmas --- instabilities --- gamma-rays: galaxies --- galaxies: evolution --- ISM: structure
\end{keywords}

\section{Introduction}
\label{sec:intro}

Understanding the propagation or bulk transport of cosmic rays (CRs) through the inter-stellar, circum-galactic, and inter-galactic medium (ISM, CGM, IGM) remains a fundamental and unsolved problem of critical importance for high-energy particle physics, plasma physics, and the astrophysics of star and galaxy formation. In the Milky Way (MW), and (probably) most dwarf and star-forming galaxies, the CR energy density and pressure are dominated by relatively low-energy $\sim$\,GeV protons, which are likely accelerated in supernovae [SNe] remnants (with $\sim 10\%$ of the ejecta kinetic energy going into CRs; \citealt{Bell.cosmic.rays}). These $\sim$\,GeV CRs are therefore the most important population governing the interaction of CRs with gas dynamics, heating and cooling of the ISM, gamma-ray emissivities of galaxies, star and galaxy formation, and the excitation of various ``streaming instabilities'' and resonant \Alf\ waves in the plasma \citep{kulsrud.1969:streaming.instability,Mann94,Enss07,guo.oh:cosmic.rays}. There has been a tremendous amount of both analytic \citep{Socr08,Ever08,Dorf12,Mao18} and numerical \citep{jubelgas:2008.cosmic.ray.outflows,uhlig:2012.cosmic.ray.streaming.winds,Wien13,Sale14,Simp16,Pakm16,Rusz17,Giri18} work studying these effects. Recent work on galactic scales has argued $\sim$\,GeV CRs can play an important role, in particular, in the CGM, by suppressing accretion onto low-redshift $\sim L_{\ast}$ galaxies, launching or re-accelerating galactic outflows in these systems, and strongly modifying the phase structure of cool and warm absorption systems \citep{2016MNRAS.456..582S,chan:2018.cosmicray.fire.gammaray,Buts18,su:turb.crs.quench,hopkins:cr.mhd.fire2,ji:fire.cr.cgm}.

\begin{footnotesize}
\ctable[caption={{\normalsize Subset of CR transport models studied. All models include star formation, stellar feedback, MHD, anisotropic conduction and viscosity.}\label{tbl:transport}},center,star
]{=r +l +c +c +c +c}{
\tnote[ ]{Summary of the different CR transport models (models for the effective transport coefficients $\kappa_{\|}$ and $v_{\rm st}$ in Eq.~\ref{eqn:flux}). 
Column include: (1) Name. (2) Description. (3) References where previously studied. (4) $\langle \kappa_{\rm eff}^{\rm iso} \rangle_{29}^{\nu}$: time (redshifts $z<0.1$, sampled each $\sim10\,$Myr) and space (galacto-centric radii $<10\,$kpc) and angle (isotropic-equivalent) averaged, {\em scattering-rate-weighted} effective diffusivity $\kappa_{\rm eff}^{\rm iso} \equiv |{\bf F}_{\rm cr}|/|\nabla e_{\rm cr}|$ (in units of $10^{29}\,{\rm cm^{2}\,s^{-1}}$) in our MW-like ({\bf m12i}) simulations. (5) $L_{\gamma}$, $X_{s}$: qualitative comparison of the predicted $\gamma$-ray luminosity and MW grammage to observational constraints, for dwarf ({\bf m11i}), intermediate ({\bf m11f}), and MW-mass ({\bf m12i}) galaxies. A $\checkmark$ indicates consistency with observations, ``high'' or ``low'' indicates the prediction is too high or low. (6) $\langle e_{\rm cr} \rangle$, the time-and-space averaged, {\em volume-weighted} mean CR energy density (in ${\rm eV\,cm^{-3}}$) in our MW-like ({\bf m12i}) simulations at $z<0.1$ at approximately the solar position (averaged in the thin disk in a galacto-centric radial annulus from $7-9$\,kpc with height $\pm 250\,$pc). Models are grouped by categories (labeled). Models in \textcolor{red}{red} produce excessive confinement and are ruled out by $\gamma$-ray observations and MW constraints. Models in \textcolor{cyan}{cyan} produce less confinement than observed: these are allowed, but cannot dominate scattering. Models in \textcolor{black}{black} produce reasonable agreement with the observations. References are: $a$ \paperone, $b$ \citet{farber:decoupled.crs.in.neutral.gas}, $c$ \citet{chandran00}, $d$ \citet{yan.lazarian.02}, $e$ \citet{yan.lazarian.04:cr.scattering.fast.modes,yan.lazarian.2008:cr.propagation.with.streaming}, $f$ \citet{jokipii:1966.cr.propagation.random.bfield}, $g$ \citet{lazarian:2016.cr.wave.damping}. Different turbulent power spectra include:  GS95 \citep{GS95.turbulence}, K41 \citep{kolmogorov:turbulence}, ``dynamically aligned'' \citep{Boldyrev2006}, B73 \citep{burgers1973turbulence}.}
}{
\hline\hline
Name & Description & Ref. & $\langle \kappa^{\rm iso}_{\rm eff} \rangle_{29}^{\nu}$ & $L_{\gamma}$, $X_{s}$? & $\langle e_{\rm cr} \rangle$ \\
\hline\hline
CD:  & \multicolumn{5}{l}{{\bf Constant-Diffusivity Models (\S~\ref{sec:constant.diffusivity}; Eq.~\ref{eqn:F.effective})}: $\kappa_{\|}=\kappa_{29}\,10^{29}\,{\rm cm^{2}\,s^{-1}}$, 
 varied $v_{\rm st}\sim v_{A}$}  \\
\hline
\toohigh \hfill $\kappa_{29}=0$ & $\kappa_{29}=0$, $v_{\rm st}=(0,\,1,\,3,\,4,\,1+\beta^{1/2},\,3\,[1+\beta^{1/2}])\,v_{A}$ (\S~\ref{sec:advection.streaming}) & $a$ & $\lesssim$0.01 & $\times$ (high) & 40 \\ 
\toohigh \hfill $\kappa_{29}=0.03$ & $\kappa_{29}=0.03$, $v_{\rm st}=(1,3)\,v_{A}$ & $a$ & 0.015 & $\times$ (high) & 50 \\ 
\toohigh \hfill $\kappa_{29}=0.3$ & $\kappa_{29}=0.3$, $v_{\rm st}=(0,1,3)\,v_{A}$ & $a$ & 0.1 & $\times$ (high) & 8 \\ 
\justright \hfill $\kappa_{29}=3$ & $\kappa_{29}=3$, $v_{\rm st}=(0,1,3)\,v_{A}$ (favored models in \paperonetwo) & $a$ & 1 & $\checkmark$ & 1 \\ 
\justright $\kappa_{29}=30$ & $\kappa_{29}=30$, $v_{\rm st}=v_{A}$ & $a$ & 10 &  $\checkmark$ & 0.4 \\ 
\toolow \hfill $\kappa_{29}=300$ & $\kappa_{29}=300$, $v_{\rm st}=v_{A}$ & $a$ & 100 & $\circ$ (low) & 0.04 \\ 
\toohigh \hfill $\kappa_{\rm ion-neutral}$ & $\kappa_{29}=3$ in neutral gas, $=0.1$ in ionized gas (\S~\ref{sec:fast.slow}; Eq.~\ref{eqn:kappa.fastslow}) & $b$ & 0.05 & $\times$ (high) & 20 \\ 
\hline\hline
ET:  & \multicolumn{5}{l}{{\bf Extrinsic Turbulence Models (\S~\ref{sec:extrinsic}, Eq.~\ref{eqn:model.extrinsic})}: $\kappa_{\|}=\mathcal{M}_{A}^{-2}\,c\,\ell_{\rm turb}\,f_{\rm turb}$,  varied $f_{\rm turb}$}  \\
 \hline 
\toolow \hfill \Alf-C00 & $f_{\rm turb} = 0.14\,(c_{s}/v_{A})/\ln(\ell_{\rm turb}/r_{\rm L})$: anisotropic GS95 spectrum of \Alf\ modes & $c$ & 1500 & $\circ$ (low) & 0.2 \\ 
\toolow \hfill \Alf-C00-Vs & as  \Alf-C00, adding additional ``streaming'' $v_{\rm st} = v_{A}$ or $v_{A}^{\rm ion}$ & -- & 1500 & $\circ$ (low) & 0.2 \\ 
\toolow \hfill \Alf-YL02 & $f_{\rm turb} = 70\,(c/v_{A})^{5/11}\,(\ell_{\rm turb}/r_{\rm L})^{9/11}$: modified non-resonant \Alf\ scattering  & $d$ & >$10^{4}$ & $\circ$ (low) & 0.001 \\ 
\toolow \hfill \Alf-Hi & $f_{\rm turb} = 1000$: arbitrarily changed $f_{\rm turb}$ & -- & 400 & $\circ$ (low) & 0.02 \\ %
\justright \hfill \Alf-Max & $f_{\rm turb} = 1$: GS95 \Alf\ scattering ignoring gyro-averaging/anisotropy & -- & 1 &  $\checkmark$ & 2 \\ 
\toolow \hfill Fast-YL04 & $f_{\rm turb} = f(\lambda_{\rm damp})$: non-resonant fast-modes, damped below $\lambda_{\rm damp}$  & $e$ & 80 & $\circ$ (low) & 0.006 \\ 
\justright \hfill Fast-Max & as YL04, neglect ion-neutral and $\beta>1$ viscous damping & $e$ & 6 & $\checkmark$ & 1 \\ 
\toolow \hfill Fast-Mod & $f_{\rm turb} \sim 1000\times$ the ``Fast-Max'' value (different spectrum, broadening) & -- & 700 & $\circ$ (low) & 0.04 \\ 
\toohigh \hfill Fast-NoDamp & $f_{\rm turb} = (r_{\rm L}/\ell_{\rm turb})^{1/2}$: Fast-YL04, ignoring any fast-mode damping  & -- & 0.003 & $\times$ (high) & 3 \\ 
\toohigh \hfill Fast-NoCDamp & $f_{\rm turb}$ given by Fast-Max with viscous damping only  & -- & 0.03 & $\times$ (high) & 5 \\ 
\toohigh \hfill Iso-K41 & $f_{\rm turb} = (r_{\rm L}/\ell_{\rm turb})^{1/3}$: {isotropic}, {undamped} K41 cascade down to $<r_{\rm L}$ & $f$ & 0.004 & $\times$ (high) & 0.4 \\ 
\justright \hfill Fast-Max+Vs & as Fast-YL04, adding additional ``streaming'' $v_{\rm st} = v_{A}$ or $v_{A}^{\rm ion}$  & -- & 7 & $\checkmark$ & 1 \\ 
\hline\hline
SC:  & \multicolumn{5}{l}{{\bf Self-Confinement Models (\S~\ref{sec:self.confinement}, Eq.~\ref{eqn:kappa.self.confinement})}: $\kappa_{\|}\propto \Gamma$ (damping), $v_{\rm st}=v_{A}^{\rm ion}$,  varied $\Gamma$}  \\
\hline 
\toohigh \hfill Default & default scalings for $\Gamma=\Gamma_{\rm in}+\Gamma_{\rm turb}+\Gamma_{\rm LL} + \Gamma_{\rm NLL}$, Appendix~\ref{sec:damping} & -- & 0.02 & $\times$ (high) & 10 \\ 
\toohigh \hfill Non-Eqm & replace $\kappa_{\|}$, $v_{\rm st}$ with evolved gyro-resonant $\dBprl$ (\S~\ref{sec:non.equilibrium}) & -- & 0.03 & $\times$ (high) & 4 \\
\toohigh \hfill $10\,$GeV & adopt $\CRegy=10$ instead of $=1$ (typical $E_{\rm cr}/Z\sim10\,$GeV; \S~\ref{sec:cr.energy}) & -- & 0.03 & $\times$ (high) & 15 \\ 
\toohigh \hfill $v_{A}^{\rm ideal}$ &  adopt $v_{A}=v_{A}^{\rm ideal}$ instead of $v_{A}^{\rm ion}$ in Eq.~\ref{eqn:kappa.self.confinement} (\S~\ref{sec:self.confinement.alfven.speed})  & -- & 0.007 & $\times$ (high) & 15 \\ 
\toohigh \hfill $f_{\rm QLT}$-6 & multiply $\kappa_{\|}$ in Eq.~\ref{eqn:kappa.self.confinement} by $f_{\rm QLT}$ (weaker growth or stronger damping; \S~\ref{sec:define.fQLT}) & -- & 0.05 & $\times$ (high) & 10 \\ 
\toohigh \hfill $f_{\rm QLT}$-6,\,10\,GeV & combines ``$f_{\rm QLT}$-6'' and ``10\,GeV'' models & -- & 0.1 & $\times$ (high) & 8 \\ 
\toohigh \hfill $f_{\rm QLT}$-6,\,$v_{A}^{\rm ideal}$ & combines ``$f_{\rm QLT}$-6'' and ``$v_{A}^{\rm ideal}$'' models & -- & 0.04 & $\times$ (high) & 10 \\ 
\justright \hfill $f_{\rm QLT}$-100 & multiply $\kappa_{\|}$ in Eq.~\ref{eqn:kappa.self.confinement} by $f_{\rm QLT}=100$  & -- & 5 & $\checkmark$ & 0.3 \\ 
\toohigh \hfill $f_{\rm cas}$-5 & $f_{\rm cas}=5$ in $\Gamma_{\rm turb}$ \&\ $\Gamma_{\rm LL}$ & -- & 0.06 & $\times$ (high) & 8 \\ 
\justright \hfill $f_{\rm cas}$-50 & $f_{\rm cas}=50$ in $\Gamma_{\rm turb}$ \&\ $\Gamma_{\rm LL}$ & -- & 2 & $\checkmark$ & 0.3 \\ 
\justright \hfill $f_{\rm cas}$-500 & $f_{\rm cas}=500$ & -- & 10 & $\checkmark$ & 0.4 \\ 
\toohigh \hfill $f_{\rm cas}$-DA & $f_{\rm cas}=(\ell_{\rm turb}/r_{\rm L})^{1/10}$, for a ``dynamically aligned'' perpendicular spectrum ($\sim k_{\perp}^{-3/2}$) & -- & 0.02 & $\times$ (high) & 10 \\ 
\toohigh \hfill $f_{\rm cas}$-B73 & $f_{\rm cas}={\rm MIN}(1,\,\mathcal{M}_{A}^{-1/2})$, for a B73 spectrum above $\ell_{A}$ & -- & 0.005 & $\times$ (high) & 20 \\ 
\toohigh \hfill $f_{\rm cas}$-L16 & $f_{\rm cas}$ follows a multi-component cascade model from L16 & g & 0.004 & $\times$ (high) & 15 \\ 
\justright \hfill $f_{\rm cas}$-K41 & $f_{\rm cas}=\mathcal{M}_{A}^{-1/2}\,(\ell_{\rm turb}/r_{\rm L})^{1/6}$ for an isotropic, undamped K41 cascade & -- & 15 & $\checkmark$ & 0.3 \\ 
\toohigh \hfill NE,\,$f_{\rm cas}$-L16 & as ``Non-Eqm'' but with $f_{\rm cas}$ following $f_{\rm cas}$-L16 model & -- & 0.01 & $\times$ (high) & 4 \\ 
\justright \hfill NE,\,$f_{\rm QLT}$-100 & as ``Non-Eqm'' but with $f_{\rm QLT}=100$  & -- & 7 & $\checkmark$ & 0.3 \\ 
\hline\hline
ET+SC:  & \multicolumn{5}{l}{{\bf Combined Extrinsic-Turbulence \&\ Self-Confinement (\S~\ref{sec:combo})}: $\nu_{\rm total} = \sum \nu_{i}$ (sum ET+SC terms), $v_{\rm st}=v_{A}^{\rm ion}$} \\ 
\hline 
A+F+SC100 & ET:\Alf-C00 + ET:Fast-Max + SC:$f_{\rm turb}=100$ & -- & 2 & $\checkmark$ & 1 \\
A+SC100 & ET:\Alf-C00 + SC:$f_{\rm turb}=100$ & -- & 5 & $\checkmark$ & 0.3 \\
\hline
}
\end{footnotesize}

The transport of these low-energy CRs is {\em especially} uncertain because (1) there are limited direct observational constraints; (2) the gyro-radii of such CRs are extremely small ($\lesssim 1\,{\rm au}$), much smaller than observationally resolved scales in most of the MW ISM (let alone other galaxies); (3) the ``back-reaction'' of the magnetic fields and gas from CRs (e.g.\ excitation of \Alf\ waves via gyro-resonant instabilities) is maximized around this energy scale because this is where the CR energy density is maximized, and can strongly non-linearly alter the propagation of the CRs, i.e.\ they are ``self-confined''; and (4) the structure of the ISM/CGM in which the CRs propagate is uncertain. 

For example, in most of the previous literature, constraints on CR propagation have been inferred assuming a constant (spatially-universal and time-independent) and isotropic diffusivity $\kappa_{\rm iso}$, along with an analytic time-independent model of the MW gas distribution that ignores any small-scale phase structure. Most constraints are also based on ``leaky box'' or ``flat halo'' diffusion models where CRs ``escape'' if they go outside a specified volume (historically, a thin disk with height $\sim 200\,$pc). But all these assumptions can be orders-of-magnitude incorrect. Small gyro-radii mean diffusion is strongly anisotropic, and MW star formation and ISM structure is strongly time-variable on timescales well below the CR residence time and spatially-variable on scales $\lesssim$\,kpc. Perhaps most problematic, it is now firmly established that essentially all galaxies are embedded in massive, extended CGM gaseous halos containing {\em most} of the baryons, with smooth, shallow density profiles extending to $\gtrsim 200\,$kpc (with scale-lengths $\sim 20-50\,$kpc; see e.g.\ \citealt{tumlinson:2017.cgm.review}, and references therein). In analytic or idealized numerical ``leaky box'' or ``flat halo diffusion'' CR transport models when a toy-model ``halo'' is added (usually a cylinder of height $H_{\rm halo} \sim 1-10\,$kpc), the inferred $\kappa_{\rm iso}$ increases with $\sim H_{\rm halo}$ \citep{strong:2001.galprop,vladimirov:cr.highegy.diff,gaggero:2015.cr.diffusion.coefficient,2016ApJ...819...54G,2016ApJ...824...16J,2016ApJ...831...18C,2016PhRvD..94l3019K,evoli:dragon2.cr.prop,2018AdSpR..62.2731A}, so this effect alone can increase the ``required'' diffusivities by factors of $\sim 100$.

    Making matters more complicated, recent work has shown the properties of the gaseous halo itself can depend strongly on the $\sim $\,GeV CR transport \citep{Buts18,ji:fire.cr.cgm}. Moreover, in physically-motivated CR transport models, the local diffusivity is typically a strong function of the local plasma properties (strength of turbulence, magnetic field strength, density, ionization level), which vary by {\em orders of magnitude} on $\sim 0.1-100\,$pc scales within the ISM.

\begin{footnotesize}
\ctable[
  caption={{\normalsize Zoom-in simulation volumes (details in \papertwo). All units are physical.}\label{tbl:sims}},center,star
  ]{lcccccr}{
\tnote[ ]{Properties of the ``primary'' galaxy in each zoom-in volume at $z=0$, including: virial mass ($M_{\rm halo}^{\rm vir}$), stellar mass $M_{\ast}$ in the our reference ``No CRs'' run ($M_{\ast}^{\rm (NoCR)}$) from \papertwo, and full range of stellar masses in our runs here with CRs but different transport physics ($M_{\ast}^{\rm (CR)}$), mass resolution ($m_{i,\,1000}$), Plummer-equivalent force softening at the mean density of star formation ($\langle \epsilon_{\rm gas} \rangle^{\rm sf}$; note the actual softening is adaptive and varies accordingly).}
}{
\hline\hline
Simulation & $M_{\rm halo}^{\rm vir}$ & $M_{\ast}^{\rm (NoCR)}$ & $M_{\ast}^{\rm (CR)}$ & $m_{i,\,1000}$ &  $\langle \epsilon_{\rm gas} \rangle^{\rm sf}$ & Notes \\
Name \, & $[\msun]$ &  $[\msun]$  &   $[\msun]$  &$[1000\,\msun]$ & $[{\rm pc}]$ & \, \\ 
\hline 
{\bf m11i} & 6.8e10 & 6e8 & (2-7)e8 & 7.0 & 1.3 & dwarf galaxy ($\sim$\,SMC-mass), with episodic ``bursty'' star formation \\
{\bf m11f} & 5.2e11 & 4.0e10 & (1.5-4)e10 & 12 & 1.8 &  late-type galaxy, with intermediate surface densities \\
{\bf m12i} & 1.2e12 & 7.0e10 & (2.5-8)e10 & 7.0 & 1.4 &   $\sim L_{\ast}$ galaxy in a ``massive'' halo, dense CGM and higher surface density \\ 
\hline\hline
}
\end{footnotesize}

However, several recent breakthroughs have made real progress possible. (1) Recent $\gamma$-ray observations (mostly from Fermi) have established strong constraints on $\sim$\,GeV CRs in a number of nearby galaxies, complementing the classical Solar-neighborhood constraints on inferred CR grammage, residence times, and energy density. Surprisingly, while the most dense starburst systems observed appear to be proton calorimeters, all ``normal'' $\sim L_{\ast}$ and dwarf galaxies observed (the MW, Andromeda/M31, SMC, LMC, M33) have robust upper limits or detections indicating that at least $\sim 95-99\%$ of the $\sim$\,GeV CRs must escape {\em without} hadronic collisions, requiring large diffusivities \citep{lacki:2011.cosmic.ray.sub.calorimetric,tang:2014.ngc.2146.proton.calorimeter,griffin:2016.arp220.detection.gammarays,fu:2017.m33.revised.cr.upper.limit,wjac:2017.4945.gamma.rays,wang:2018.starbursts.are.proton.calorimeters,lopez:2018.smc.below.calorimetric.crs}. (2) Analytic and numerical work explicitly following transport and scattering of CRs on ``micro-scales'' \citep[e.g.][]{bai:2015.mhd.pic,bai:2019.cr.pic.streaming,lazarian:2016.cr.wave.damping,holcolmb.spitkovsky:saturation.gri.sims,
2019MNRAS.tmp.2249V}, coupled to improved intermediate-scale ``effective fluid'' theories \citep[e.g.][]{zank:2014.book,zweibel:cr.feedback.review,thomas.pfrommer.18:alfven.reg.cr.transport}, has begun to yield more detailed prescriptions for the ``effective'' transport coefficients of CRs as a function of local plasma properties (appropriate on scales much larger than the CR gyro-radius, but much smaller than the scales of e.g.\ ISM phases where these properties change dramatically), for both extrinsic-turbulence and self-confinement scenarios. (3) Cosmological galaxy simulations can now self-consistently model the time-and-space dependent phase structure of the ISM together with extended CGM halos, while explicitly following CR populations \citep{chan:2018.cosmicray.fire.gammaray,Buts18,su:turb.crs.quench,hopkins:cr.mhd.fire2,ji:fire.cr.cgm}. 

In this paper, we synthesize these three advances, to directly constrain proposed micro-physical models of $\sim $\,GeV CR transport. 
To properly model observables like grammage, residence time, and $\gamma$-ray emission, we need to forward-model CR production and transport self-consistently in cosmological simulations which can actually model the ISM/CGM gaseous halos and phase structure (since these strongly influence the observables). The Feedback In Realistic Environments (FIRE)\footnote{\FIREurl} simulations we use here have been shown to reproduce MW and dwarf galaxies with CGM phase structure and gas mass profiles \citep{vandevoort:sz.fx.hot.halos.fire,hafen:2018.cgm.fire.origins,su:turb.crs.quench,ji:fire.cr.cgm}, outflow properties \citep{muratov:2015.fire.winds,hopkins:stellar.fb.winds,hopkins:2013.merger.sb.fb.winds,hayward.2015:stellar.feedback.analytic.model.winds}, ISM phases and detailed molecular cloud properties \citep{hopkins:fb.ism.prop,guszejnov:imf.var.mw,guszejnov:fire.gmc.props.vs.z}, morphologies \citep{elbadry:fire.morph.momentum,elbadry:HI.obs.gal.kinematics,wheeler.2015:dwarfs.isolated.not.rotating,garrisonkimmel:fire.morphologies.vs.dm}, star formation histories and masses \citep{hopkins:2013.fire,hopkins:fire2.methods,garrisonkimmel:local.group.fire.tbtf.missing.satellites}, and magnetic field strengths/morphologies \citep{su:2016.weak.mhd.cond.visc.turbdiff.fx,su:fire.feedback.alters.magnetic.amplification.morphology,su:2018.stellar.fb.fails.to.solve.cooling.flow,guszejnov:fire.gmc.props.vs.z}, all consistent with state-of-the-art observations. 
These simulations reach $\sim $\,pc resolution, which is much larger than the gyro-radii $r_{\rm L}$ of $\sim$\,GeV CRs, so we cannot {\em a priori} predict the CR scattering rates (or diffusivity/streaming speeds). However, this resolution {\em is} sufficient to begin to resolve two crucial scales: (1) the scales of the dominant ISM/CGM phase structures and driving scales of ISM turbulence, and (2) the CR ``mean free path'' or deflection length $\lambda_{\rm mfp} \sim c/\nu$ (where $\nu$ is the CR scattering rate), for the observationally-favored values of $\nu$. This means that if we have a model for the effective diffusion coefficient or ``streaming speed''  of CRs as a function of local plasma properties (or for the more complicated hybrid transport parameters that arise in self-confinement theories), we can self-consistently resolve the full end-to-end CR transport and the observables above on galactic scales. In our previous work \citep{chan:2018.cosmicray.fire.gammaray,su:turb.crs.quench,hopkins:cr.mhd.fire2,ji:fire.cr.cgm}, we did this assuming a simplified anisotropic streaming+diffusion model with a constant parallel diffusivity $\kappa_{\|}$ and parallel streaming at $v_{\rm st}=v_{A}$ (the \Alf\ speed). These works showed that one can obtain converged solutions that reproduce the observed $\gamma$-ray constraints as well as MW grammage/residence-time constraints. We now extend this to a variety of detailed physical models for CR propagation, motivated by both extrinsic turbulence and self-confinement models for scattering.

In \S~\ref{sec:methods} we briefly review the simulation numerical methods, and in \S~\ref{sec:models} we review the different micro-physical CR transport models surveyed. \S~\ref{sec:results} presents the results and compares to present observational constraints. \S~\ref{sec:discussion} discusses and compares these in more detail, considers which models are ruled out and discusses what missing physics might reconcile these with observational constraints, and compares simple analytic or order-of-magnitude expectations for various quantities. \S~\ref{sec:comparison.previous} briefly compares to historical simulation and analytic models. We summarize in \S~\ref{sec:conclusions}.

\section{Methods}
\label{sec:methods}

\subsection{Overview \&\ Non-CR Physics}
\label{sec:methods:overview}

The simulations here extend those in \citet{chan:2018.cosmicray.fire.gammaray} (\paperone) and \citet{hopkins:cr.mhd.fire2} (\papertwo), where numerical details are described. We only briefly summarize these and the non-CR physics here. The simulations are run with {\small GIZMO}\footnote{A public version of {\small GIZMO} is available at \gizmourl} \citep{hopkins:gizmo}, in its meshless finite-mass MFM mode (a mesh-free finite-volume Lagrangian Godunov method). All simulations include ideal magneto-hydrodynamics (MHD), solved as described in \citep{hopkins:mhd.gizmo,hopkins:cg.mhd.gizmo}, and fully-anisotropic Spitzer-Braginskii conduction and viscosity \citep[implemented as in \papertwo; see also][]{hopkins:gizmo.diffusion,su:2016.weak.mhd.cond.visc.turbdiff.fx}. Gravity is solved with adaptive Lagrangian force softening (matching hydrodynamic and force resolution). We treat cooling, star formation, and stellar feedback following the FIRE-2 implementation of the Feedback In Realistic Environments (FIRE) physics \citep[all details in][]{hopkins:fire2.methods}. We follow 11 abundances \citep{colbrook:passive.scalar.scalings,escala:turbulent.metal.diffusion.fire}; cooling chemistry from $\sim 10-10^{10}\,$K accounting for a range of processes including metal-line, molecular, fine-structure, photo-electric, and photo-ionization, including local sources and the \citet{faucher-giguere:2009.ion.background} meta-galactic background (with self-shielding) and tracking detailed ionization states; and star formation in gas which is dense ($>1000\,{\rm cm^{-3}}$), self-shielding, thermally Jeans-unstable, and locally self-gravitating \citep{hopkins:virial.sf,grudic:sfe.cluster.form.surface.density}. Once formed, stars evolve according to standard stellar evolution models accounting explicitly for the mass, metal, momentum, and energy injection via individual SNe (Ia \&\ II) and O/B or AGB-star mass-loss \citep[for details see][]{hopkins:sne.methods}, and radiation (including photo-electric and photo-ionization heating and radiation pressure with a five-band radiation-hydrodynamic scheme; \citealt{hopkins:radiation.methods}). Our models are fully-cosmological ``zoom-in'' simulations, evolving a large box from redshifts $z\gtrsim 100$, with resolution concentrated in a $\sim 1-10\,$Mpc co-moving volume centered on a ``target'' halo of interest. While there are many smaller galaxies in that volume, for the sake of clarity we focus just on the properties of the ``primary'' (i.e.\ best-resolved) galaxies in each volume. The galaxies studied are summarized in Table~\ref{tbl:sims}. 

\subsection{CR Physics \&\ Basic Equations}
\label{sec:methods:crs}

All simulations here also include CRs as described in \paperonetwo. We evolve a single-bin ($\sim\,$GeV) of CRs, or (equivalently) a constant spectral distribution,  as a relativistic fluid (energy density $e_{\rm cr}$, pressure $P_{\rm cr}=(\gamma_{\rm cr}-1)\,e_{\rm cr}$ with $\gamma_{\rm cr}=4/3$), with a fixed fraction $\epsilon_{\rm cr}=0.1$ of the initial SNe ejecta kinetic energy in each explosion injected into CRs. CRs contribute to the total pressure which appears in the gas momentum equation according to the local strong-coupling approximation. Throughout, we denote the CR gyro/Larmor  radius $r_{\rm  L}\equiv c/\Omega$  with $c$ the speed of light and $\Omega = Z\,e\,c\,|{\bf B}|/E_{\rm cr}$ the gyro frequency of the CRs (where $e$ is the electron charge and $E_{\rm cr}/Z \equiv \CRegy\,$GeV, with $\CRegy\sim1-10$ for the CR protons of interest here).

Following \paperonetwo, CRs then obey a  standard energy and flux equation \citep[see e.g.][]{mckenzie.voelk:1982.cr.equations}: 
\begin{align}
\label{eqn:ecr} \frac{\partial e_{\rm cr}}{\partial t}  + \nabla\cdot \left( {\bf u}\,h_{\rm cr} + {\bf F} \right)
&= {\bf u}\cdot \nabla P_{\rm cr} - \Lambda_{\rm st} - \Lambda_{\rm coll} + S_{\rm in} \\  
\label{eqn:flux} 
-\frac{(\gamma_{\rm cr}-1)}{\kappa_{\ast}}\,{\bf F} &= \nabla_{\|} P_{\rm  cr} + \frac{\mathbb{D}_{t}{\bf F}}{\tilde{c}^{2}}
\end{align}
In Eq.~\ref{eqn:ecr}, ${\bf  u}$ is the gas fluid velocity, ${\bf F}$ the CR flux in the fluid frame, $h_{\rm cr} \equiv e_{\rm cr}+P_{\rm cr}$ the CR enthalpy, $S_{\rm  in}$ the CR source injection, and $\Lambda_{\rm st}={\rm MIN}(v_{A},\,v_{\rm st})\,|\nabla_{\|}P_{\rm cr}|$ represents ``streaming losses,'' which arise because gyro-resonant \Alf\ waves (unresolved wavelengths $\sim r_{\rm L}$) are excited by CR streaming (with speed $v_{\rm st}$, defined below) and rapidly damp \citep{wentzel:1968.mhd.wave.cr.coupling,kulsrud.1969:streaming.instability}. These losses are limited to the \Alf\ speed $v_{A}$, as we show below (see also \paperone\ and \citealt{Rusz17}). The $\Lambda_{\rm coll}$ term represents collisional (hadronic and Coulomb) losses with $\Lambda_{\rm coll}=5.8\times10^{-16}\,{\rm s^{-1}\,cm^{3}}\,(n_{\rm n}+0.28\,n_{e})\,e_{\rm cr}$ (with $n_{\rm n}$ and $n_{e}$ the nucleon and free electron number densities), following \citet{guo.oh:cosmic.rays}. Of $\Lambda_{\rm coll}$, all Coulomb (the $n_{e}$ term) and $\sim 1/6$ of the hadronic ($n_{\rm n}$) losses are thermalized; $\Lambda_{\rm st}$ is thermalized as well. In Eq.~\ref{eqn:flux}, $\nabla_{\|} P_{\rm cr} \equiv (\bhat\otimes\bhat)\cdot \nabla P_{\rm cr}= \bhat\,(\bhat\cdot \nabla P_{\rm cr})$ is the parallel derivative, $\tilde{c}$ is the maximum (physical or numerical) CR free-streaming/signal speed ($\ge1000\,{\rm km\,s^{-1}}$ here), $\kappa_{\ast}$ is a local effective diffusivity (defined below), and $\mathbb{D}_{t}{\bf F} \equiv \hat{\bf F}\,[\partial |{\bf F}|/\partial t + \nabla\cdot({\bf u}\,|{\bf F}|) +  {\bf F}\cdot\{ (\hat{\bf  F}\cdot \nabla)\,{\bf u}\} ]$ is  the derivative operator derived in \citet{thomas.pfrommer.18:alfven.reg.cr.transport} from a two-moment expansion of the relativistic Vlasov equation for CRs (assuming a locally gyrotropic CR distribution in the  fluid frame and vanishingly small gyro radii, to  $\mathcal{O}(v^{2}/c^{2})$).\footnote{As discussed in Appendix~\ref{sec:alternative.flux.eqns}, the operator $\mathbb{D}_{t}$ in Eq.~\ref{eqn:flux} is very slightly different from that adopted in \papertwo, but the difference enters at $\mathcal{O}(1/\tilde{c}^{2})$ and has no effect on our conclusions.} Because the gyro radii of GeV CRs are vastly smaller than resolved scales, they move along the field lines,  with $\hat{\bf F}=\bhat$ by construction.

As shown in \paperonetwo\ and below, the overwhelmingly dominant uncertainty in CR transport on these scales comes from the form of $\kappa_{\ast}$, which we will explore extensively. Variations to other choices above, e.g.\ turning off the sink terms $\Lambda_{\rm st}$ or $\Lambda_{\rm coll}$, otherwise altering the functional form of the flux Eq.~\ref{eqn:flux} (or simply solving a single energy equation, specifying some equilibrium ${\bf F}$), varying $\tilde{c}$ widely, or varying $\epsilon_{\rm cr} \sim 0.05-0.2$, all have minor or negligible effects on our results. These are reviewed in Appendix~\ref{sec:alternative.flux.eqns}.

\subsection{Effective CR ``Transport Parameters''}
\label{sec:methods:cr.transport}

We explicitly evolve ${\bf F}$ according to Eq.~\ref{eqn:flux}. However because the bulk CR flux,  by construction, always points along the magnetic field direction ($\hat{\bf F} = \hat{\bf v}_{\rm st} = -\hat{\nabla}_{\|} P_{\rm cr} =  \pm \bhat$), one can always write the instantaneous flux in terms  of an {\em effective} local scalar diffusion and/or streaming coefficient, i.e.:
\begin{align}
\label{eqn:F.effective} {\bf F} &\equiv -\kappa_{\rm eff}\,\nabla_{\|} e_{\rm cr} \equiv  \bar{{\bf v}}_{\rm st,\,eff}\,h_{\rm  cr} 
\equiv -\bar{\kappa}_{\|}\,\nabla_{\|} e_{\rm cr} + \bar{{\bf v}}_{\rm st}\,h_{\rm  cr} 
\end{align}
where $\bar{{\bf v}}_{\rm st} = -\bar{v}_{\rm  st}\,(\nabla_{\|} P_{\rm cr}) / |\nabla_{\|} P_{\rm cr}|$ is the streaming velocity, defined to point along the ${\bf B}$-field down the CR pressure gradient. In  other words, we can always simply {\em define} $\kappa_{\rm eff} \equiv |{\bf F}|/|\nabla_{\|} e_{\rm cr}|$, or re-write pure-diffusion ($v_{\rm st}=0$) as pure-streaming with $\bar{v}_{\rm st} \rightarrow \bar{\kappa}_{\|}  / (\gamma_{\rm cr}\,\ell_{\rm cr})$ (where for convenience we define  the parallel CR pressure gradient scale-length $\ell_{\rm cr} \equiv P_{\rm cr}  / |\nabla_{\|} P_{\rm cr}|$), or vice-versa ($\bar{\kappa}_{\|} \rightarrow \gamma_{\rm cr}\,\bar{v}_{\rm st}\,\ell_{\rm cr}$).

In quasi steady-state ($\mathbb{D}_{t}{\bf F}\rightarrow 0$),  the Newtonian limit ($\tilde{c}$ sufficiently large),  on scales large compared to the CR mean free  path/time ($\sim \kappa_{\ast}/\tilde{c}$), or in the ``pure streaming+diffusion'' approximation for the flux ($\mathbb{D}_{t}\rightarrow 0$), Eq.~\ref{eqn:flux} gives ${\bf F} \rightarrow \kappa_{\ast}\nabla_{\|} e_{\rm cr}$, so $\kappa_{\ast} \rightarrow \kappa_{\rm eff} = \bar{\kappa}_{\|} + \gamma\,v_{\rm st}\,\ell_{\rm cr}$ {\em exactly}. For this and other physical reasons (see \paperone\ and \citealt{jiang.oh:2018.cr.transport.m1.scheme}), we therefore write $\kappa_{\ast} = \kappa_{\|} + \gamma_{\rm cr}\,v_{\rm st}\,\ell_{\rm cr}$ in Eq.~\ref{eqn:flux}, where we refer to the coefficients $\kappa_{\|}({\bf x},\,t,\,...)$ and $v_{\rm st}({\bf x},\,t,\,...)$ as the local ``diffusivity''  and ``streaming speed,'' respectively. But we emphasize that these can be {\em  arbitrary  functions} of the local plasma properties and their derivatives, so Eq.~\ref{eqn:ecr} does not necessarily behave like a traditional streaming or diffusion equation. 

 We will explore variations in the functions $\kappa_{\|}$ and $v_{\rm st}$ below, and we will write  and refer to both $\kappa_{\|}$ and $v_{\rm st}$,  even though once  they are arbitrary functions, their individual values are irrelevant to the CR propagation (only  the combined function $\kappa_{\ast}$ is meaningful). Our reason for making this  distinction  between diffusion and streaming is largely historical, and  we stress that the traditional differences in ``diffusive-like''  vs.\ ``streaming-like'' behavior only apply  when $\kappa_{\|}$ and/or $v_{\rm st}$ are constants. This is explored further in Appendix~\ref{sec:deriv:behavior}.

\subsection{The \Alf\ Speed \&\ Gyro-Resonant Wavelengths}
\label{sec:alfven.speed}

Ideal or Braginskii MHD, in which the \Alf\ speed is $v_{A} = v_{A}^{\rm ideal}\equiv (|{\bf B}|^{2}/4\pi\,\rho)^{1/2}$,  is an excellent approximation on  all {\em resolved} scales in the  simulations here  (even when $f_{\rm ion} \ll 1$ in e.g.\ GMCs),\footnote{Formally, the ion-neutral ``strong-coupling'' approximation (ion-neutral collision times are short compared to resolved timescales) applies on all simulated scales ($\sim$\,pc or larger).}  But self-confinement models often refer specifically to the \Alf\ speed of gyro-resonant \Alf\ waves, which are vastly shorter-wavelength (parallel wavenumbers $k_{\|} \sim k_{\rm L} = 2\pi/\lambda_{\rm L} \sim 1/r_{\rm L}$) and therefore can have frequencies much larger than the collision frequency between ions and neutrals in GMCs, and so propagate at the ``ion-\Alf'' speed  $v_{A}^{\rm ion} \equiv (|{\bf B}|^{2}/4\pi\,\rho_{\rm i})^{1/2}  = f_{\rm ion}^{-1/2}\,v_{A}^{\rm ideal}$ \citep{1975MNRAS.172..557S}. Such short-branch waves are rapidly damped when $f_{\rm ion}\ll1$, but the models can account for this. So in general when we refer to $v_{A}$, we take $v_{A}=v_{A}^{\rm ideal}$, but we explicitly note when we consider $v_{A}^{\rm ion}$.

Anisotropic viscosity in hot, dilute gas formally modifies the Alfv\'en speed as well (e.g., \citealt{Kempski2020}), but the fractional change in Alfv\'en speed is small for the hot ISM and CGM.

\begin{figure*}
    \plotside{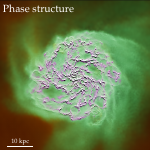}{0.33}\plotside{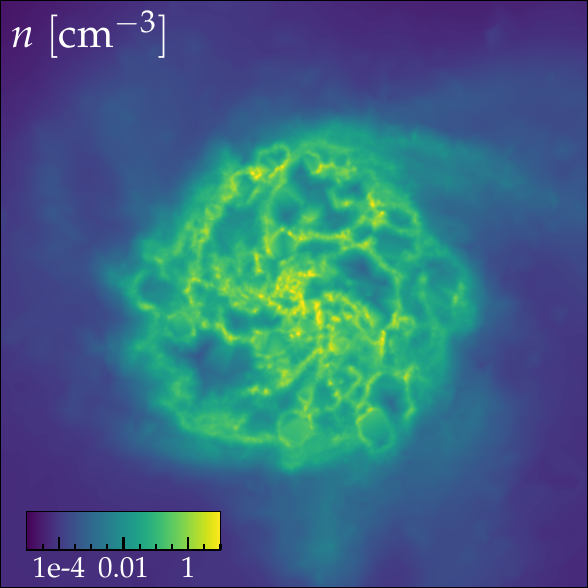}{0.33}\plotside{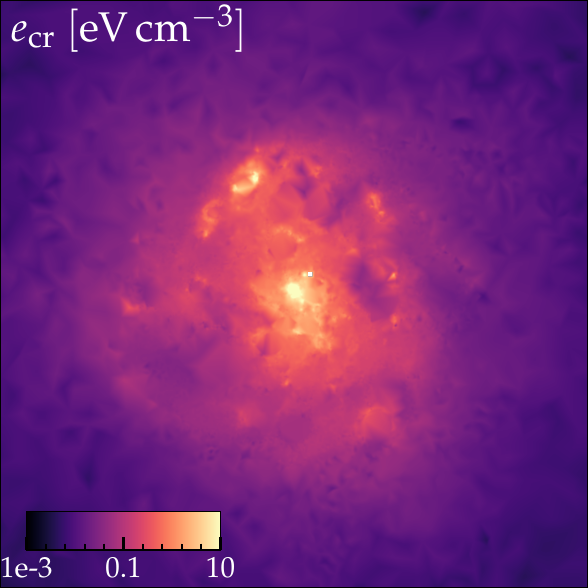}{0.33}\\
    \plotside{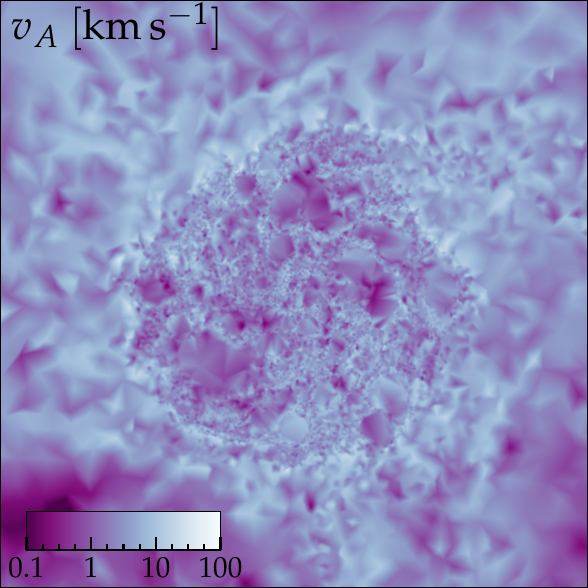}{0.33}\plotside{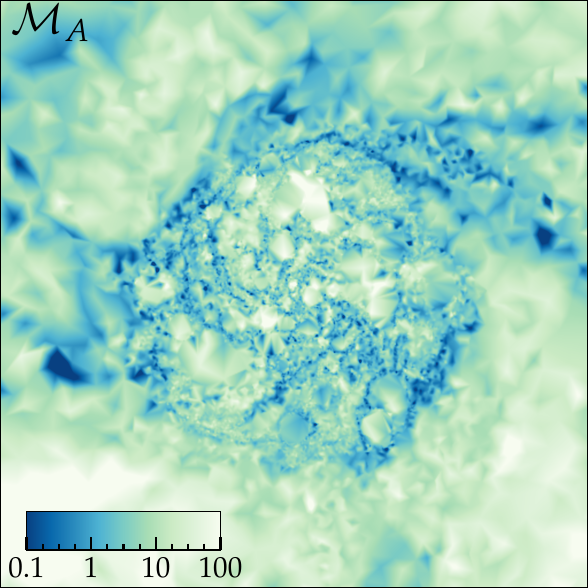}{0.33}\plotside{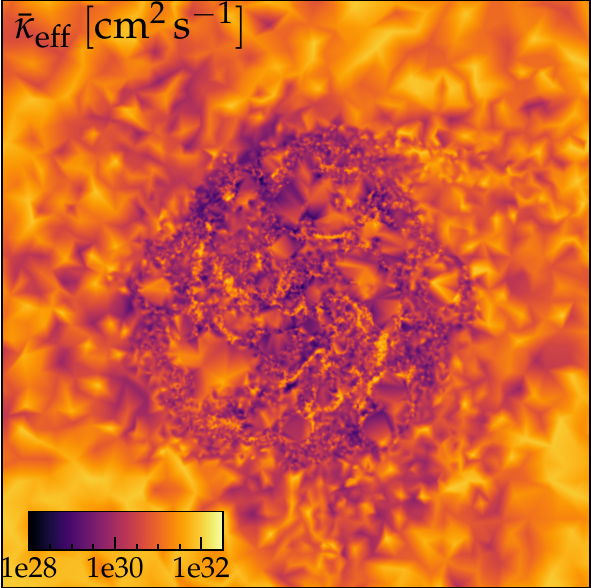}{0.33}\\
    \vspace{-0.25cm}
    \caption{Images of one of our simulated galaxies ({\bf m11f}) at present-day ($z=0$), in a mid-plane slice with box $\sim 60\,$kpc on a side (see scale-bar), viewed face-on. We show the self-confinement (SC)-motived model ``$f_{\rm QLT}$-100.''
    {\em Top left:} Phase map showing cold neutral ({\em magenta}, $T\lesssim 8000\,$K), warm ionized ({\em green}; $10^{4} \lesssim T \lesssim 10^{5}$\,K) and hot ionized ($T\gtrsim 10^{5}\,$K) gas. {\em Top center:} Gas density $n$. {\em Top right:} CR energy density $e_{\rm cr}$. {\em Bottom left:} Ideal MHD \Alf\ speed $v_{A}$. {\em Bottom center:}  \Alf\ Mach number $\mathcal{M}_{A}$. {\em Bottom right:} Effective diffusivity $\bar{\kappa}_{\rm eff} \equiv |{\bf F}|/|\nabla_{\|}e_{\rm cr}|$, where ${\bf F}$ is the local CR flux. 
    Multi-phase structure with large fluctuations in turbulent dissipation rates and $v_{A}$ are evident on scales $\ll$\,kpc, while galactic outflows give rise to large $\mathcal{M}_{A}$ in the CGM and in ``superbubbles'' within the disk. These give rise to orders-of-magnitude fluctuations in $\bar{\kappa}_{\rm eff}$ on small-scales, though $\bar{\kappa}_{\rm eff}$ generally rises outside the galactic disk. The CR energy $e_{\rm cr}$ is smoother, following a radial gradient to first order (as expected), though with a notable ``hot spots'' surrounding clustered SNe.
    \label{fig:image}}
\end{figure*}

\section{Different CR Transport Models Considered}
\label{sec:models}

Here we describe the different CR transport models considered in this paper, summarized in Table~\ref{tbl:transport}. For each of these models, we have run a suite of cosmological simulations with at least galaxies {\bf m11i}, {\bf m11f}, {\bf m12i}, chosen because these span a range of masses and, at each mass, show representative effects and scalings of CRs on galaxy dynamics in \paperonetwo. An illustration of the galaxies and their properties is shown in Fig.~\ref{fig:image}.

\subsection{Constant-Diffusivity Models}
\label{sec:constant.diffusivity}

Lacking a physical model, we can simply {\em assume} $\kappa_{\|}=$\,constant. This is commonly done in empirical models  for CR transport, and we  explored such models extensively in \paperonetwo. For the relatively large diffusion coefficients favored by observations ($\kappa_{\|}\sim 3\times10^{29-30}\,{\rm cm^{2}\,s^{-1}}$, see \S~\ref{sec:results}), we showed in \paperonetwo\ that adding or neglecting an ``additional'' CR streaming at trans-\Alf{ic} or trans-sonic speeds made only a very small difference to our conclusions. This follows from  our discussion in \S~\ref{sec:methods:cr.transport}: what matters on large scales is not $\kappa_{\|}$ or $v_{\rm st}$ individually but the total transport function $\kappa_{\ast} = \kappa_{\|} + \gamma_{\rm cr}\,v_{\rm  st}\,\ell_{\rm cr}$,  where the second (streaming) term is $\sim 4\times 10^{27}\,{\rm cm^{2}\,s^{-1}}\,(v_{\rm st}/10\,{\rm km\,s^{-1}})\,(\ell_{\rm cr}/{\rm kpc})$. Thus, even factor of $\sim 10$ variations  in $v_{\rm st}$ around typical trans-\Alf{ic} values amount to $\sim 0.1-10\%$ variations in $\kappa_{\ast}$ (for $\kappa_{\|} \sim 10^{30}\,{\rm cm^{2}\,s^{-1}}$), compared to the order-of-magnitude variations in $\kappa_{\ast} \sim \kappa_{\|}$ which fall within the ``allowed'' range. 

We stress that these models have no particular physical motivation: they simply provide an empirical reference point for the transport speeds ``needed'' (in the ISM and near-field CGM where e.g.\ $\gamma$-ray emission originates) to reproduce observational constraints. 

\subsubsection{Model Variant: ``Fast'' Transport in Neutral  Gas, ``Slow'' in Ionized Gas}
\label{sec:fast.slow}

In self-confinement scenarios, strong ion-neutral damping can produce rapid transport in primarily-neutral gas. In \citet{farber:decoupled.crs.in.neutral.gas}, the authors attempt to approximate this effect with a ``two-$\kappa$'' model,  with a constant-but-different diffusivity in neutral and ionized gas.\footnote{They adopted $\kappa_{\|}=10^{29}$ or $3\times10^{27}\,{\rm cm^{2}\,s^{-1}}$ in gas below/above $T=10^{4}\,$K, using temperature as a proxy for ionization state.} We therefore consider a similar model, parameterized as:
\begin{align}
\label{eqn:kappa.fastslow} \kappa_{\|} &= 3\times10^{29}\,{\rm cm^{2}\,s^{-1}}\,\left( 1-f_{\rm ion} + \frac{f_{\rm ion}}{30} \right)
\end{align}
(with $v_{\rm st}=v_{A}$), so  $\kappa_{\|}=3\times10^{29}$ or $\kappa_{\|}=10^{28}\,{\rm cm^{2}\,s^{-1}}$ in neutral or ionized gas, respectively. This is a useful reference model because it allows us to explore whether CR diffusion must be relatively ``fast'' in {\em both} neutral and ionized gas, or just the densest (neutral) gas.

\subsubsection{Model Variant: Pure-Advection \&\ \Alf{ic}/Sonic Streaming-Only}
\label{sec:advection.streaming}

If $\kappa_{\ast} \rightarrow 0$ (i.e.\ $\kappa_{\|}\rightarrow0$ and $v_{\rm st}\rightarrow0$), then ${\bf F}\rightarrow \mathbf{0}$ and CRs are purely advected with gas. It is well-established that this cannot possibly reproduce observations in the MW and nearby galaxies. If the only CR transport beyond advection were streaming with trans-\Alf{ic} or trans-sonic speeds, this is identical to our default constant-$\kappa_{\|}$ models with $\kappa_{\|}\rightarrow 0$ (and $v_{\rm st} \sim v_{A}$). In the MW warm ISM, with $v_{A} \sim c_{s} \sim 10\,{\rm km\,s^{-1}}$, this gives effective diffusivities $\kappa_{\rm eff} \sim v_{\rm A}\,\ell_{\rm cr} \sim 10^{27}\,{\rm cm^{2}\,s^{-1}}$, much lower than our preferred $\kappa_{\rm eff}$. These cases are considered explicitly in \paperonetwo, with $v_{\rm  st} \sim 0,\, v_{A},\,3\,v_{A},\,10\,v_{A},\,v_{\rm fast},\,3\,v_{\rm fast}$ (where $v_{\rm fast}^{2}=c_{s}^{2}+v_{A}^{2}$ is the fastest ideal-MHD wavespeed), where we showed all produce far too-slow CR transport and over-predict observed $\gamma$-ray fluxes from nearby galaxies by $\sim 1-2$\,dex. So we do not consider these cases further, except as the obvious limit when $\kappa_{\|}\rightarrow 0$.

\subsection{Extrinsic Turbulence Scenarios}
\label{sec:extrinsic}

The CR diffusivity is $\kappa_{\rm eff} \sim c^{2}/3\,\nu$, where $\nu$ is the scattering rate ($\lambda_{\rm mfp} \sim c/\nu$ is the CR mean free path). In the standard picture, CRs scatter off of magnetic-field fluctuations $\delta{\bf B}$, with a strong preference for ``resonant'' fluctuations $\dBprl$, i.e. fluctuations with parallel wavenumber $k_{\|} \sim k_{\rm L} \sim 1/r_{\rm L}$. Simple quasi-linear theory calculations give the scattering rate $\nu \sim \Omega\,|\dBprl|^{2}/|{\bf B}|^{2}$ \citep[e.g.][]{jokipii:1966.cr.propagation.random.bfield,wentzel:1968.mhd.wave.cr.coupling,skilling:1971.cr.diffusion}.

In the simplest possible ``extrinsic turbulence'' model \citep[e.g.][]{jokipii:1966.cr.propagation.random.bfield,1975RvGSP..13..547V}, we can estimate $\kappa_{\rm eff}$ by extrapolating $|\dBprl|$  from a turbulent power spectrum with (1D) \Alf\ Mach number $\mathcal{M}_{A} = \mathcal{M}_{A}[\ell_{\rm turb}] \equiv |\delta {\bf B}[\ell_{\rm turb}]|/|{\bf B}|\approx |\delta {\bf v}[\ell_{\rm turb}]|/v_{A}^{\rm ideal}$ on some resolved scale $\ell_{\rm turb}$. While very high energy CRs (with large $r_{\rm L}$) may scatter significantly  on $\ell_{\rm turb}$ scales  directly,  we are interested in low-energy CRs with $r_{\rm L} \sim 10^{-6}\,{\rm pc}$. Such scales are smaller than the damping/viscous scale for fast/acoustic modes, while \Alf{ic} modes, although not strongly damped, are highly anisotropic on these scales, which must be taken into account for estimates of $\nu$ (as we do below). Nonetheless, as a reference model, let us assume a  \citet{GS95.turbulence}-type (GS95) cascade ($E_{\|} \propto k_{\|}^{-2}$), giving:
\begin{align}
\label{eqn:model.extrinsic} \frac{\kappa_{\|}}{c\,r_{\rm L}} &\sim \frac{|{\bf B}|^{2}}{|\delta {\bf B}[k_{\|}\sim 1/r_{\rm L}]|^{2}}\,{f}_{\rm turb} \sim \mathcal{M}_{A}^{-2}\, \frac{\ell_{\rm turb}}{r_{\rm L}} \,{f}_{\rm turb} \\
\nonumber \kappa_{\|} &\sim 10^{32}\,{\rm cm^{2}\,s^{-1}}\,\mathcal{M}_{A}^{-2}\,\ell_{\rm turb,\,{kpc}}\,{f}_{\rm turb}
\end{align}
where $\ell_{\rm turb,\,kpc} \equiv \ell_{\rm turb}/{\rm kpc}$, and we absorb all the microphysics of turbulence and scattering into ${f}_{\rm turb}$.

\begin{figure*}
    \plotside{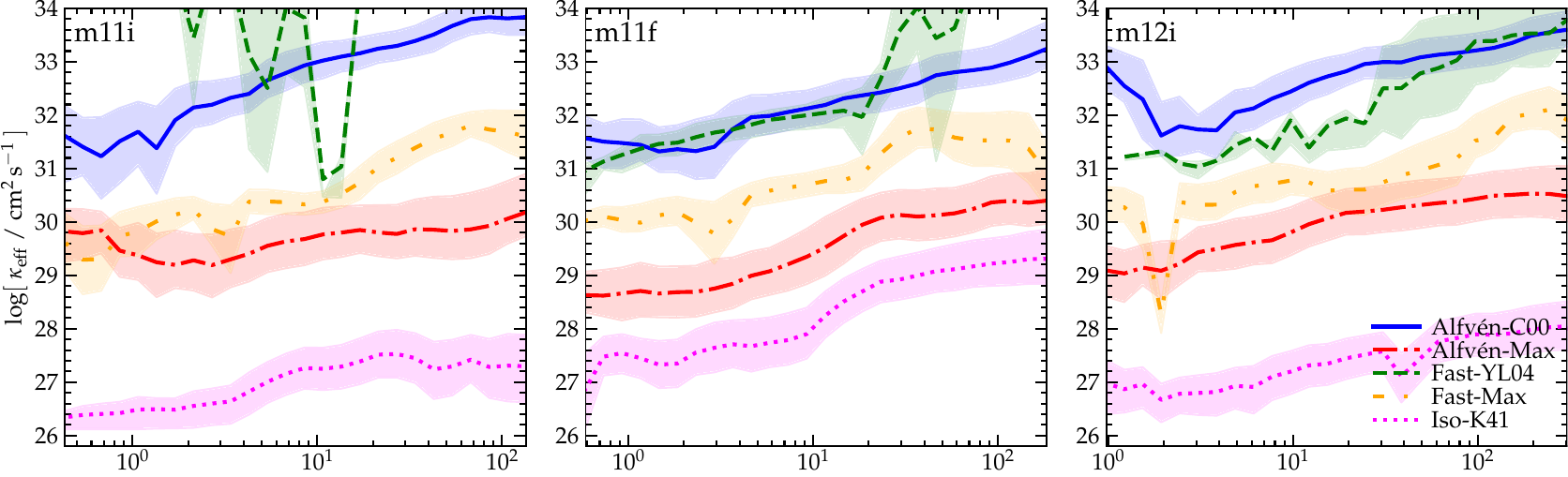}{0.99}
    \plotside{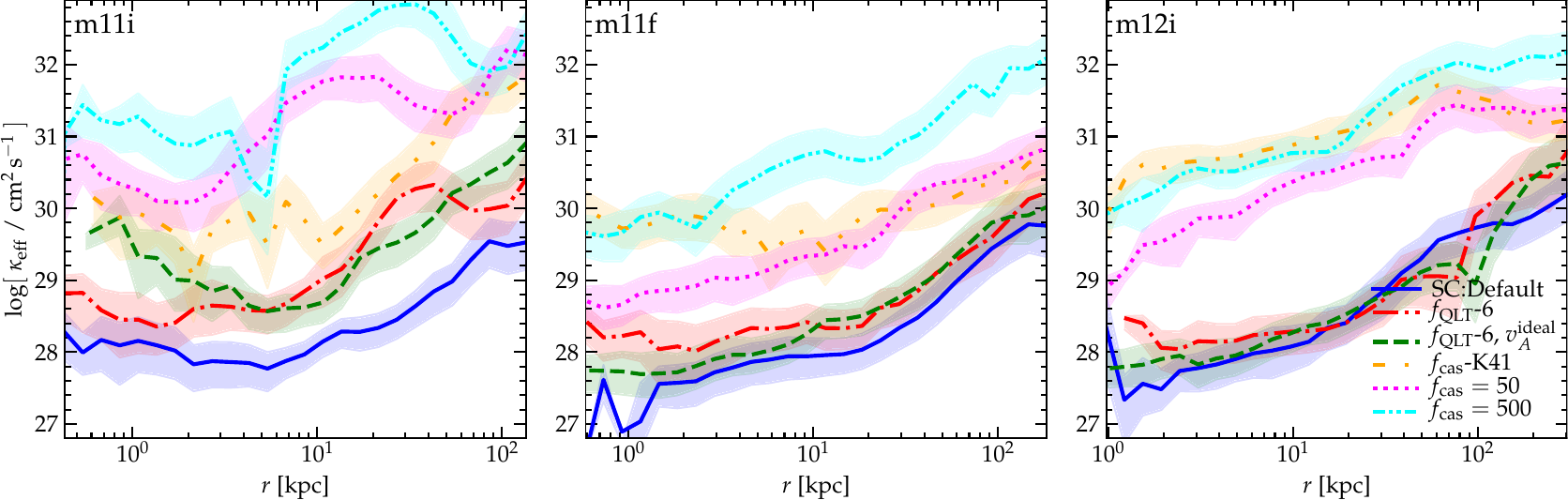}{0.99}
    \vspace{-0.25cm}
    \caption{Effective {\em scattering-weighted} mean parallel CR diffusivity $\kappa_{\rm eff} \equiv |{\bf F}|/|\nabla_{\|}e_{\rm cr}|$ (\S~\ref{sec:methods:cr.transport}), as a function of galacto-centric radius $r$, in galaxies {\bf m11i} (dwarf), {\bf m11f} (intermediate/MW/NGC 253-mass), {\bf m12i} (M31-mass) at $z=0$ (Table~\ref{tbl:sims}).
    We compare some representative models from Table~\ref{tbl:transport} for CR scattering via extrinsic turbulence (ET; {\em top}; \S~\ref{sec:extrinsic}) and self-confinement (SC; {\em bottom}; \S~\ref{sec:self.confinement}). 
    Our definition of $\kappa_{\rm eff}$ means this includes both traditional ``diffusion'' and ``streaming'' terms.
    Solid lines show the mean $\kappa_{\rm eff}$ in spherical shells at each $r$, weighted by the contribution of each resolution element to the scattering rate (shaded shows weighted $25-75\%$ range).
     Diffusivities $\kappa_{\rm eff}$ generally rise with radius $r$ around a given galaxy, or in lower-mass dwarf galaxies, 
      as densities $\rho$ and field strengths $|{\bf B}|$ decrease. Different models considered here produce up to factor $\sim 10^{8}$ systematic differences in $\kappa_{\rm eff}$ -- far larger than any other physical/numerical uncertainties in the models here (see Appendix~\ref{sec:alternative.flux.eqns}). 
      {\em Top:} Theoretically-preferred scattering rates from ET from \Alf\ waves (``\Alf-C00'') or fast modes (``Fast-YL04'') give large $\kappa_{\rm eff}$: models ``\Alf-Max'' and ``Fast-Max'' artificially make the scattering rate much larger ($\kappa_{\rm eff}$ smaller) by neglecting some damping/anisotropy terms, while ``Iso-K41'' neglects {\em all} damping or anisotropy in the turbulence down to $\sim r_{L}$. 
      {\em Bottom:} Our ``SC:Default'' model (accounting for ion-neutral, turbulent, linear and non-linear Landau damping) produces low $\kappa_{\rm eff}$: multiplying the diffusivity by a factor ``$f_{\rm QLT}=6$`` makes little difference owing to non-linear effects (increasing $\kappa$ produces lower $e_{\rm CR}$, which then re-increases $\kappa$ in SC models); using the ideal-MHD \Alf\ speed $v_{A}^{\rm ideal}$ instead of the ion speed $v_{A}^{\rm ion}$ also has weak effects, but $\kappa_{\rm eff}$ can be made larger if $f_{\rm QLT}$ or $f_{\rm cas}$ (turbulent damping rates) are increased by $\sim 100$.
    \label{fig:kappa.vs.model}}
\end{figure*}

\subsubsection{Model Variant: Turbulent Structure Assumptions}
\label{sec:extrinsic.turb.model}

There is an extensive literature regarding the ``correct'' form of Eq.~\ref{eqn:model.extrinsic} (or, equivalently, $f_{\rm turb}$) for extrinsic turbulence \citep[see e.g.][and references therein]{Zwei13}. We cannot possibly be comprehensive here, so we focus on a few models chosen to bracket a range of possibilities. Note that the expressions proposed for $f_{\rm turb}$ or $\kappa$ are often very complicated: we simplify these to order-of-magnitude scalings for the parameter space of interest ($\sim$\,GeV CRs, etc.).

\begin{enumerate}

\item{{\bf \Alf-C00}}: \citet{chandran00} attempt to self-consistently derive $\kappa_{\ast}$ in a \citet{GS95.turbulence} cascade, accounting for anisotropy. For all limits relevant here, their result (Eq.~16 therein) gives $f_{\rm turb} \approx (0.14/\ln{(\ell_{\rm turb}/r_{\rm L})})\,(c/v_{A}) \sim 1000\,n_{1}^{1/2}\,B_{\rm \mu G}^{-1}$ (where $n_{1}=\rho/(m_{p}\,{\rm cm^{-3}})$). Here $f_{\rm turb} \gg 1$ arises because the GS95 cascade has power at $k_{\rm L}$ only for $k_{\bot} \gg k_{\|}$, which leads to an effective ``reduction factor'' in scattering from gyro-averaging.

\item{{\bf \Alf-YL02}}: \citet{yan.lazarian.02} dismiss the dominant non-resonant pitch-angle scattering term from \citet{chandran00} as spurious, and argue that one should include only the much weaker resonant scattering term (Eq.~17 in \citealt{chandran00}), modified slightly by the factor $\sim \Gamma[13/2,\,( \ell_{\rm turb} /r_{\rm L})^{1/3}\,(c_{s}/c)^{2/3}]$ owing to their different assumed form of the cross-correlation tensor (Eq.~8 in \citealt{yan.lazarian.02}). This gives $f_{\rm turb} \sim 7\times10^{-4}\,(c/v_{A})^{5/11}\,(\ell_{\rm turb}/r_{\rm L})^{9/11}\sim 3\times10^{6}\,n_{1}^{0.2}\,B_{\rm \mu G}^{0.4}\,\ell_{\rm turb}^{0.8}\,\CRegy^{-0.8}\,\mathcal{M}_{A}^{-2.5}$. This is so large that it produces totally negligible confinement/scattering.

\item{{\bf Fast-YL04}}: \citet{yan.lazarian.04:cr.scattering.fast.modes,yan.lazarian.2008:cr.propagation.with.streaming} argue that fast magnetosonic modes could dominate CR scattering despite most mode angles $\hat{\bf k}$ being strongly damped below wavelengths $\lambda_{\rm damp} \gg r_{\rm L}$, if (1) they are isotropic with a shallow power spectrum, (2) non-resonance broadening enhances transit-time damping (TTD), and (3) gyro-resonant ($k\approx k_{\|} \approx 1/r_{\rm L}$)  parallel fast modes  with $\hat{\bf k} \approx \bhat$ are undamped. Using their assumptions (see Appendix~\ref{sec:fast}), $\lambda_{\rm damp}$ is then set by the maximum of either collisionless (Landau) or viscous damping: when collisionless dominates we can approximate $f_{\rm turb} \sim 2\,(\pi\,m_{e}\,\beta/4\,m_{p})^{1/2} \sim 0.04\,\beta^{1/2}$, and when viscous dominates we have $f_{\rm turb} \sim \mathcal{M}_{A}^{5/3}\,{\rm Re}^{-1/3}\,(\ell_{\rm turb}/r_{\rm L})^{1/6}$, where ${\rm Re}\equiv (\mathcal{M}_{A}\,v_{A}\,\ell_{\rm turb})/\nu_{\rm v}$ is the Reynolds number with $\nu_{\rm v}$ the kinematic viscosity.\footnote{We take $\nu_{\rm v} \sim 10^{18}\,{\rm cm^{2}\,s^{-1}}\,T_{4}^{1/2}\,\rho_{-24}^{-1}\,(0.6\,f_{\rm ion}\,T_{4}^{2} + 300\,f_{\rm neutral})$ to be the sum of Braginskii (dominant in ionized gas) and atomic collisional (dominant in neutral gas) viscosities \citep{spitzer:conductivity}. To interpolate between collisionless/viscous regimes we simply take the maximum $f_{\rm turb}$ defined by either.} However, even given these assumptions, efficient confinement by fast modes requires near fully-ionized gas ($f_{\rm neutral} \ll f_{\rm n,\,0} \approx 0.001\,(n_{1}\,\beta)^{-3/4}\,T_{4}^{1/4}\,(\ell_{\rm turb,\,kpc}\,\CRegy)^{-1/2}$) {\em and}  low $\beta < 1$, otherwise damping of the gyro-resonant fast modes gives extremely large $\kappa$.\footnote{See e.g.\ \citealt{yan.lazarian.04:cr.scattering.fast.modes} who show that any models with $\beta\ge1$, such as their ``hot ionized medium'' (HIM) model, or with non-negligible neutrals, such as their warm neutral (WNM) or cold cloud (CNM or DC) models, give $\kappa_{\|} \gg 10^{33}\,{\rm cm^{2}\,s^{-1}}$.} We approximate these ``cutoffs'' by multiplying $f_{\rm turb}$ by a factor $f_{\rm cut}=\exp{\{ (f_{\rm neutral}/f_{\rm n,\,0})^{4} + (\beta/0.1)^{1.5}\} }$ (see Appendix \ref{sec:fast}). 
 
\item{{\bf Fast-Max}}: If we make the ad-hoc assumption that some other physics contributes large scattering rates at small pitch angles, or simply neglect any damping of gyro-resonant parallel fast modes, then we  approximately obtain the ``Fast-YL04'' model but without the ``cutoff'' terms suppressing scattering where $f_{\rm neutral} \gtrsim 10^{-3}$ or $\beta \gtrsim 1$. We consider this model ($f_{\rm cut}=1$) for the sake of reference, if the fast-mode scattering rates for well-ionized, low-$\beta$ gas were simply applied everywhere in the ISM.

\item{{\bf Fast-Mod}}: \citet{yan.lazarian.04:cr.scattering.fast.modes,yan.lazarian.2008:cr.propagation.with.streaming} make a number of uncertain assumptions in deriving the effect of fast modes. For example, they assume a fast-mode spectrum $\propto k^{-3/2}$, but the simulations in \citet{cho.lazarian:2003.mhd.turb.sims} used to justify this choice are in several cases more consistent with \citet{kolmogorov:turbulence} (K41; $k^{-5/3}$) or even \citet{burgers1973turbulence} (B73; $k^{-2}$) spectra \citep[as others have argued for fast modes in the ISM, e.g.][]{boldyrev:2002.sf.cloud.turb,schmidt:2008.turb.structure.fns,kritsuk:2007.isothermal.turb.stats,pan:boldyrev.structfn.tests,burkhart:2009.mhd.turb.density.stats,hopkins:2012.intermittent.turb.density.pdfs}, the latter of which would give $f_{\rm turb}\sim 1$. They also assume the non-linear TTD terms are ``broadened`` with the maximum possible broadening (given by the driving-scale $\delta{\bf B}/|{\bf B}|$, despite $r_{\rm L} \ll \lambda_{\rm damp} \ll \ell_{\rm turb}$); modifying this would increase $f_{\rm turb}$ by a large (exponential) factor \citep{1975RvGSP..13..547V}. Lacking a more detailed model, we consider a case with $f_{\rm turb}$ equal to the ``Fast-Max'' model times $1000$.

\item{{\bf Iso-K41}}: If we {\em entirely} ignore anisotropy and damping, and extrapolate an isotropic \citet{kolmogorov:turbulence} spectrum from $\ell_{\rm turb}$ to $r_{\rm L}$, we obtain $f_{\rm turb} \sim (r_{\rm L}/\ell_{\rm turb})^{1/3} \sim 0.001\,(\CRegy/B_{\rm \mu G}\,\ell_{\rm turb,\,kpc})^{1/3}$.  This model is not physically motivated, since the anisotropy of magnetized turbulence is well understood and observed in the solar wind \citep{Chen2016a}, but it provides a useful reference.

\end{enumerate}

We have also run a number of additional variations to gain further insight: (vii) assuming fixed $f_{\rm turb}=1$ (i.e.\ assume a GS95 cascade, but {\em ignore} the effect of anisotropy on scattering calculated by \citet{chandran00} and \citet{yan.lazarian.02}); (viii) fixed $f_{\rm turb}=1000$ (not motivated by a specific model, but for reference); (ix) variations of model ``Fast-YL04'' neglecting all damping (even more extreme than ``Iso-K41''), so $f_{\rm turb} \sim (r_{\rm L}/\ell_{\rm turb})^{1/2} \sim 10^{-5.5}\,(\CRegy/B_{\rm \mu G}\,\ell_{\rm turb,\,kpc})^{1/2}$; (x) variation of ``Fast-YL04''/``Fast-Max'' neglecting all but collisionless damping (similar to ``Iso-K41''); (xi) several variants of ``Iso-K41'' as proposed in the literature, e.g.\ that in \citet{snodin:cr.diffusion.in.tangled.b.fields} which gives $f_{\rm turb} \sim 0.003 + 0.3\,(r_{\rm L}/\ell_{\rm turb})^{1/3}$; (xii) versions of models (i)-(v) with an additional streaming with both $v_{\rm st}=v_{A}^{\rm ideal}$ and $v_{A}^{\rm ion}$; (xiii) versions of (i)-(v) where we assume a \citet{kolmogorov:turbulence} or \citet{burgers1973turbulence} spectrum on large (simulation-resolved) scales of $\mathcal{M}_{A}>1$, down to the scale $\ell_{A}$ where $\mathcal{M}_{A}[\ell_{A}]=1$, then the specified spectrum below this scale (as opposed to a single spectrum on all scales), which modifies $f_{\rm turb}$ by, at most, one power of $\mathcal{M}_{A}[\ell_{\rm turb}] \sim 1$. 

Note that in all of the models in this section except ``Fast-YL04,'' we neglect ion-neutral damping/ambipolar diffusion in gas with $f_{\rm ion} \ll 1$, which will suppress scattering (increasing $f_{\rm turb}$) substantially in molecular clouds. However, we do consider ``fast transport in neutral gas'' elsewhere, and in some of the variants here.

\subsection{Self-Confinement Scenarios}
\label{sec:self.confinement}

In the self-confinement picture, $|\dBprl|$ is dominated by fluctuations from plasma instabilities self-excited by the CR flux. CRs stream down their number density/pressure gradient with speed $\bar{v}_{\rm st}$, but this excites gyro-resonant \Alf\ waves ($k_{\|}\sim k_{\rm L}$) with growth rate  $\Gamma_{\rm grow} \sim \Omega\,(\CRegy\,n_{\rm  cr}/n_{i})\,(\bar{v}_{\rm st}/v_{A}-1) \sim v_{A}\,[|{\bf F}|-v_{A}\,h_{\rm cr}] / (e_{\rm B}\,c\,r_{L})$,\footnote{\changedtext{Crudely, the \citet{kulsrud.1969:streaming.instability} gyro-resonant streaming instability has linear-theory growth rate: 
\begin{align}
\Gamma_{\rm grow} 
\nonumber &\sim \Omega\,\left(\frac{\CRegy\,n_{\rm  cr}}{n_{i}}\right)\,\left(\frac{\bar{v}_{\rm st}}{v_{A}}-1\right)   \sim \Omega\,\left(\frac{e_{\rm cr}}{m_{p}\,c^{2}}\right)\,\left(\frac{m_{p}}{\rho}\right)\,\left(\frac{|{\bf F}|-v_{A}\,h_{\rm cr}}{v_{A}\,e_{\rm cr} }\right) \\
&\sim \Omega\,\left( \frac{v_{A}}{c} \right)\,\left( \frac{|{\bf F}|-v_{A}\,h_{\rm cr}}{e_{\rm  B}\,c} \right)
 \sim v_{A}\,\left(\frac{|{\bf F}|-v_{A}\,h_{\rm cr}}{e_{\rm B}\,c\,r_{L}} \right),
\end{align} 
using $e_{\rm cr} \sim n_{\rm cr}\,\CRegy\,m_{p}\,c^{2}$ with $\rho \sim n_{i}\,m_{p}$, $e_{B} \sim \rho\,v_{A}^{2}$, and $\bar{v}_{\rm st} \sim |{\bf F}| /h_{\rm cr}$.}} which in turn scatter the CRs (suppressing ${\bf F}$). A local quasi-steady-state arises in which this growth is balanced by damping of these gyro-resonant waves with rate $\Gamma_{\rm damp}$, giving $\Gamma_{\rm grow}\approx \Gamma_{\rm damp}$ or $|{\bf F}| - v_{A}\,h_{\rm cr} = \kappa_{\|}\,|\nabla_{\|} e_{\rm cr}| \sim \Gamma_{\rm damp}\,(e_{\rm B}\,c\,r_{\rm L}/v_{A})$, i.e.\ CR transport with: 
\begin{align}
\label{eqn:kappa.self.confinement} \frac{\kappa_{\|}}{c\,r_{\rm L}} &\approx \frac{16}{3\pi}\,\left( \frac{\ell_{\rm cr}\,\Gamma_{\rm eff}}{v_{A}} \right)\,\left( \frac{e_{\rm B}}{e_{\rm cr}} \right)\,f_{\rm QLT}\ \ \ \ , \ \ \ \ \ v_{\rm st}\approx v_{A} \\
\nonumber \kappa_{\|} &\sim 6\times10^{26}\,{\rm cm^{2}\,s^{-1}}\,\frac{\CRegy\,\Gamma_{-11}\,\ell_{\rm cr,\,kpc}\,f_{\rm ion}^{1/2}\,n_{1}^{1/2}\,f_{\rm QLT}}{e_{\rm cr,\,eV}}
\end{align}
where $e_{\rm B}\equiv |{\bf B}|^{2}/8\pi$ is  the magnetic energy density, $f_{\rm QLT}$ is a factor we insert to parameterize any deviations from the quasi-linear derivation above, and $\Gamma_{\rm eff} \approx \Gamma_{\rm in} + \Gamma_{\rm turb} +  \Gamma_{\rm LL} + \langle \Gamma_{\rm NLL} \rangle + \Gamma_{\rm other}$ represents the damping rate of gyro-resonant \Alf\ waves (i.e.\ $\partial |\delta{\bf B}|^{2}/\partial t \sim -\Gamma_{\rm eff}\,|\delta {\bf B}|^{2}$), here de-composed  into ion-neutral ($\Gamma_{\rm in}$), turbulent ($\Gamma_{\rm turb}$),  linear Landau ($\Gamma_{\rm LL}$), non-linear Landau ($\Gamma_{\rm NLL}$), and ``other'' ($\Gamma_{\rm other}$) terms 
\citep[see e.g.][]{skilling:1971.cr.diffusion,holman:1979.cr.streaming.speed,kulsrud:plasma.astro.book,yan.lazarian.2008:cr.propagation.with.streaming,ensslin:2011.cr.transport.clusters,wiener:cr.supersonic.streaming.deriv,wiener:2017.cr.streaming.winds}. A derivation of Eq.~\ref{eqn:kappa.self.confinement} is given in Appendix~\ref{sec:deriv}, and expressions for each of the $\Gamma$ are given in Appendix~\ref{sec:damping}. In the latter equality, $\ell_{\rm cr,\,kpc}\equiv \ell_{\rm cr}/{\rm kpc}$, $e_{\rm cr,\,eV}\equiv  e_{\rm cr}/{\rm eV\,cm^{-3}}$, $\Gamma_{-11}\equiv \Gamma_{\rm eff}/10^{-11}\,s^{-1}$. Per \S~\ref{sec:methods:cr.transport} we can combine the streaming+diffusion terms into a ``pure streaming'' expression\footnote{It is also common to see Eq.~\ref{eqn:vstream.self.confinement} written in the form 
\begin{align}
\bar{v}_{\rm st} &\rightarrow v_{A}\,\left[ 1 + \frac{4\,c\,r_{\rm L}\,\Gamma_{\rm eff}\,e_{\rm B}}{\pi\,v_{A}^{2}\,e_{\rm cr}}\right] = v_{A}\,\left[ 1 + \frac{2}{\CRegy\,\pi}\,\frac{\Gamma_{\rm eff}}{\Omega}\,\frac{n_{\rm ion}}{n_{\rm cr}}  \right]
\end{align}
where $e_{\rm cr} \equiv \CRegy\,\mu\,n_{\rm cr}\,c^{2}$, $\rho_{\rm ion} = \mu\,n_{\rm ion}$, $n_{\rm ion}$ and $n_{\rm cr}$ are the ion and CR number densities. This form is less useful for our purposes, however.}
 with $v_{\rm st}  \rightarrow \bar{v}_{\rm st} = v_{A} + \kappa_{\|}/(\gamma_{\rm cr}\,\ell_{\rm cr})$:
\begin{align}
\label{eqn:vstream.self.confinement} \bar{v}_{\rm st} &\rightarrow v_{A}\,\left[ 1 + \frac{4\,c\,r_{\rm L}\,\Gamma_{\rm eff}\,e_{\rm B}\,f_{\rm QLT}}{\pi\,v_{A}^{2}\,e_{\rm cr}}\right] \\ 
\nonumber & \sim v_{A}\,\left[ 1 + \frac{0.4\,\CRegy\,\Gamma_{-11}\,f_{\rm ion}\,n_{1}\,f_{\rm QLT}}{B_{\rm \mu G}\,e_{\rm cr,\,eV}} \right]
\end{align}
Now our uncertainty in $\kappa_{\ast}$ is encapsulated in the damping rates $\Gamma$.

We stress that although we can (per \S~\ref{sec:methods:cr.transport}) write the CR transport equations in terms of ``diffusion+streaming'' coefficients (Eq.~\ref{eqn:kappa.self.confinement}) or ``pure (super-\Alf{ic}) streaming'' (Eq.~\ref{eqn:vstream.self.confinement}), the behavior of Eqs.~\ref{eqn:kappa.self.confinement}-\ref{eqn:vstream.self.confinement} is distinct from either a traditional ``pure diffusion'' (constant-$\kappa$) or ``pure-streaming'' (constant-$v_{\rm st}$) equation, because the coefficients themselves depend on $e_{\rm cr}$ and its gradient (see \S~\ref{sec:deriv:behavior}).

\subsubsection{Model Variant: Choice of \Alf\ Speed}
\label{sec:self.confinement.alfven.speed}

The \Alf\ speed of interest in Eqs.~\ref{eqn:kappa.self.confinement}-\ref{eqn:vstream.self.confinement} is that of the gyro-resonant modes, which as noted in  \S~\ref{sec:methods:cr.transport} should naively follow the {\em ion} \Alf\ speed $v_{A}^{\rm ion}=f_{\rm ion}^{-1/2}\,v_{A}^{\rm ideal}$ in partially-neutral gas. In our ``default'' self-confinement model we therefore adopt $v_{A}=v_{A}^{\rm ion}$ in Eq.~\ref{eqn:kappa.self.confinement} (consistency requires the same $v_{A}$ appear in the ``streaming loss'' term $\Lambda_{\rm st}=v_{A}\,|\nabla_{\|}P_{\rm cr}|$). But while the gyro-resonant wave frequencies are un-ambiguously larger than ion-neutral collision frequencies in GMCs, other aspects of the assumptions used to derive Eqs.~\ref{eqn:kappa.self.confinement}-\ref{eqn:vstream.self.confinement} (e.g.\ how to treat gas advection terms and boosts to/from the frame of the fluid, and how CRs enter the gas momentum equation) implicitly assume the ``gas frame'' and ``magnetic-field frame'' are the same (which is true on large scales even in GMCs, but breaks down at the gyro-resonant scales if $v_{\rm A}^{\rm ion} \gg v_{\rm A}^{\rm ideal}$). Also other timescales (like the CR travel and scattering times) are much longer than ion-neutral collision times. At a fundamental level, knowing how different terms are modified in this limit requires re-deriving CR fluid models such as \citet{thomas.pfrommer.18:alfven.reg.cr.transport} for a three-fluid (CR, ion, neutral) system. Lacking this, we simply compare model variants where we assume  ideal MHD scalings, so $v_{A}=v_{A}^{\rm ideal}$ in Eq.~\ref{eqn:kappa.self.confinement} and $\Lambda_{\rm st}$.

\subsubsection{Model Variant: Non-Equilibrium Description}
\label{sec:non.equilibrium}

Recently, \citet{zweibel:cr.feedback.review} and \citet{thomas.pfrommer.18:alfven.reg.cr.transport} attempted to derive non-equilibrium ``macroscopic'' dynamical equations for $|\dBprl|$, $\kappa$, and $v_{\rm st}$, accounting for un-resolved gyro-resonant waves by explicitly evolving a sub-grid energy density ($e_{A\pm} \sim |\dBprl|^{2}/4\pi$) or wave spectrum propagating in the $\pm \bhat$ directions. We have implemented the full set of equations from \citet{thomas.pfrommer.18:alfven.reg.cr.transport} and compare it to our default ``local equilibrium'' assumption here. Appendix~\ref{sec:deriv} details the complete set of modifications to our default equations, but the important difference is that $\kappa_{\ast}$ is replaced with the {\em explicitly-evolved} diffusivities $\kappa_{\pm}/(c\,r_{\rm L}) \approx (16/9\pi)\,(e_{\rm B}/e_{A\pm}) \sim (1/3)\,|{\bf B}|^{2}/|\dBprl|^{2}$, and the scattering term ${\bf F}/3\kappa_{\ast}$   becomes ${\bf g}_{+}+{\bf g}_{-}$ in the CR flux equation (Eq.~\ref{eqn:flux}). The \Alf-wave energy densities evolve as   $\partial{e}_{A\pm}/\partial t = \pm {\bf v}_{A}\cdot{\bf g}_{\pm} - \Gamma_{\rm eff}\,e_{A\pm}$, where ${\bf g}_{\pm} \equiv ({\bf F}\mp{\bf v}_{A}\,h_{\rm cr})/3\,\kappa_{\pm}$ and ${\bf v}_{A}\cdot {\bf g}_{\pm}$ represents growth from the gyro-resonant instability. In Appendix~\ref{sec:deriv}, we show that when the \Alf\ energy subsystem reaches local steady-state ($\partial e_{A\pm}/\partial t \rightarrow 0$), which occurs on short timescales $\sim \Gamma^{-1}$, the non-equilibrium system reduces to our default CR evolution equations, with $\kappa_{\|}$ and $v_{\rm st}$ following Eq.~\ref{eqn:kappa.self.confinement}.

\subsubsection{Model Variant: CR Energy}
\label{sec:cr.energy}

We can also vary the effective CR energy $\CRegy$ ($=1$\,GeV in our default) assumed in our single-bin approximation. This should represent an effective energy containing most of the CR pressure, but that could vary between $\sim 0.5-10\,$GeV, in principle, given present observational and theoretical constraints. We have run several variants assuming $\CRegy=0.1$ or $10$. However, note that given the damping rates in \S~\ref{sec:damping}, $\kappa$ and $v_{\rm st}$ are either independent of $\CRegy$ (depending only on $e_{\rm cr}$), or scale as $\CRegy^{1/2}$ at most. Thus, even order-of-magnitude variation in $\CRegy$ produces only factor $\sim2-3$ differences in $\kappa_{\rm eff}$.

\subsubsection{Model Variant: Different Growth or Scattering Rates}
\label{sec:define.fQLT}

In deriving Eq.~\ref{eqn:kappa.self.confinement} (see also \S~\ref{sec:deriv}), if we either (a) multiply the gyro-resonant \Alf-wave damping rates $\Gamma_{\rm eff}$ by a factor $f$; (b) divide the effective scattering rate $\nu$ for a given $|\dBprl|$ by $f$ (or equivalently multiply the timescale for those waves to isotropize the CR distribution function by $f$); or (c) divide the growth rate of the gyro-resonant modes $\Gamma_{\rm grow}$ by $f$, then $\kappa_{\|}$ in Eq.~\ref{eqn:kappa.self.confinement} is multiplied by $f$. We call this `fudge factor'' $f_{\rm QLT}$,  which could have its  physical origins in any (or a combination) of the aforementioned effects. Lacking any particular model for $f_{\rm QLT}$, we have simply run simulations with $f_{\rm QLT} = 1,\,6,\,100,\,1000$ ($=1$ is our default).

\subsubsection{Model Variant: Turbulent Cascade Assumptions}
\label{sec:define.fcas}

While there is relatively little ambiguity in the ion-neutral damping rate $\Gamma_{\rm in}$, and we will show the non-linear Landau damping $\Gamma_{\rm NLL}$ only dominates in the ISM in models which are excluded by observations, both the ``turbulent'' ($\Gamma_{\rm turb}$) and  ``linear Landau'' ($\Gamma_{\rm LL}$) damping rates scale with the turbulent dissipation/cascade timescale $t_{\rm cas}$ at wavelengths $\sim r_{\rm L}$, which is not well-constrained. In \S~\ref{sec:damping}, we detail the default model, which, following \citet{farmer.goldreich.04}, assumes a K41 cascade on super-\Alf{ic} scales and a GS95 cascade  on scales $<\ell_{\rm A}$ ($\ell_{\rm A}$ is the \Alf\ scale where $\delta v_{\rm turb}(\ell_{A}) \sim v_{A}$). This gives $\Gamma_{\rm turb}=v_{A}^{\rm ideal}/(r_{\rm L}\,\ell_{A})^{1/2}\,f_{\rm cas}$ (with $\Gamma_{\rm LL}\approx 0.4\,\beta\,\Gamma_{\rm turb}$ scaling proportionally), where $f_{\rm cas}=1$ for these default assumptions. However if we consider different cascade models, we obtain correspondingly different $f_{\rm cas}$; moreover the exact damping rates will depend on the specific temporal and spatial structure of the turbulent field on these micro-scales, so any analytic model for $\Gamma_{\rm turb}$ is an order-of-magnitude average estimate (where $f_{\rm cas}$ parameterizes our ignorance). 

Our default model assumes $f_{\rm cas}=1$. We consider several variant assumptions, including: (1-3) arbitrarily increasing $f_{\rm cas}=5,\,50,\,500$; (4) assuming a supersonic \citet{burgers1973turbulence} spectrum at scales $>\ell_{\rm A}$ instead of K41, giving $f_{\rm cas}={\rm MIN}(1,\,\mathcal{M}_{A}^{-1/2})$; (5) assuming a ``dynamically aligned'' $\sim k^{-3/2}$ spectrum (\citealt{Boldyrev2006}; see also \citealt{iroshnikov:1963.ik.aniso.turb,kraichnan:1965.ik.aniso.turb}) instead of GS95 below $\ell_{A}$, giving $f_{\rm cas}=(\ell_{\rm turb}/r_{\rm L})^{1/10}$; (6) assuming a pure (isotropic) K41 cascade from the driving scale to $r_{\rm L}$, giving $f_{\rm cas} \approx \mathcal{M}_{A}^{-1/2}\,(\ell_{\rm turb}/r_{\rm L})^{1/6}$ (this is not well-motivated but provides a useful ``upper limit''); (7) assuming the multi-component cascade model from \citet{lazarian:2016.cr.wave.damping} which adopts isotropic K41 for $\ell>\ell_{A}$ with a transition between a ``weak'' cascade with form following \citet{1981PhFl...24..825M,1994ApJ...432..612S} on large scales to a GS95 cascade on smaller scales, giving $f_{\rm cas}={\rm MIN}[\mathcal{M}_{A}^{1/2},\,\mathcal{M}_{A}^{7/6}\,(\ell_{\rm turb}/r_{\rm L})^{1/6}]$ when $\mathcal{M}_{A}<1$ and $f_{\rm cas} = {\rm MIN}[1,\,\mathcal{M}_{A}^{-1/2}\,(\ell_{\rm turb}/r_{\rm L})^{1/6}]$ when $\mathcal{M}_{A}\ge1$.

\subsection{Combined Extrinsic Turbulence and Self-Confinement Models}
\label{sec:combo}

Scattering by self-excited and extrinsic fluctuations are not mutually exclusive. Their non-linear interplay is poorly understood, but in quasi-linear theory the scattering rates should add linearly \citep[see][]{zweibel:cr.feedback.review}, giving $\kappa_{\|}^{-1} \sim \kappa_{\rm self}^{-1} + \kappa_{\rm extrinsic}^{-1}$. We have therefore also run simulations adopting $v_{\rm st}=v_{A}$, $\kappa_{\|}^{-1}=\kappa_{\|,\,{\rm self}}^{-1} + \kappa_{\|,\,{\rm turb}}^{-1}$ where $\kappa_{\|,\,{\rm self}}$ follows Eq.~\ref{eqn:kappa.self.confinement} and $\kappa_{\|,\,\rm turb}$ follows Eq.~\ref{eqn:model.extrinsic}, with several combinations of the ``variant'' model assumptions. Usually, one model (typically the extrinsic turbulence model) has much-larger $\kappa$ (much lower scattering rate), so the prediction simply becomes identical to that of the model with the lower $\kappa$ (higher $\nu$). Even in the rare cases where the two contribute comparably (e.g.\ using ``Fast-Max'' for $f_{\rm turb}$ and $f_{\rm cas}=500$), this simply gives similar behavior to both ``individual'' models and so  does not change any of our conclusions regarding which scattering processes are observationally allowed. We therefore discuss these only briefly and defer a more detailed study to the future work.

\section{Results}
\label{sec:results}

\begin{figure*}
    \plotside{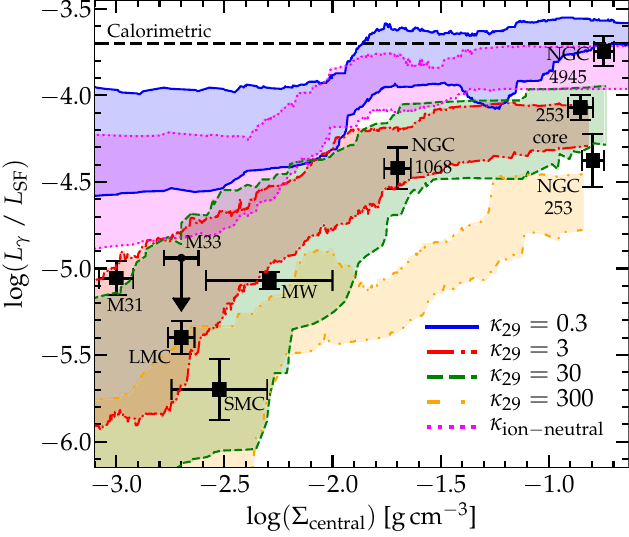}{0.34}\plotside{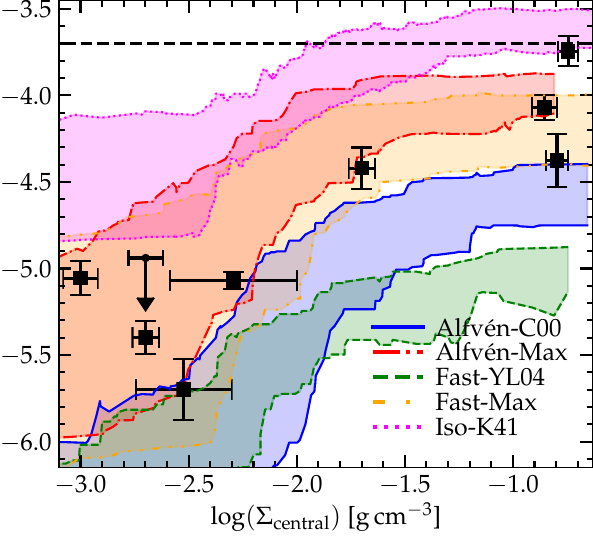}{0.32}\plotside{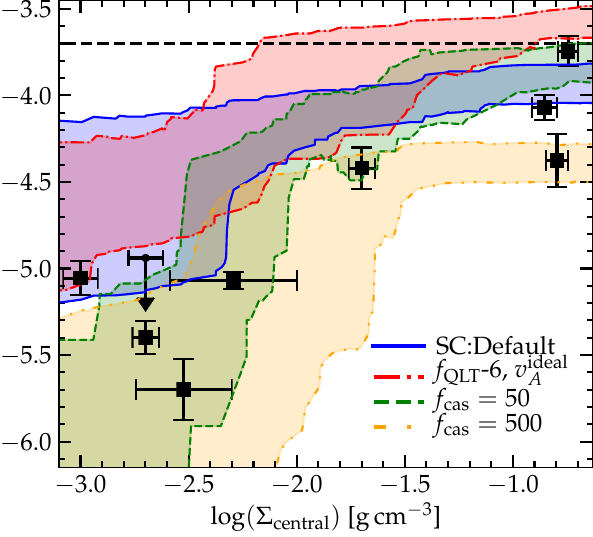}{0.32}
    \vspace{-0.2cm}
        \caption{Predicted ratio of $\gamma$-ray luminosity from hadronic collisions ($L_{\gamma}$; see \S~\ref{sec:Lgamma.obs.compare}) to luminosity from star formation/massive stars ($L_{\rm SF}$), as a function of galaxy central gas surface density ($\Sigma_{\rm central}$). Shaded range shows $1\,\sigma$ ($\sim68\%$) inclusion interval of all points measured at uniform time intervals at $z<1$ (for all {\bf m11i}, {\bf m11f}, {\bf m12i}). Dashed horizontal line is the steady-state calorimetric limit. Black squares compare observations (upper limit is M33). Panel compare subsets of transport models (Table~\ref{tbl:transport}). 
        {\em Left:} Constant-diffusivity (CD; \S~\ref{sec:constant.diffusivity}) models. Models with $\kappa_{29} = \kappa_{\|}/10^{29}\,{\rm cm^{2}\,s^{-1}} \sim 3-30$ agree well with observations. Lower (higher) $\kappa$ over (under) predicts $L_{\gamma}$. Model ``$\kappa_{\rm ion-neutral}$''  with $\kappa_{29}=3$ ($0.1$) in neutral (ionized) gas only slightly decreases $L_{\gamma}$, relative to models with $\kappa_{\rm 29}<1$ everywhere.
        {\em Center:} ET models. Expected scattering by \Alf{ic} or fast-mode ET (\Alf-C00, Fast-YL04) is sub-dominant (under-predicting $L_{\gamma}$), although scattering by fast modes could be important ($L_{\gamma}$ similar to observed) under some extreme assumptions (\Alf-Max, Fast-Max). Model ``Iso-K41'' ignores anisotropy and damping of ET, and over-predicts $L_{\gamma}$.
        {\em Right:} SC models. ``Default'' SC assumptions over-predict $L_{\gamma}$; this is only weakly-influenced by the assumed CR energy ($\sim1-10\,$GeV), choice of \Alf\ speed (\S~\ref{sec:alfven.speed}), and other details. Multiplying the turbulent damping rates by factors $f_{\rm cas}\sim 50-500$, gives good agreement with the observed $L_{\gamma}$. 
    \label{fig:Lgamma}}
\end{figure*}

\begin{figure}
    \plotone{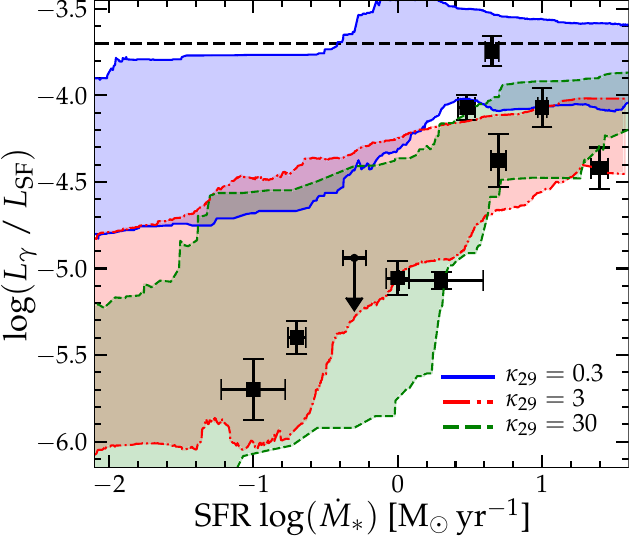}{0.49}\plotone{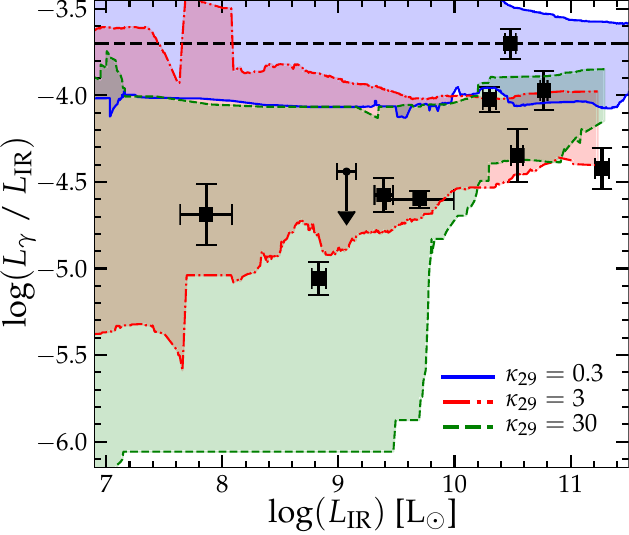}{0.49}
    \vspace{-0.2cm}
    \caption{As Fig.~\ref{fig:Lgamma}, comparing $L_{\gamma}/L_{\rm SF}$ versus the galaxy-integrated SFR $\dot{M}_{\ast}$ ({\em left}) or infrared (IR; $8-1000\,\mu{\rm m}$) luminosity $L_{\gamma}/L_{\rm IR}$ versus $L_{\rm IR}$ ({\em right}; obtained by ray-tracing from each star to a mock observer at infinity assuming a MW-like extinction curve with a constant dust-to-metals ratio equal to the MW value, following \citealt{hopkins:lifetimes.letter}). Comparing $L_{\gamma}/L_{\rm SF}$ versus SFR shows essentially identical behavior to $L_{\gamma}/L_{\rm SF}$ versus $\Sigma_{\rm central}$ in Fig.~\ref{fig:Lgamma}. 
    Comparing $L_{\gamma}/L_{\rm IR}$ is less useful: in dwarfs, $L_{\rm IR}/L_{\rm SF}$ declines proportional to the optical/UV attenuation $\tau_{\rm OUV} \approx \kappa_{\rm OUV}\,\Sigma_{\rm central}$, itself proportional to $\Sigma_{\rm central}$, while $L_{\gamma}/L_{\rm SF}$ similarly scales with $\sim \Sigma_{\rm central}$, so their ratio varies more weakly ($\propto L_{\rm IR}^{0.3}$) and models overlap more heavily. These diagnostics do not rule out any models not already ruled out by the comparison in Fig.~\ref{fig:Lgamma}.
    \label{fig:LgammaVsSFRLIR}}
\end{figure}

\begin{figure*}
    \plotside{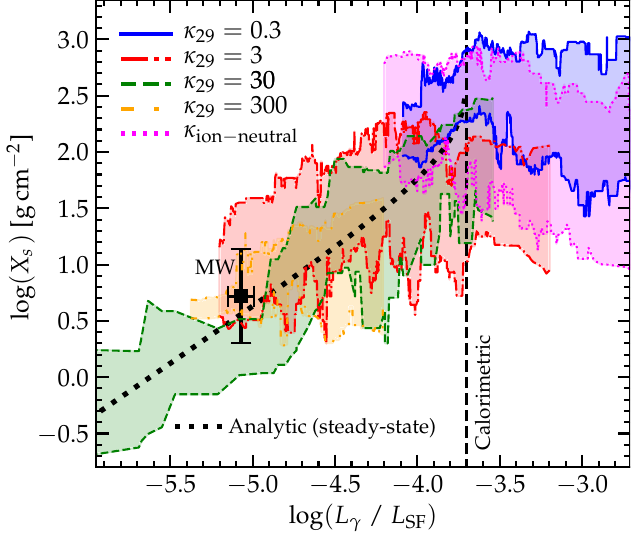}{0.34}\hspace{-0.3cm}\plotside{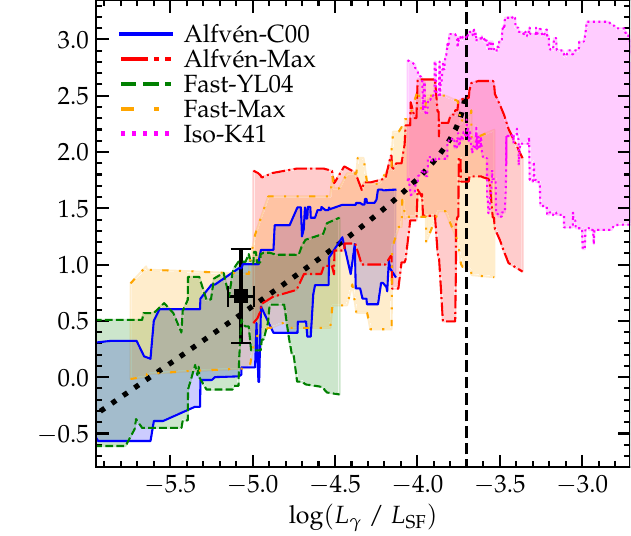}{0.34}\hspace{-0.3cm}\plotside{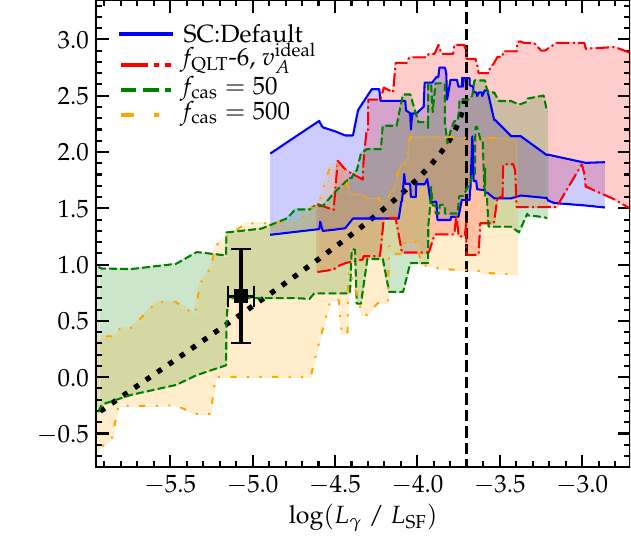}{0.34}
    \vspace{-0.2cm}
        \caption{$\gamma$-ray luminosity relative to star formation ($L_{\gamma}/L_{\rm SF}$, models and shaded ranges as Fig.~\ref{fig:Lgamma}) versus CR grammage $X_{s}$ calculated for an observer far from the galaxy center, at all simulation times $z<3$. We label the calorimetric limit and the analytic relation between $X_{s}$ and $L_{\gamma}/L_{\rm SF}$ for a homogeneous, steady-state system (Eqs.~\ref{eqn:grammage.vs.lcalor}-\ref{eqn:grammage.vs.lcalor.exact}). Regardless of the CR transport model, the simulations follow $X_{s} \sim 100\,{\rm g\,cm^{-2}}\,(L_{\gamma}/L_{\rm SF})$ for $L_{\gamma} < L_{\rm SF}$, consistent with the MW observations (square labeled). At $L_{\gamma} > L_{\rm calor}$, $X_{s}$ saturates (any CRs with higher grammage are lost to collisions before escaping to reach the ``observer''). The scatter is primarily driven by short-timescale ($\sim 10\,$Myr) variations in SFR (i.e.\ $L_{\rm SF}$) and (to a lesser extent) in $L_{\gamma}$ and $X_{s}$ driven by ISM clumpiness.
    \label{fig:Lgamma.vs.grammage}}
\end{figure*}

\subsection{Effective Diffusivities \&\ Observational Constraints}

\subsubsection{Effective Diffusivities}

Fig.~\ref{fig:kappa.vs.model} compares the effective diffusivities $\kappa_{\rm eff} \equiv |{\bf F}|/|\nabla_{\|}e_{\rm cr}|$ from a representative subset of the models in \S~\ref{sec:models}, at $z=0$ in a dwarf ({\bf m11i}), intermediate-mass ({\bf m11f}), and MW-mass ({\bf m12i}) galaxy. Among the ET models, as expected, models with larger $f_{\rm turb}$ produce larger $\kappa_{\rm eff}$. Some (e.g.\ model ``\Alf-YL02'') produce such high $\kappa_{\rm eff} \gg 10^{34}\,{\rm cm^{2}\,s^{-1}}$ they fall above the plot. Models which ignore anisotropy and/or damping (e.g.\ ``Iso-K41'') produce very low $\kappa_{\rm eff}$; the ``Fast-NoDamp'' variant ignoring damping entirely produces $\kappa_{\rm eff} \ll 10^{26}\,{\rm cm^{2}\,s^{-1}}$, well below the plotted range. In the SC models, $\kappa_{\rm eff}$ is not strongly sensitive to model variations such as the choice of \Alf\ speed or equilibrium vs.\ non-equilibrium description, but varies systematically with the strength of turbulent damping (increasing with $f_{\rm cas}$), in an analogous (inverted) manner to the ET models. 

There are few other universal systematic trends: (1) $\kappa_{\rm eff}$ tends to rise with galacto-centric radius, but the strength of this rise varies widely. (2) There are some radial fluctuations at a given time in $\kappa_{\rm eff}$: there is actually considerably more small-scale scatter than this plot suggests, which depends on how we weight the ``mean'' $\kappa_{\rm eff}$, explored below (Fig.~\ref{fig:kappa.ecr.vs.weight}). (3) In many self-confinement (but not extrinsic turbulence) models, the diffusivities are systematically higher in lower-mass dwarf galaxies (with lower $\rho$, $|{\bf B}|$, $e_{\rm cr}$, etc.).

Some models run are not plotted in Fig.~\ref{fig:kappa.vs.model}, as they simply interpolate between the models shown or give nearly-identical results. For example, increasing $\CRegy$ to $\sim10$ in the self-confinement models (\S~\ref{sec:cr.energy}) simply increases $\kappa_{\rm eff}$ by a factor $\sim 1.5-3$ at large radii (and less at $\lesssim$\,kpc, where ion-neutral damping dominates).

\subsubsection{$\gamma$-Ray Luminosities}
\label{sec:Lgamma.obs.compare}

Fig.~\ref{fig:Lgamma} compares the predicted $\sim$\,GeV $\gamma$-ray emission from each simulation. This was studied in \paperonetwo\ in detail and we follow their methodology, mimicking the compiled (plotted) observations from \citet{lacki:2011.cosmic.ray.sub.calorimetric,tang:2014.ngc.2146.proton.calorimeter,griffin:2016.arp220.detection.gammarays,fu:2017.m33.revised.cr.upper.limit,wjac:2017.4945.gamma.rays,wang:2018.starbursts.are.proton.calorimeters,lopez:2018.smc.below.calorimetric.crs}. Briefly, we assume $5/6$ of the collisional hadronic losses go to pions, with branching ratio of $1/3$ to $\pi^{0}$ that decay to $\gamma$-rays with a spectrum giving $\sim 70\%$ of the energy at $>1\,$GeV \citep{guo.oh:cosmic.rays,chan:2018.cosmicray.fire.gammaray}, and integrate this within apertures ($\sim5-10\,$kpc) matched to the observations. We similarly compute the central ($\lesssim 2-5$\,kpc, taken as $1/2$ the half-mass radius) projected gas surface density $\Sigma_{\rm central}$, and the luminosity from young/massive stars $L_{\rm SF}$ (using all stars $<100\,$Myr old, convolved with appropriate stellar population synthesis for their ages and metallicities). The ``calorimetric limit'' line denotes the ratio $L_{\gamma}/L_{\rm sf} = L_{\rm calor}/L_{\rm sf} \sim 2\times10^{-4}$, which corresponds to the assumption that all CR energy injected by SNe is lost collisionally in steady state with a uniform time-constant SFR and SNe rate. 

First, let us consider the constant-diffusivity models. These models and variants are the main focus of \paperonetwo\ (with additional simulations and more widely-varied assumptions related to streaming and numerics).  We echo their conclusion: $\kappa_{29} \sim 3-30$ is required to reproduce the observations, with lower-$\kappa_{29} \lesssim 1$ producing near-calorimetric predictions even in dwarfs, and $\kappa_{29} \gtrsim 100$ under-predicting $L_{\gamma}$. We also see model $\kappa_{\rm ion-neutral}$ rather severely over-predicts $L_{\gamma}$, comparable to models with constant $\kappa_{29}\sim 0.5$. We also note (see \paperonetwo\ for further discussion) that adding additional trans-sonic streaming (with $v_{\rm st} \sim v_{A}^{\rm ideal}$ or $\sim v_{A}^{\rm ion}$) makes only a small $\sim 10\%$ difference to $L_{\gamma}$.

Next, compare ET models: as expected, those with systematically higher $\kappa_{\rm eff}$ in Fig.~\ref{fig:kappa.vs.model} produce lower $L_{\gamma}$. Model ``\Alf-C00'' ((i) in \S~\ref{sec:extrinsic}) and others with $f_{\rm turb} \gtrsim 100$ in the WIM ($\kappa_{29} \gg 100$) under-predict $L_{\gamma}$: this includes models ``\Alf-YL02'' (ii) and ``Fast-Mod'' (iv), which are not shown but fall below the plotted range, and $f_{\rm turb}=1000$ (vii), which is similar to ``\Alf-C00'' (as expected). Models with $f_{\rm turb} \ll 0.01$, on the other hand, over-produce $L_{\gamma}$, with $\kappa_{29} \lesssim 0.1$ within the galaxy (although $\kappa_{29}$ varies widely in dwarfs). This includes models ``Iso-K41'' (v) and its variants assuming different turbulent spectra or geometries (e.g.\ models (viii), (ix), (x), (xii), not shown but all similar to ``Iso-K41''), which neglect both the dominant turbulent damping terms and anisotropy of small-scale turbulence in the ISM. For $f_{\rm turb}\sim 0.1-10$, $L_{\gamma}$ is broadly similar to observations: this occurs in the ad-hoc ``Fast-Max'' (iii) and ``\Alf-Max'' ($f_{\rm turb}=1$; vi) models. 

We also see that the ``default'' SC model produces excessive $L_{\gamma}$, compared to observations. Varying $v_{\rm st}=v_{A}^{\rm ion}$ versus $v_{A}^{\rm ideal}$ has relatively little effect on this conclusion, as does varying the assumed CR energy from $\CRegy\sim1-10$\,GeV, or adopting non-equilibrium models for $\kappa$ and $v_{\rm st}$. Increasing the turbulent damping rate $f_{\rm cas}$ decreases $L_{\gamma}$, with models where $f_{\rm cas} \sim 30-300$ in agreement with the observations. This includes models that increase $f_{\rm cas}$ by a similar factor assuming a different turbulent spectrum (e.g.\ ``$\Gamma_{\rm damp}$-K41'').

Fig.~\ref{fig:LgammaVsSFRLIR} also plots $L_{\gamma}/L_{\rm SF}$ versus absolute SFR, and $L_{\gamma}/L_{\rm IR}$ versus $L_{\rm IR}$, the total infrared luminosity ($8-1000\,\mu{\rm m}$) computed self-consistently in our simulations by ray-tracing $\sim 100$ lines-of-sight from every star particle (with an input spectrum following the \citealt{starburst99} stellar population models for the same age, metallicity, and mass) through the resolved gas and dust in the simulation, assuming a MW-like extinction curve (adopting SMC-like extinction makes little difference) with constant dust-to-metals ratio $=0.4$ (see \citealt{hopkins:lifetimes.letter}). These give somewhat redundant constraints: the same models are (in)consistent with the data in these projections, but they generally show more overlap in the model predictions and are less theoretically well-motivated (see \S~\ref{sec:params}), so they are less useful for distinguishing models.

\subsubsection{Grammage and Residence Time}
\label{sec:cr.grammage.calc}

As discussed in \paperonetwo, our comparison to the MW point in Fig.~\ref{fig:Lgamma} is essentially equivalent to comparing to the observed grammage in the Galaxy. Specifically, for the MW, quantities like the inferred diffusion coefficient are model dependent: what is most directly constrained by observations like the secondary-to-primary ratios is the effective column density or grammage $X_{s} \equiv \int_{\rm CR\,path}\,\rho_{\rm nuclei}\,d\ell_{\rm CR} = \int_{\rm CR\,path}\,\rho_{\rm gas}\,c\,dt$ integrated over the path of individual CRs from their source locations to the Earth (with $X_{s} \sim 5\,{\rm g\,cm^{-2}}$, or $\sim 3\times10^{24}\,{\rm nucleons\,cm^{-2}}$, measured).\footnote{Note that the measured grammage we compare to is an energy-weighted average around $\sim 1-10\,$GeV, for which typical estimates in the MW give $\sim 2-10\,{\rm g\,cm^{-2}}$ \citep{2014ApJ...786..124C,2016PhRvD..94l3019K,evoli:dragon2.cr.prop,2018AdSpR..62.2731A,2019PrPNP.10903710K}.} If the galaxy is in quasi-steady state with some CR injection rate $\dot{E}_{\rm cr} \propto \dot{E}_{\rm SNe} \propto L_{\rm sf}$ and losses are small ($L_{\gamma} \ll L_{\rm calor}$), then $e_{\rm cr}({\bf x}) \approx \dot{E}_{\rm cr}\,(dt/d^{3}{\bf x})$ at some position $x$ (where $dt/d^{3}{\bf x}$ is the residence time of individual CRs in a differential volume element). Using this and the fact that $L_{\gamma}/L_{\rm calor} = \dot{E}_{\rm coll}/\dot{E}_{\rm cr}$, where $\dot{E}_{\rm coll} = \int d^{3}{\bf x}\,\Lambda_{\rm coll} = \alpha \int n_{\rm n}\,e_{\rm cr}\,d^{3}{\bf x}$ (with $\alpha=5.8\times10^{-16}\,{\rm cm^{3}\,s^{-1}}$ and $n_{\rm n} = \rho_{\rm nuclei}/m_{p}$), we  obtain 
\begin{align}
\label{eqn:grammage.vs.lcalor} X_{s}^{\infty} &\approx 130\,{\rm g\,cm^{-2}}\,\left( \frac{L_{\gamma}}{L_{\rm calor}} \right) &\hfill (L_{\gamma} \ll L_{\rm calor})
\end{align}
or $X_{s}^{\infty} \approx 6\times10^{5}\,{\rm g\,cm^{-2}}\,(L_{\gamma}/L_{\rm sf})$ (where $X_{s}^{\infty}$ is the grammage integrated to infinity or ``escape'').\footnote{As $X_{s}^{\infty}\rightarrow \infty$, obviously $L_{\gamma}/L_{\rm calor} \rightarrow 1$, losses become significant, and the linear scaling $X_{s}^{\infty} \propto L_{\gamma}/L_{\rm calor}$ in Eq.~\ref{eqn:grammage.vs.lcalor} breaks down. If we consider a simple  slab model we can extend this further, giving
\begin{align}
\label{eqn:grammage.vs.lcalor.exact} X_{s}^{\infty} &\approx 130\,{\rm g\,cm^{-2}}\,\ln{\left\{ \frac{1}{1 - L_{\gamma}/L_{\rm calor}} \right\}}.
\end{align}
The simulations do follow this correlation reasonably well for $L_{\gamma}/L_{\rm calor} \lesssim 1$, but owing to clumpiness (non-``slab'' geometric effects) and time variability effects there is no tight correlation once $L_{\gamma}\gtrsim L_{\rm calor}$. However these near-calorimetric systems almost always have $X_{s} \gtrsim 100\,{\rm g\,cm^{-2}}$.}

We have directly confirmed that this is an excellent approximation in any of our simulations which is remotely consistent with the observational constraints, by calculating $X_{s}^{\infty}$ following Lagrangian CR trajectories (Fig.~\ref{fig:Lgamma.vs.grammage}).\footnote{\changedtext{Specifically, we re-run the simulation for a short time $\sim 300\,$Myr near $z\approx 0$, with CR tracer particles probabilistically injected every time a SNe injects CR energy (expected number proportional to CR energy injected), each recording its time of injection. Tracers are deleted stochastically with probability equal to the ratio of total catastrophic losses to total CR energy in a cell each timestep, or can ``jump'' to neighbor gas cells with probability equal to the fractional CR energy flux from their parent cell to the neighbor (similar to the scheme in \citealt{genel:tracer.particle.method}).}} To match the constraints at Earth more directly, we have also explicitly calculated $X_{s}^{(8.1)}$ (or $X_{s,\,\oplus}$), the grammage from sources to random star particles at the solar circle ($8.1\pm0.1\,$kpc in the thin disk midplane, at $z=0$) in several of our transport models (for galaxies {\bf m11f} and {\bf m12i}) and in almost all cases find $X_{s}^{(8.1)} \approx (0.7-0.9)\,X_{s}^{\infty}$ (since this is well outside the effective radius of star formation in our Milky Way) -- a negligible correction compared to other uncertainties here.

We also calculate the {\em true} ``residence time'' $\Delta t_{\rm res}$ of CRs in our simulations by following a random subset of tracer CRs which end up in this mock solar circle at $z=0$, tracing them back to their time of injection. Note that residence time is only well-defined with respect to an observer at a specific location in the galaxy (so we only consider this for our MW-like systems {\bf m11f} and {\bf m12i}), as it diverges for any CRs that escape the galaxy. It also becomes artificially limited by the hadronic loss timescale $\sim 270\,{\rm Myr}\,(0.1\,{\rm cm^{-3}}/n_{\rm gas})$ when collisional losses become dominant (as $L_{\gamma} \rightarrow L_{\rm calor}$): indeed, we confirm that all our models with $\Delta t_{\rm res} \gtrsim (1-2)\times10^{8}\,{\rm yr}$ (consistent with loss times for $n \gtrsim 0.1\,{\rm cm^{-3}}$) have $L_{\gamma} \sim L_{\rm calor}$, and vice versa.\footnote{For example, our ``Iso-K41'' and ``SC:Default'' models (in {\bf m12i}) give estimated median $\Delta t_{\rm res} \sim 2-3\times10^{8}\,{\rm yr}$, but this is primarily limited by hadronic losses in both cases (both have $L_{\gamma} \sim L_{\rm calor}$). If we ignore the losses for our tracer CRs, we obtain the order-of-magnitude larger $\Delta t_{\rm res} \sim 1-4\times10^{9}\,{\rm yr}$.} 

By definition, $\Delta t_{\rm res} = \int_{\rm emission}^{\oplus} dt = X_{s} / (\langle n \rangle\,m_{p}\,c)$ where $\int_{\rm emission}^{\oplus}$ represents the integral from emission to observation at ``Earth'' at $z=0$, $dt$ is the time along an individual CR trajectory, and $\langle n \rangle \equiv m_{p}^{-1}\,(\int \rho\,dt)/(\int dt)$ is a residence-time-weighted average. But in a highly inhomogeneous medium, there is no single $\langle n \rangle$ (and its ``effective'' value depends on the transport model). As a result, there is (as one might expect) a broad range of residence times for CRs at the mock observer (with non-trivial ``tails'' worth further investigation in future work). Considering just the median at each time, we find that for otherwise ``favored'' models (\Alf-Max, Fast-Max, $f_{\rm cas}$-50, $f_{\rm QLT}$-100) we obtain median $\Delta t_{\rm res} \sim 3-50\,{\rm Myr}$ (and for $f_{\rm cas}$-500, $f_{\rm cas}$-K41 we find $\Delta t_{\rm res} \sim 0.5-15\,{\rm Myr}$) in galaxies {\bf m11f} and {\bf m12i} at times where their $\Sigma_{\rm gas}$ is similar to that of the MW in Fig.~\ref{fig:Lgamma}, matching roughly our expectation given the predicted $X_{s}$ and a mean $\langle n \rangle \sim 0.1-1\,{\rm cm^{-3}}$ typical of the ISM dominating the grammage. But in each of these cases a significant (few percent or more) fraction of the population seen at the ``observer'' has had residence times $<1\,$Myr or $>50\,$Myr. All of this is broadly within the range allowed by MW constraints \citep{2007ARNPS..57..285S,2010A&A...516A..66P,2011ApJ...729..106T,2016PhRvL.117i1103A,2018PhRvL.120b1101A,2017PhRvD..95h3007Y,2019PrPNP.10903710K}. On the other hand (as noted above) the models with $L_{\gamma} \sim L_{\rm calor}$ all have $\Delta t_{\rm res} \gtrsim 100$\,Myr (clearly ruled out), while those with $L_{\gamma}$ much less than observed (e.g.\ ``\Alf-C00'') all have $\Delta t_{\rm res} \lesssim 1\,$Myr.

\begin{figure*}
    \plotside{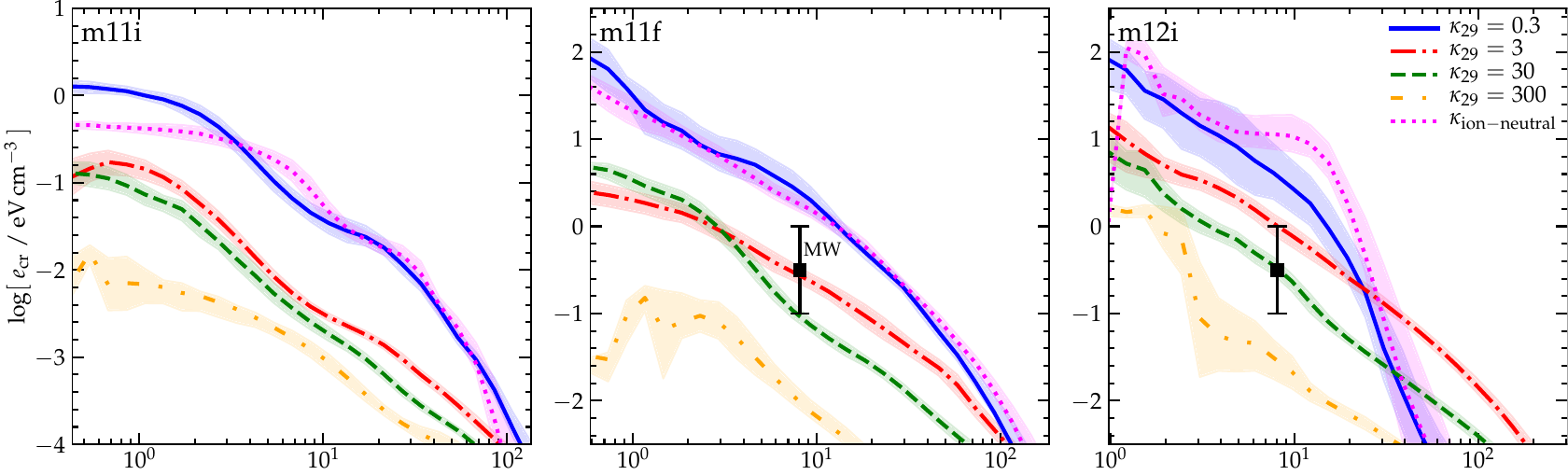}{0.95}
    \plotside{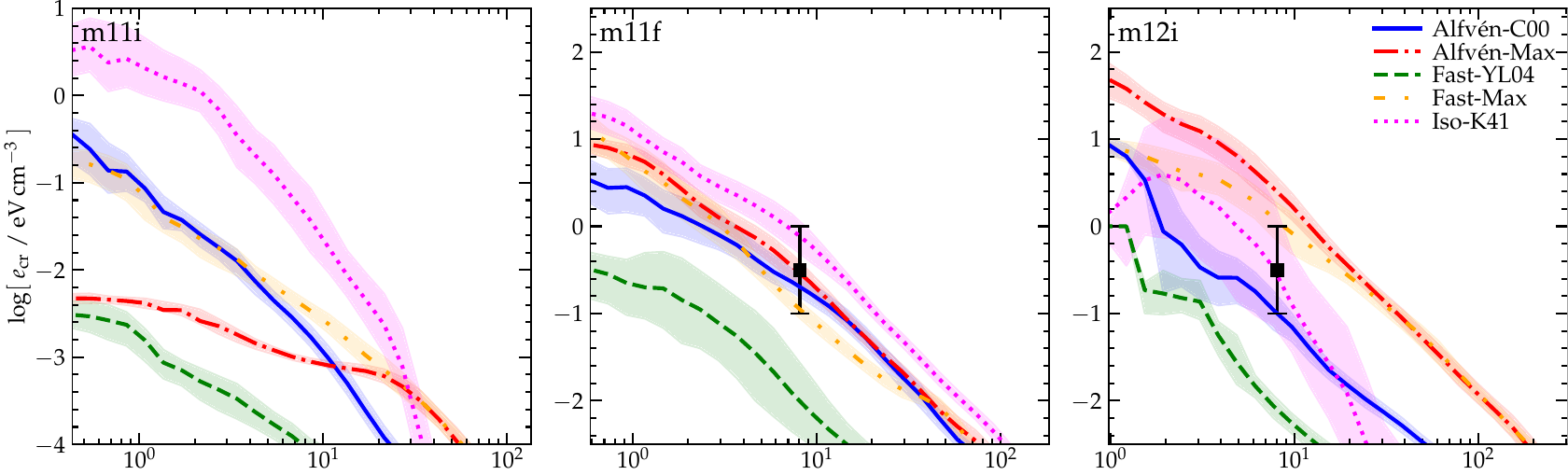}{0.95}
    \plotside{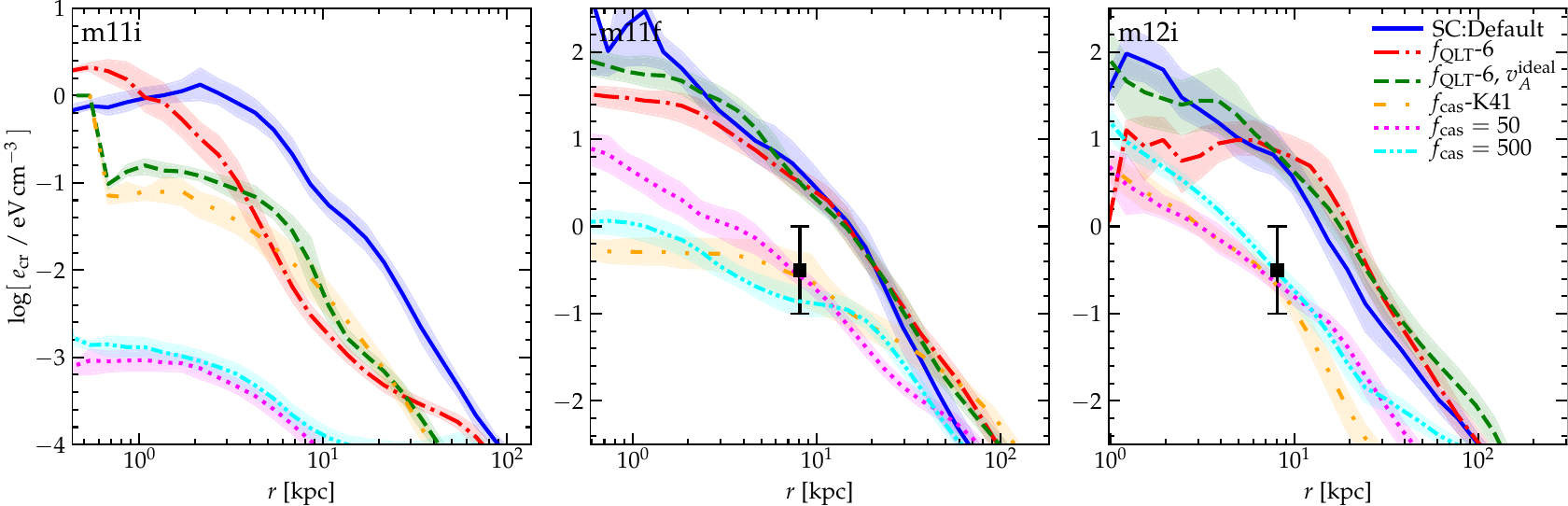}{0.95}
    \vspace{-0.2cm}
    \caption{Volume-weighted CR energy density $e_{\rm cr}$ vs.\ galacto-centric radius in different transport models (as Fig.~\ref{fig:kappa.vs.model}; see \S~\ref{sec:cr.egy.dens.compare}). In {\bf m11f} and {\bf m12i}, we note the location and order-of-magnitude observed $e_{\rm cr}$ at the solar circle (error bar). Crudely, $e_{\rm cr}$ decreases as $\kappa_{\rm eff}$ increases in different models. 
        {\em Top:} CD models. Low (high) $\kappa_{29} \ll 0.3$ ($\gg 30$) produce too much (too little) CR confinement and so over (under) predict $e_{\rm cr}$ in MW-like galaxies, consistent with their over (under) prediction of $L_{\gamma}$ in Fig.~\ref{fig:Lgamma}. Model $\kappa_{\rm ion-neutral}$ produces an $e_{\rm cr}$ profile similar to a model with the ``low'' ionized-gas $\kappa_{29}=0.1$ everywhere. 
        {\em Middle:} ET models. Qualitative trends with $\kappa_{\rm eff}$ are similar except model ``Iso-K41'' in {\bf m12i} which can produce such efficient CR confinement that CRs lose their energy collisionally, lowering $e_{\rm cr}$. 
        {\em Bottom:} SC models. These give almost bimodal results in the MW-mass systems, owing to the SC ``runaway'' or ``bottleneck'' effect where higher $e_{\rm cr}$ produces lower $\kappa_{\rm eff}$ (\S~\ref{sec:fast.cosmo}). Transport is ``too slow'' in default SC models causing CRs to ``pile up'' in excess of observations; $f_{\rm QLT}\,f_{\rm cas} \sim 100$ produces good agreement. 
    \label{fig:ecr.vs.model}}
\end{figure*}

\begin{figure}
    \plotone{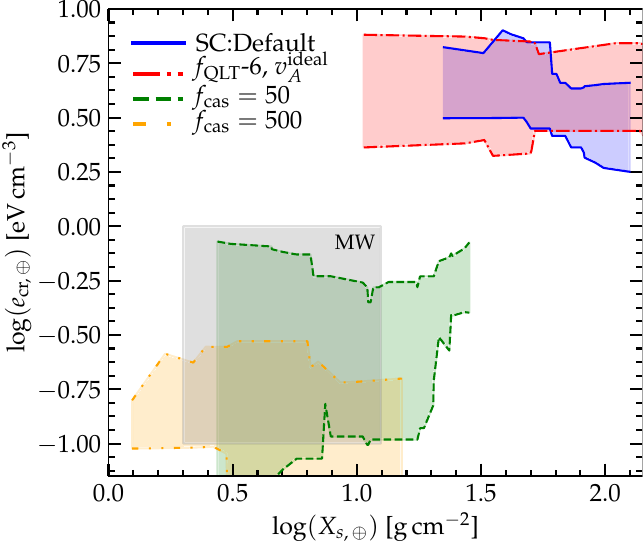}{0.85}
    \caption{Grammage $X_{s,\,\oplus}$ calculated by following a Monte Carlo subset of CRs from emission to a mock ``observer'' at the Solar circle (galacto-centric $r = 8.1\,$kpc) versus CR energy density in the disk midplane at the same location (see \S~\ref{sec:cr.grammage.calc}), sampled over different locations and times at $z<0.5$ in our {\bf m11f} and {\bf m12i} models selecting times at $z<1$ where the gas density $\Sigma_{\rm central}$ is similar to the MW value observed (Fig.~\ref{fig:Lgamma}; $\sim 0.002-0.01\,{\rm g\,cm^{-2}}$). Shaded grey range shows observationally allowed values for $\sim$\,GeV CRs. The same models which are consistent with $L_{\gamma}/L_{\rm SF} \propto X_{s}$ in Fig.~\ref{fig:Lgamma} and $e_{\rm cr}$ in Fig.~\ref{fig:ecr.vs.model} are consistent with the grammage/residence time constraints, for galaxies at times similar to the MW. We show a subset of SC models but have considered additional ET and CD models and reach the same conclusion.
    \label{fig:Xs.vs.model}}
\end{figure}

\subsubsection{CR Energy Densities}
\label{sec:cr.egy.dens.compare}

Fig.~\ref{fig:ecr.vs.model} compares the radial CR energy density profile averaged in spherical shells,\footnote{Because of rapid diffusion, the CR energy density is very similar in cylindrical annuli within the thin disk; see also Fig.~\ref{fig:Xs.vs.model}.} again at $z=0$, for the same galaxies and models as Fig.~\ref{fig:Lgamma}. For {\em otherwise fixed galaxy properties}, we expect $e_{\rm cr} \sim \dot{E}_{\rm cr}/(4\pi\,r\,\kappa_{\rm eff}) \propto \kappa_{\rm eff}^{-1}$ in steady-state, since the CR flux and hadronic losses must balance the injection by SNe $\dot{E}_{\rm cr}$, on average.  In a rough sense, we do see $e_{\rm cr}$ decrease with larger $\kappa_{\rm eff}$ (especially in the constant-$\kappa$ models), but the trend is weaker and occasionally non-monotonic, owing to the non-linear changes in galaxy properties (e.g.\ SNe rates) with different $\kappa$ (see below). 

Unlike $L_{\gamma}$, there are no direct observational constraints on $e_{\rm cr}$, except in the solar neighborhood (galacto-centric $r\sim 8\,$kpc) of the MW, where the most current observations indicate $e_{\rm cr} \sim 0.5-1.2\,{\rm eV\,cm^{-3}}$ in the diffuse ISM, integrating {\em all} CRs with energies $\gtrsim 5\,$MeV \citep{1998ApJ...506..329W,2009A&A...501..619P,2012ApJ...745...91I,2016ApJ...831...18C}. This corresponds to $e_{\rm cr} \sim 0.1-1\,{\rm eV\,cm^{-3}}$ integrated within a factor of $\sim 10$ of $1\,$GeV. We therefore compare these values to the MW-mass simulations: there are some models which can be ruled out by this constraint, but they are all models {\em already} ruled out by $L_{\gamma}$ or grammage constraints (Fig.~\ref{fig:Lgamma}). Fig.~\ref{fig:Xs.vs.model} shows this explicitly: we compare more detailed calculations of both $e_{\rm cr}$ and $X_{s}$ as measured by a mock observer at a random Solar-neighborhood star, selecting only low-redshift times where the broad galaxy properties (mass and $\Sigma_{\rm central}$ and, as a consequence SFR) are similar to the MW. 

For a given CR model, lower-mass galaxies exhibit systematically smaller $e_{\rm cr}$ at all radii, as expected given their lower SFRs (hence SNe rates and CR injection rates $\dot{E}_{\rm cr}$), and similar-or-larger $\kappa_{\rm eff}$.

\subsubsection{Rigidity-Dependence of Grammage and Other Properties}

It is worth commenting on how the implied grammage and residence time depend on the CR energy $E_{\rm cr}=\CRegy\,{\rm GeV}$ or rigidity $\mathcal{R} = \CRegy\,{\rm GV}$. Because our simulations only follow a single bin (so we do not directly evolve high-$\mathcal{R}$ CRs while evolving the $\sim$\,GeV CRs that dominate $e_{\rm cr}$) we cannot make detailed predictions for this. However, if we assume that higher-energy CRs behave as tracers (containing relatively little CR energy) that do not dynamically perturb the galaxies, and neglect losses (valid for  $\mathcal{R} \gtrsim 1\,$GV), we can predict how $\tilde{\kappa}_{\rm eff}$ and $X_{s}^{\infty}$ depend on $\mathcal{R}$ in the different models here.\footnote{We do this by calculating $X_{s}^{\infty}$ for tracer particles (as above) with different $\mathcal{R}$, using the expressions for $\kappa_{\|}(\CRegy)$ in the text, then fitting the power-law dependence $X_{s}^{\infty} \propto \mathcal{R}^{-\delta}$.} If all else is equal and $\tilde{\kappa}_{\rm eff} = \tilde{\kappa}_{\rm eff}(1\,{\rm GV})\,(\mathcal{R}/{\rm GV})^{\delta}$ then we simply have $X_{s}^{\infty} \propto \mathcal{R}^{-\delta}$. Most analyses of MW observations of, e.g.\ the B/C ratio, favor $X_{s} \approx 5\,{\rm g\,cm^{-2}}\,(\mathcal{R}/{\rm GV})^{-(0.5-0.6)}$ (i.e.\ $\delta \sim 0.5-0.6$) at energies $\sim 1-100$\,GeV \citep{2006ApJ...642..902P,2010A&A...516A..66P,2017PhRvD..95h3007Y,2017MNRAS.471.1662B,2018PhRvL.120b1101A}, although systematically varying assumptions about anisotropy, advection/winds, ``halo'' size, and source spectral shape can lead to values in the range $\delta \sim 0.3-0.8$ \citep{2010A&A...516A..67M,2011ApJ...729..106T,2017MNRAS.471.1662B}.

Although it is commonly assumed that ET models give $\delta = 1/3$ (or $\delta=1/2$ for a dynamically aligned or Iroshnikov-Kraichnan spectrum), this is only true if anisotropy and damping are totally ignored (as in e.g.\ our ``Iso-K41'' model), which is un-ambiguously ruled out by all other observational constraints. Almost {\em all} the ET models considered here, give $\delta \lesssim 0$: \Alf-C00, \Alf-C00-Vs, \Alf-Hi, \Alf-Max all predict $\delta=0$, while the \Alf-YL02 model gives negative $\delta=-0.8$. Model Fast-YL04 gives $\kappa_{\|} \propto \mathcal{R}^{0}$ when collisionless damping dominates and $\propto \mathcal{R}^{-1/6}$ when viscous damping dominates: since viscous damping dominates throughout the ISM and inner CGM, which dominate the residence time, we find, by integrating test particles, an effective $\delta \approx -0.12$ in this model and the related Fast-Mod/Fast-Max/Fast-NoCDamp variations. In short, at energies $\lesssim$\,TeV (where anisotropy and damping are important), ET models predict the wrong {\em qualitative} sense of $\delta$, {\em regardless} of the turbulent spectrum assumed.

On the other hand, in the default SC models here (or those with constant $f_{\rm QLT}$ or $f_{\rm cas}$), $\kappa_{\|} \propto \mathcal{R}^{1/2}$ if turbulent, linear or nonlinear Landau damping dominate and $\kappa_{\|} \propto \mathcal{R}^{0-1}$ when ion-neutral damping dominates ($0$ if $v_{A}=v_{A}^{\rm ion}$ dominates over $\kappa_{\|}$, as it often does when ion-neutral damping dominates, $1$ otherwise). Since we show below that the grammage and residence times are dominated by the regimes where ion-neutral damping is sub-dominant, we predict an effective $\delta \approx 0.5 \pm 0.1$ for almost all of these models (even models $f_{\rm cas}$-DA and $f_{\rm cas}$-K41, with different turbulent spectra, give $\delta=0.42$ and $=0.36$ respectively).

\begin{figure}
    \plotone{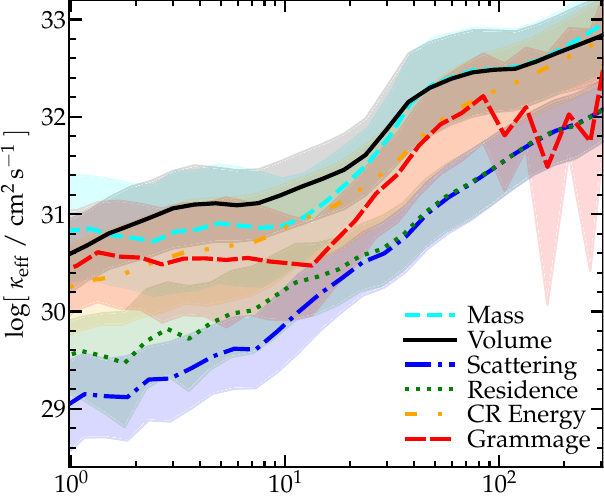}{0.8}\\
    \plotone{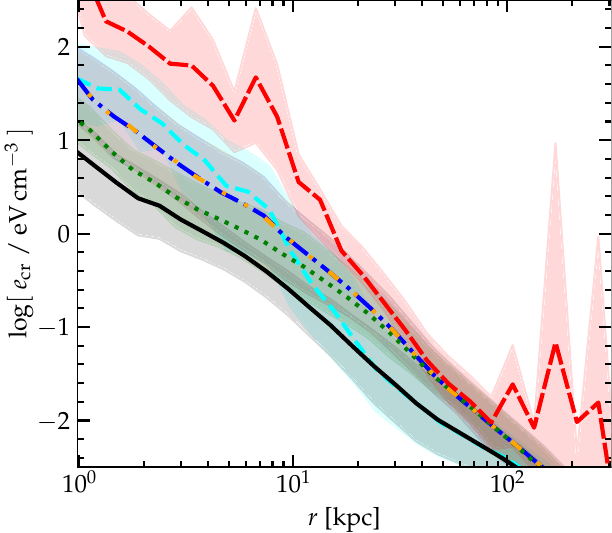}{0.8}
    \caption{Radial profile of $\kappa_{\rm eff}$ ({\em top}; as Fig.~\ref{fig:kappa.vs.model}) and $e_{\rm cr}$ ({\em bottom}; as Fig.~\ref{fig:ecr.vs.model}), in one example consistent with observations ({\bf m12i} in SC model ``$f_{\rm QLT}$-100''). We calculate the profiles weighting each resolution element by different quantities in each radial annulus (\S~\ref{sec:defining.typical.params}, averaged over all times $z<0.5$): gas mass, volume, CR scattering rate, CR residence time, CR energy, grammage (or equivalently contribution to $L_{\gamma}$). 
    {\em Top:} The ``mean'' $\kappa_{\rm eff}$ (at a fixed radius and time) can vary systematically by factors up to $\sim 100$ based on weight, owing to the very large local variations in the ISM/CGM (Fig.~\ref{fig:image}). 
     Weighting by scattering rate or residence time ($\propto 1/\kappa$) biases towards the lowest-$\kappa$ regions, where CRs can be ``trapped,'' while volume-weighting gives the highest $\kappa$ and others lie in-between. Differences are smaller in the CGM (where e.g.\ density differences in phases are less extreme), but still factor $\sim 10$. 
    {\em Bottom:} Because of rapid diffusion, differences in $e_{\rm cr}$ are smaller (it is smoother; see Fig.~\ref{fig:image}), but still significant, as weighting by e.g.\ total grammage ($\propto e_{\rm cr}\,\rho\,d^{3}{\bf x}$) biases to the densest gas with the highest $e_{\rm cr}$. 
    \label{fig:kappa.ecr.vs.weight}}
\end{figure}

\begin{figure}
    \plotone{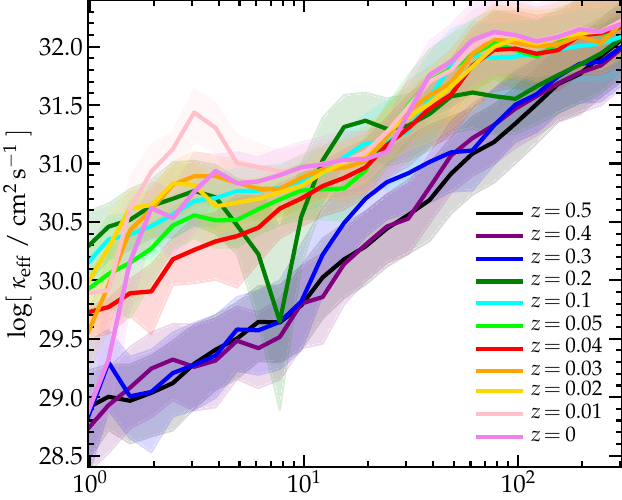}{0.8}\\
    \plotone{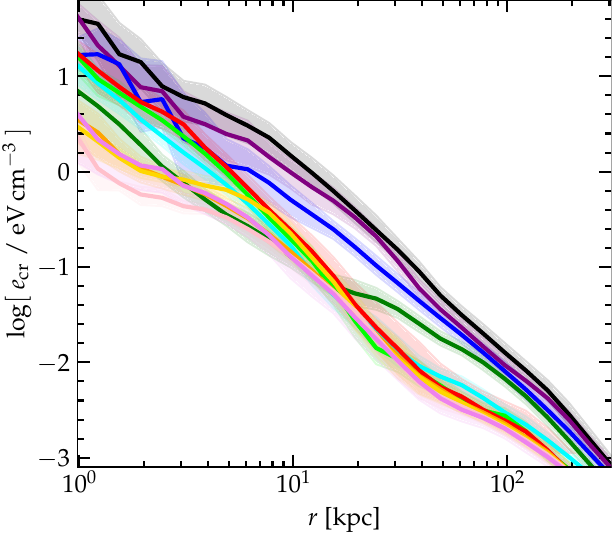}{0.8}
    \caption{Time-dependence of $\kappa_{\rm eff}$ (scattering-rate-weighted) and $e_{\rm cr}$ (volume-weighted). We plot profiles of both in {\bf m12i} SC model ``$f_{\rm QLT}$-100'' as Fig.~\ref{fig:kappa.ecr.vs.weight}, but sampling different times at $z<0.5$ (different colors; note the time/redshift spacing is not uniform). 
    There is considerable variation in time, which is not simply a continuous systematic evolution but reflects substantial changes over time as bar and spiral arms, phase structure and presence/absence of super-bubbles, and periods of elevated star formation  (e.g.\ associated with higher $e_{\rm cr}$ at $z\sim 0.3-0.5$) and galactic outflow appear and recede. 
    \label{fig:kappa.ecr.vs.redshift}}
\end{figure}

\subsection{Local Variations in Transport Parameters \&\ the ``Effective'' Diffusivity or Streaming Speed}

Having narrowed down the observationally-allowed range of ET and SC models, we now explore the distribution of transport parameters in these systems. 

\subsubsection{Defining ``Typical'' Parameters}
\label{sec:defining.typical.params}

Fig.~\ref{fig:kappa.ecr.vs.weight} shows $\kappa_{\rm eff}(r)$ and $e_{\rm cr}(r)$, for a representative example of both an ET model (``Fast-Max'') and SC model (``SCx100'') which produce $L_{\gamma}$ and grammage similar to observations (meaning they could, in principle, represent the dominant CR scattering). We determine the median and scatter in each annulus  with various different weights, e.g.\ weighting each cell by the local gas mass ($\rho\,d^{3}{\bf x}$), volume ($d^{3}{\bf x}$), CR energy ($e_{\rm cr}\,d^{3}{\bf x}$), grammage or contribution to $L_{\gamma}$ ($\propto e_{\rm cr}\,\rho_{\rm gas}\,d^{3}{\bf x}$), CR scattering rate ($\propto (e_{\rm cr}/\kappa)\,d^{3}{\bf x}$), or CR residence time ($\propto (e_{\rm cr}\,d^{3}{\bf x})\,(e_{\rm cr}\,d r/|{\bf F}|)$). Fig.~\ref{fig:image} highlights local variations in $e_{\rm cr}$ and $\kappa_{\rm eff}$ by showing a 2D map of their local values, in a slice through the galaxy.

Within the galaxy, we see the resulting ``typical'' $\kappa_{\rm eff}$ differs by as much as $\sim 2\,$dex (in the CGM, the differences are $\sim0.5-1\,$dex). This owes to inhomogeneity in the plasma properties inside the ISM, discussed below (\S~\ref{sec:fast.cosmo}) and which, in these CR transport models, directly translates to large (orders-of-magnitude) local variations in $\kappa_{\rm eff}$ and $v_{\rm st}$. Weighting by, e.g.\ volume, favors diffuse ISM. Weighting by scattering rates or residence times, $\propto 1/\kappa_{\rm eff}$, selects the {\em lowest} local values of $\kappa_{\rm eff}$, as relevant to  the ``residence'' or ``escape'' time in an inhomogeneous medium, which is dominated by the regions with the slowest CR propagation. Fundamentally, different ``weights'' correspond to different questions: observational constraints on $L_{\gamma}$ and grammage are sensitive to  residence-time-weighted transport parameters, while  the median CR energy density and effects of CRs on pressure support of the CGM and ISM are sensitive to the ISM mass and volume-weighted parameters. 

We also see this inhomogeneity reflected in significant time-variation in Fig.~\ref{fig:kappa.ecr.vs.redshift}, even averaging within annuli.  Relatively large-scale structure in $\kappa_{\rm eff}$ at a given radius (dominated by spiral arms or large cloud complexes or super-bubbles) can still be somewhat transient, producing factor $\sim3-10$ changes in the mean $\kappa_{\rm eff}$ within an annulus over a galactic dynamical time (while smaller structures vary on smaller timescales). Galactic-scale ``events'' (a burst of star formation and associated outflow) can produce large coherent changes in $e_{\rm cr}$ and $\kappa_{\rm eff}$.  

This explains much of why there is not a trivial one-to-one linear relation between $\kappa_{\rm eff}$ and $L_{\gamma}$ in Figs.~\ref{fig:kappa.vs.model}-\ref{fig:Lgamma}, in the SC and ET models. Some of these models can produce very large volume or $L_{\gamma}$-weighted $\kappa_{\rm eff}$, but in the central few kpc of the galaxy (which dominate $L_{\gamma}$) the residence-time or scattering-rate weighted $\kappa_{\rm eff}$ is much lower, producing larger $L_{\gamma}$. Some of this variation also translates to $e_{\rm cr}$, although the diffusive nature of CR transport reduces the variations here.

\begin{figure*}
    \plotside{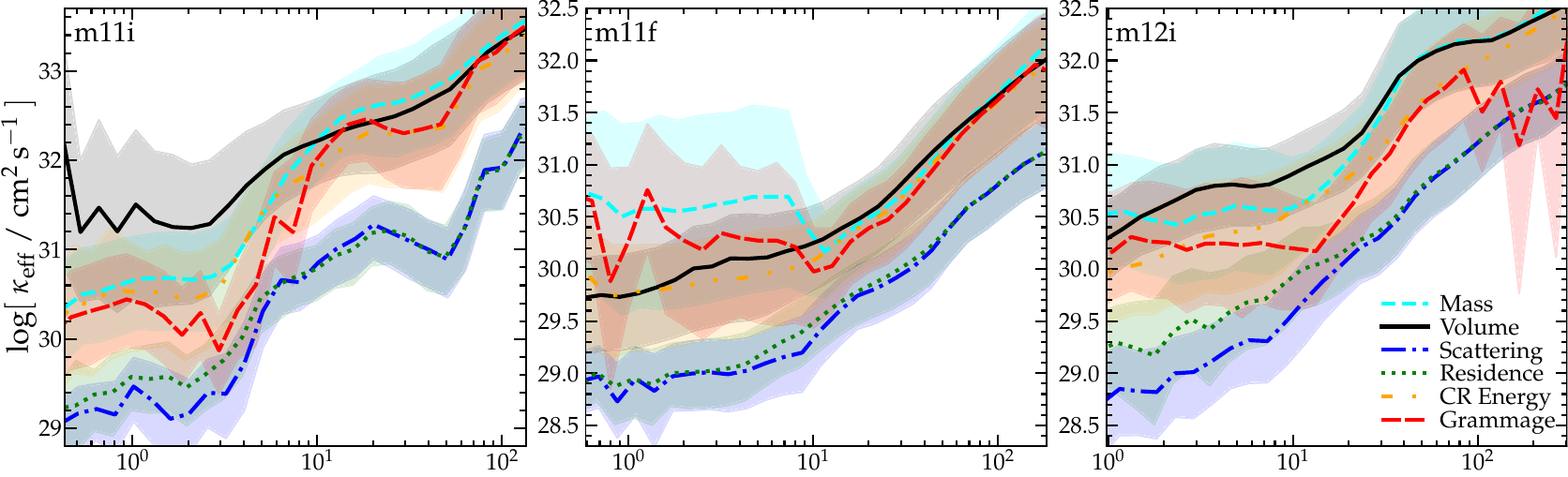}{0.96}
    \plotside{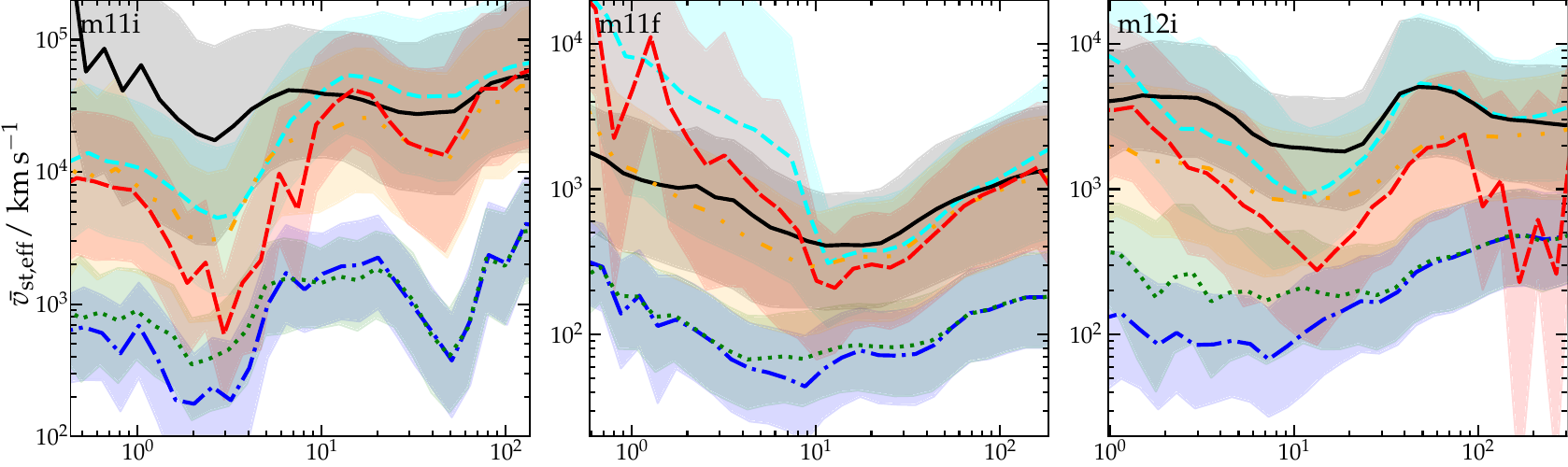}{0.96}
    \plotside{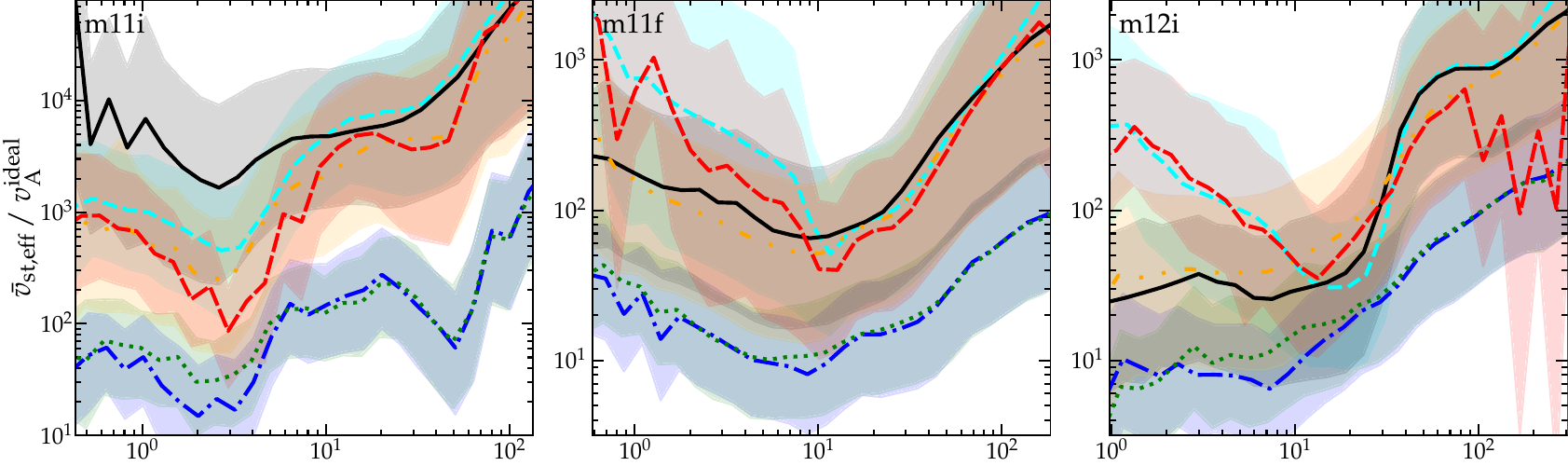}{0.96}
    \plotside{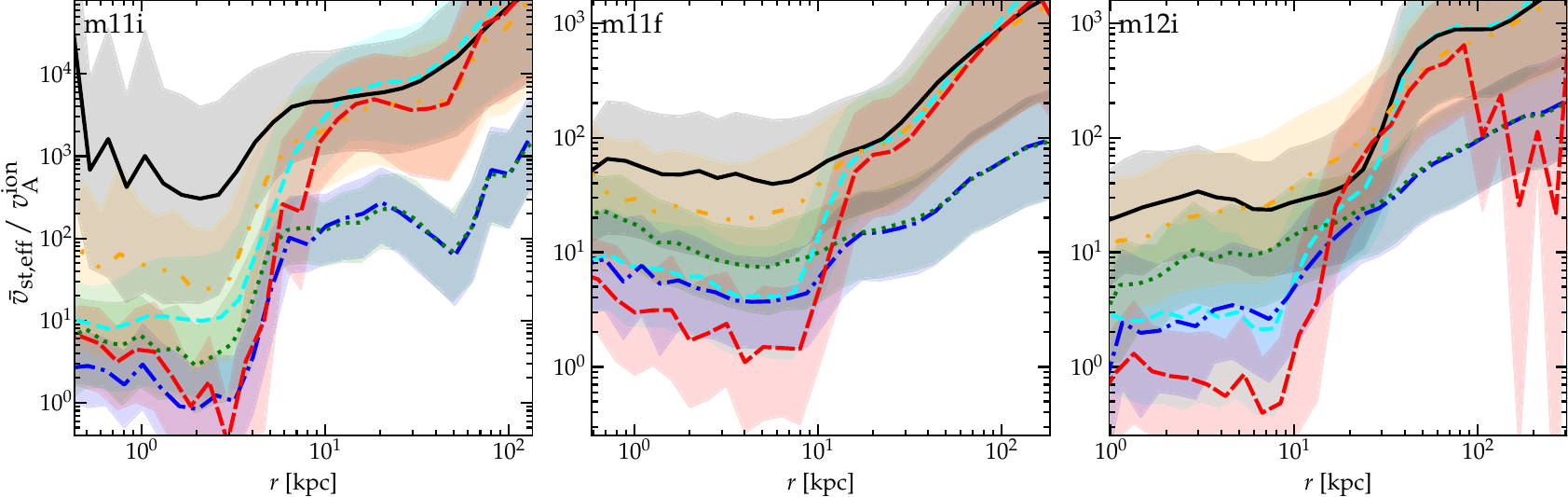}{0.96}
    \vspace{-0.2cm}
    \caption{Radial profile of transport parameters in SC model ``$f_{\rm QLT}$-100'' in different galaxies vs.\ weight (as Fig.~\ref{fig:kappa.ecr.vs.weight}), see \S~\ref{sec:diffusion.vs.streaming}.
    {\em Top:} Effective ``diffusivity'' $\kappa_{\rm eff} \equiv |{\bf F}|/|\nabla_{\|} e_{\rm cr}|$. 
    {\em Second:} Effective ``streaming speed'' $\bar{v}_{\rm st,\,eff} \equiv |{\bf F}|/h_{\rm cr}$.
    {\em Third:} Effective streaming speed in units of local ideal-MHD \Alf-speed $\bar{v}_{\rm st,\,eff}/v_{A}^{\rm ideal}$.
    {\em Bottom:} Effective streaming speed in units of local ion \Alf-speed $\bar{v}_{\rm st,\,eff}/v_{A}^{\rm ion}$ ($v_{A}^{\rm ion} \approx f_{\rm ion}^{-1/2}\,v_{A}^{\rm ideal}$).
    In all cases, the choice of weight has similar (large) effects: this reflects genuine inhomogeneity, not the particular diagnostic.
    Diffusivity $\kappa_{\rm eff}$ is reasonably constant within a single galaxy ($r\lesssim$\,a few kpc) but rises with $r$ in the CGM (by factors $\sim 100-1000$ at the virial radius); the scattering-weighted $\kappa_{\rm eff}$ also depends surprisingly weakly on which galaxy we consider.    
    The absolute $\bar{v}_{\rm st,\,eff}$ is much closer to $r$-independent, though the scattering-rate-weighted value $\sim 100-1000\,{\rm km\,s^{-1}}$ depends more strongly systematically on galaxy type.
    Considering $\bar{v}_{\rm st,\,eff}$ in units of $v_{A}^{\rm ideal}$ or $v_{A}^{\rm ion}$ increases the scatter/radius dependence/systematic variations between galaxies: it is not accurate to simply think of ``super-\Alf{ic} streaming'' arising from SC as some multiple of $v_{A}$. 
    \label{fig:kappa.vs.vstream}}
\end{figure*}

\subsubsection{Diffusion versus Streaming}
\label{sec:diffusion.vs.streaming}

Fig.~\ref{fig:kappa.vs.vstream} compares $\kappa_{\rm eff}(r)$ with different weights like Fig.~\ref{fig:kappa.ecr.vs.weight}, but extends this to dwarf and intermediate-mass galaxies, and also compares the effective streaming speed $\bar{v}_{\rm st,\,eff}(r)$. Recall (\S~\ref{sec:methods:cr.transport}) we can freely translate locally between the two using $\bar{v}_{\rm st,\,eff} \equiv \kappa_{\rm eff}/(\gamma_{\rm cr}\,\ell_{\rm cr})$. Fig.~\ref{fig:kappa.vs.vstream} considers $\bar{v}_{\rm st,\,eff}$ in absolute units as well as relative to $v_{A}^{\rm ideal}$ and $v_{A}^{\rm ion}$.

First, we see that the local and systematic variations (weight-dependence) in $\kappa_{\rm eff}$ {\em within a single galaxy} discussed above extend to all galaxies simulated. They also do not vanish or significantly decrease if we consider $\bar{v}_{\rm st,\,eff}$ or $\bar{v}_{\rm st,\,eff}/v_{A}$ instead of $\kappa_{\rm eff}$. Likewise, systematic galaxy-to-galaxy variations in $\kappa_{\rm eff}$ (being larger in dwarfs) appear in $\bar{v}_{\rm st,\,eff}$ as well. In other words, these results are not simply an artifact of parameterizing the transport with $\kappa_{\rm eff}$ instead of $\bar{v}_{\rm st,\,eff}$.

Second, we see that, for a given model and weight (usually), $\kappa_{\rm eff}$ is approximately independent of $r$ within the galaxy (within a few kpc), but then rises at larger $r$ (in the CGM), while $\bar{v}_{\rm st,\,eff}$ depends on $r$ within the galaxy but is less-strongly $r$-dependent in the CGM. 

Third, we see that $\bar{v}_{\rm st,\,eff}$ in absolute units is actually closer to $r$-independent (and exhibits weaker systematic weight-dependence), compared to $\bar{v}_{\rm st,\,eff}/v_{A}^{\rm ideal}$ or $\bar{v}_{\rm st,\,eff}/v_{A}^{\rm ion}$, even though the SC simulations plotted assume $v_{\rm st}=v_{A}^{\rm ion}$. In other words, because $\kappa_{\|}$ is non-zero, we have $\bar{v}_{\rm st,\,eff} \approx v_{\rm st} + \kappa_{\|}/(\gamma_{\rm cr}\,\ell_{\rm cr}) \ne v_{\rm st}$.

Finally, we stress that even if the average $\kappa_{\rm eff}$ or $v_{\rm st,\,eff}$ were approximately constant across galacto-centric radius and time, the transport equations being integrated (especially for SC models) do not actually have the same form as a ``true'' diffusion or streaming/advection equation (see Appendix~\ref{sec:deriv:behavior}). Thus while $\kappa_{\rm eff}$ or $v_{\rm st,\,eff}$ are useful parameters and can guide our intuition regarding transport timescales, equilibrium fluxes, etc, care is required in their interpretation.

\subsection{Redshift Dependence and Effects on Galaxy Evolution}
\label{sec:redshift.gal.fx}

In future work, we will explore in  detail the effect of different CR models on {\em galaxy} properties, e.g.\ how they influence galactic star formation and ISM/CGM properties. Because our focus in this work is the observational constraints on CR transport models, we only briefly discuss galaxy properties here insofar as it can provide additional constraints. In \paperonetwo, we showed using ``constant diffusivity'' models that entirely turning on/off CRs, or changing $\kappa$ by factors of $\sim 1000$, makes only a modest (albeit non-negligible and potentially important) difference to global galaxy properties. We found that the strongest effects due to CRs (choosing the ``most optimal'' diffusivity) occur around MW-mass at $z\sim 0$, and even there it typically results in factor $\lesssim 2-3$ differences in e.g.\ galaxy stellar masses. This is not sufficiently large to obviously rule out a specific CR transport model or diffusivity (because, e.g.\ changing the mean mechanical energy per SNe by a similar factor, easily allowed by observations, would result in a similar effect). Among the models studied here which are allowed by $\gamma$-ray observations, we generally find effects on galaxy formation ``in between'' the ``no CR'' and ``largest CR effects'' models from \papertwo. We also find (consistent with \papertwo) that effects of CRs on galaxy properties are weaker at high redshifts (in every model considered here), owing to relatively higher ISM/CGM pressures. We therefore conclude that the indirect effects of CRs on bulk galaxy properties do not strongly constrain the CR transport models of interest.

\section{Discussion}
\label{sec:discussion}

\subsection{The Need for ``Fast'' Transport \&\ Cosmological Simulations with Resolved ISM Phases}

\subsubsection{Favored Transport Parameters: An Analytic Toy Model}
\label{sec:params}

The total (galaxy-integrated) CR collisional loss rate is $\dot{E}_{\rm coll} \equiv \int d^{3}{\bf x}\,\Lambda_{\rm coll}(n_{\rm gas},\,e_{\rm cr})$. In \papertwo, we developed a simple toy model for a constant isotropically-averaged diffusivity $\tilde{\kappa}_{\rm eff} \sim \kappa_{\rm eff}/3 \equiv \tilde{\kappa}_{29}\,10^{29}\,{\rm cm^{2}\,s^{-1}}$ (or $\tilde{v}_{\rm st,\,eff} \sim \bar{v}_{\rm st,\,eff}/3 \equiv \tilde{v}^{\rm st}_{1000}\,1000\,{\rm km\,s^{-1}}$) in a disk+halo system, with a steady-state star formation and SNe rate, hence constant $\dot{E}_{\rm cr} \approx 0.1\,\epsilon_{\rm SNe}\,\dot{M}_{\ast}$ (where $\epsilon_{\rm SNe}\sim 10^{51}\,{\rm erg}/100\,M_{\odot}$ is the energy per unit stellar mass in SNe). If the CRs are confined (not free-escaping), diffusion is relatively fast (compared to e.g.\ bulk gas motion), the SFR (hence CR injection) is centrally concentrated compared to the size of the CR halo, and collisional losses are small, then in steady-state the CR energy density should scale as $e_{\rm cr} \sim \dot{E}_{\rm cr}/(4\pi\,\tilde{\kappa}_{\rm eff}\,r)\sim \dot{E}_{\rm cr}/(4\pi\,\tilde{v}_{\rm st}\,r^{2})$.  If the disk+halo follows a realistic extended profile with most of the gas mass $M_{\rm gas}$ in a half-mass radius $\ell_{\rm gas}$ and central surface density $\Sigma_{\rm gas}$, then (performing the integrals exactly for a thin, exponential disk in a power-law halo following \papertwo): 
\begin{align}
\frac{\dot{E}_{\rm coll}}{\dot{E}_{\rm cr}} \approx \frac{L_{\gamma}}{L_{\rm calor}} &\sim \frac{0.15}{\tilde{\kappa}_{29}}\,\left( \frac{\Sigma_{\rm gas}\,\ell_{\rm gas}}{0.01\,{\rm g\,cm^{-2}\,kpc}}\right) \\
\nonumber &\sim \frac{0.06}{\tilde{v}_{1000}^{\rm st}}\,\left( \frac{\Sigma_{\rm gas}}{0.01\,{\rm g\,cm^{-2}}} \right)\ , 
\end{align}
or equivalently (using $L_{\rm calor} \approx 2\times10^{-4}\,L_{\rm sf}$)
\begin{align}
\label{eqn:lgamma.lcalor.predicted} \frac{{L}_{\gamma}}{{L}_{\rm sf}} &\sim \frac{3\times10^{-5}}{\tilde{\kappa}_{29}} \left( \frac{\Sigma_{\rm gas}\,\ell_{\rm gas}}{0.01\,{\rm g\,cm^{-2}\,kpc}}\right) \sim \frac{10^{-5}}{\tilde{v}_{1000}^{\rm st}} \left( \frac{\Sigma_{\rm gas}}{0.01\,{\rm g\,cm^{-2}}} \right) \ .
\end{align}
In terms of the grammage $X_{s}^{\infty}$, this gives
\begin{align}
\frac{X_{s}^{\infty}}{{\rm g\,cm^{-2}}}  &\sim \frac{20}{\tilde{\kappa}_{29}} \left( \frac{\Sigma_{\rm gas}\,\ell_{\rm gas}}{0.01\,{\rm g\,cm^{-2}\,kpc}}\right) \sim \frac{6}{\tilde{v}_{1000}^{\rm st}} \left( \frac{\Sigma_{\rm gas}}{0.01\,{\rm g\,cm^{-2}}} \right) 
\end{align}
The assumption that losses are small means this applies when  $\dot{E}_{\rm loss}/\dot{E}_{\rm coll}\ll1$; losses will saturate at the calorimetric limit $\dot{E}_{\rm loss}\approx \dot{E}_{\rm coll}$. This simple estimate gives a surprisingly good estimate of the full simulation prediction for $L_{\gamma}/L_{\rm sf}$ for our constant-$\kappa$ models (assuming $\tilde{\kappa}_{\rm eff} \sim \kappa_{\|}/3$) in Fig.~\ref{fig:Lgamma}.

Moreover if we assume we are in a MW-like galaxy, with a ``solar circle'' at $r_{\rm obs} \approx 8\,$kpc, we can also estimate the median CR energy density and CR residence time\footnote{For residence time, we model CR injection as a Gaussian with initial half-mass radius $r_{1/2}=5\,$kpc, motivated by the stellar (and SNe Ia) scale-length in the MW (adopting the scale-length for young-stars, for core-collapse, gives $r_{1/2}\approx 3\,$kpc), diffusing isotropically, then calculate the median time-since-injection of all CRs in a shell $r_{\rm obs}\approx 8\,$kpc in steady-state.} seen by a mock observer:
\begin{align}
\frac{e_{\rm cr}}{\rm eV\,cm^{-3}}{\Bigr|}_{\odot} &\sim \frac{2}{\tilde{\kappa}_{29}}\,\left(  \frac{R_{\rm SNe,\,MW}}{1/30\,{\rm yr}} \right) \\ 
\frac{\Delta t_{\rm res}}{\rm Myr} {\Bigr|}_{\odot} &\sim \frac{25}{\tilde{\kappa}_{29}}\,\left( \frac{r_{\rm obs}^{2} - r_{1/2}^{2}}{(8\,{\rm kpc})^{2} - (5\,{\rm kpc})^{2}} \right)
\end{align}
where $R_{\rm SNe,\,MW}$ is the MW (Galaxy-integrated) SNe rate $\sim 1/30\,{\rm yr}$.

Noting that the MW has an observed central $\Sigma_{\rm gas}\sim 20\,M_{\odot}\,{\rm pc^{-2}} \sim 0.004\,{\rm g\,cm^{-2}}$ and $\ell_{\rm gas} \sim 5\,$kpc, reproducing the observed MW grammage $X_{s}\sim 3-10\,{\rm g\,cm^{-2}}$,  $L_{\gamma}/L_{\rm sf}\sim 0.03$,  $e_{\rm cr} \sim 0.1-1\,{\rm eV\,cm^{-3}}$, or $\Delta t_{\rm res} \sim 5-20\,{\rm Myr}$ all require $\tilde{\kappa}_{\rm 29} \sim $a few. This is the median of our ``favored'' values in Table~\ref{tbl:transport}. 

This also neatly illustrates the degeneracy between inferred diffusivity and ``halo size'' in simpler leaky-box models: if the CRs escape at some height $h < \ell_{\rm gas}$ (truncating the integral above), it is roughly equivalent to replacing $\ell_{\rm gas} \rightarrow h$ in the calculation above, and for a fixed $L_{\gamma}/L_{\rm sf}$ or $X_{s}$, we have an inferred $\kappa \propto h$. As soon as we abandon the assumption of a ``leaky box'' or ``flat halo'' with $h < 1\,$kpc, all of the observations require similar, relatively ``fast'' transport speeds.

\subsubsection{Scalings of Gamma-Ray Luminosity with Galaxy Properties}

The simple model in \S~\ref{sec:params} and Eq.~\ref{eqn:lgamma.lcalor.predicted} naturally explains the trend of $L_{\gamma}/L_{\rm SF} \propto \Sigma_{\rm gas}$ at low $\Sigma_{\rm gas}$ seen in Fig.~\ref{fig:Lgamma}, as $L_{\gamma} \propto X_{s} \propto \Sigma_{\rm gas}$ -- i.e.\ for a similar transport speed, the grammage $X_{s}$ (and therefore $L_{\gamma}$ produced by collisions) simply scales with the galactic column density. 

In contrast, the trend of $L_{\gamma}/L_{\rm SF}$ with $L_{\rm SF}$ or $\dot{M}_{\ast}$ in Fig.~\ref{fig:LgammaVsSFRLIR} is closer to $L_{\gamma}/L_{\rm SF} \propto \dot{M}_{\ast}^{0.7}$. This follows from global galaxy scalings like the Schmidt-Kennicutt relation $\dot{\Sigma}_{\ast} \propto \Sigma_{\rm gas}^{1.4}$ seen in both nature and these simulations \citep{kennicutt98,orr:ks.law}, which (with Eq.~\ref{eqn:lgamma.lcalor.predicted}) gives $L_{\gamma}/L_{\rm SF} \propto \dot{M}_{\ast}^{0.7}/\tilde{\kappa}_{29}$. 

If we assume steady-state with a constant SFR, then the total IR luminosity is determined by the fraction of optical/UV light absorbed and re-emitted: $L_{\rm IR}/L_{\rm SF} \approx (1 - \exp{[-\kappa_{\rm OUV}\,\Sigma_{\rm gas}]})$ where $\kappa_{\rm OUV} \sim 1000\,{\rm cm^{2}\,g^{-1}}\,(Z/Z_{\odot})$ is the flux-averaged optical/UV opacity (scaling with galaxy metallicity $Z$). In dwarfs and the MW where $L_{\rm IR} \lesssim L_{\rm SF}$ this gives: $L_{\rm IR}/L_{\rm SF} \sim \kappa_{\rm OUV}\,\Sigma_{\rm gas}$. Combining with Eq.~\ref{eqn:lgamma.lcalor.predicted}, we have $L_{\gamma}/L_{\rm IR} \sim 3\times10^{-5}\,\tilde{\kappa}_{29}^{-1}\,(\ell_{\rm gas}/10\,{\rm kpc})\,(Z_{\odot}/Z)$, which is very weakly-dependent on galaxy properties (both $\ell_{\rm gas}$ and $Z$ scale $\propto M_{\ast}^{0.2-0.3}$, and their scalings cancel here; see \citealt{kewley:mass.metallicity.relation,hall:2011.disk.scalings}). In short, the fact that $L_{\gamma}/L_{\rm IR}$, while clearly not constant, depends only weakly on $L_{\rm IR}^{0.2-0.3}$ (Fig.~\ref{fig:LgammaVsSFRLIR}) -- i.e.\ that the $L_{\gamma}-L_{\rm IR}$ relation is closer to linear than the $L_{\gamma}$-SFR relation, trivially follows from the fact that {\em both} the grammage $X_{s}$ (which is proportional to $L_{\gamma}$) and OUV optical depth $\tau$ (proportional to $L_{\rm IR}$) scale with $\Sigma_{\rm gas}$. 

Again, reproducing any of the observed trends requires similar $\tilde{\kappa}_{29}\sim$\,a few.

\subsubsection{Importance of Cosmological Simulations \&\ Resolved ISM/CGM Phases}
\label{sec:fast.cosmo}

Although the simple analytic scalings above can explain many qualitative phenomena, we also identify in our simulations a number of important effects which can only be properly captured in cosmological simulations with {\em resolved} ISM phases. These include:
\begin{enumerate}

\item{ Extended halos:} Galaxies have extended gaseous halos reaching to $>100\,$kpc, containing most of the gas mass in relatively slowly-falling power-law density profiles (e.g.\ isothermal $\rho \propto r^{-2}$, as opposed to exponential). In {\em every} physically plausible model we consider, the $\sim$\,GeV CRs remain confined/coupled in the halo out to $\gtrsim R_{\rm vir}$ (mean free paths are $\lambda_{\rm mfp} \sim 3\,\kappa/c \sim 0.003\,\kappa_{29}\,{\rm kpc}$, compared to $\sim 100\,$kpc halo scale-lengths). The galaxy and even ``inner'' CGM halo at $\lesssim 10\,$kpc is not a ``leaky box'' or ``flat halo'' with simple escape outside some volume.

\item{ Clumpiness:} At high $\tilde{\kappa}_{\rm eff}$, ISM ``clumping'' does not strongly alter $L_{\gamma}$ because CRs rapidly move through dense gas. But if $\tilde{\kappa}_{\rm eff} \lesssim 10^{27}\,{\rm cm^{2}\,s^{-1}}$ locally, then CR diffusion/escape times ($\sim \ell^{2}/\kappa$) becomes shorter than (a) the dynamical times ($\sim 1/\sqrt{G\,\rho}$) of large ($\gtrsim 100\,$pc) GMC complexes, and (b) CR collisional loss times ($\sim 40\,n_{1}^{-1}\,{\rm Myr}$). Thus CRs get ``captured'' in dense clumps, producing order-of-magnitude higher $L_{\gamma}$.

\item{ Multi-phase neutral gas:} If the neutral gas is {\em bounded} (e.g.\ in clouds or a thin disk) by ionized gas, then even if $\tilde{\kappa}_{\rm eff}\rightarrow\infty$ in that neutral gas, the CR energy density $e_{\rm cr}$ becomes locally constant at a value $\langle e_{\rm cr}\rangle$ determined by the ``boundary condition'' value of $e_{\rm cr}$ in the ionized medium. If $\tilde{\kappa}_{\rm eff}$ is low in the ionized gas, the CRs are therefore ``trapped'' regardless of $\tilde{\kappa}_{\rm eff}$ in the cold/neutral phase. Thus the total residence time in dense gas can be large, in principle, even if the local diffusivity in said gas is also large. 

\item{ Halo ``collapse'':} As shown in \papertwo, if CRs efficiently escape the disk to $\gtrsim 10\,$kpc in intermediate and MW-mass systems, they provide substantial pressure support to the halo gas, which in turn suppresses accretion leading to significantly less dense gas {\em in the disk} at $z\approx 0$, which suppresses $L_{\gamma}$ further. But if they {\em cannot} escape to $\gtrsim 10\,$kpc, the halo ``collapses'' and produces more efficient cooling and denser gaseous disks in MW-mass systems, non-linearly raising $L_{\gamma}$.

\item{ Self-confinement ``runaway'' or ``bottleneck'':} In SC models, the diffusivity/streaming speed scales inversely with $e_{\rm cr}$ (i.e.\ the absolute CR {\em flux} is bottlenecked by the self-excited waves). Thus if $e_{\rm cr}$ builds up to large ISM values even briefly, the effect rapidly runs away, as it restricts its own transport. A number of other non-linear effects can further exacerbate this: for example, if $P_{\rm cr}$ begins to dominate pressure support in the WIM or inner CGM, then turbulence is generally weaker (as CRs suppress rapid gas cooling/collapse and star formation), hence $\Gamma_{\rm turb}$ and $\kappa$ become smaller still. These produce large local fluctuations in diffusivity/streaming speed.

\item{ Clustered supernovae:} In a resolved ISM, SNe are strongly clustered in space and time and associated with denser, star forming regions. This enhances $L_{\gamma}$ directly, but more importantly leads to {\em locally} large $e_{\rm cr}$ which can trigger the SC runaway discussed above.

\item{ Tangled fields:} Magnetic fields are highly ``tangled'' \citep{su:fire.feedback.alters.magnetic.amplification.morphology,ji:fire.cr.cgm}, reducing $\tilde{\kappa}_{\rm eff}$. And in some cases (e.g.\ strong oblique shocks), perpendicular ${\bf B}$-fields enhance CR ``trapping'' in high-density gas, which can enhance $L_{\gamma}$.

\item{ Local Turbulent Fluctuations:} Both ET and SC models depend on the local turbulent dissipation/cascade rate (as well as e.g.\ magnetic field strengths). But, even on spatial scales resolved in our simulations, which are coherent on scales comparable to CR mean-free paths and scattering times, that rate has large (order-of-magnitude) local fluctuations on $\sim0.1-100\,$pc scales. For example, if $\kappa \propto u^{2}$, where $u$ is some local ISM property (like $|\delta v_{\rm turb}|$) that is log-normally distributed with factor $\sim3$ scatter, then the residence-time or scattering-weighted mean $\kappa$ will be a factor $\sim 10$ lower than the volume-weighted $\kappa$. This means that $L_{\gamma}$ will generally be larger than assumed using just the ``median'' properties of the ISM to estimate $\kappa$.

\end{enumerate}

Clearly, one cannot fully capture these effects by post-processing CR transport in simple analytic or empirical galaxy models. The effects above produce the large systematic internal variations of $\kappa$ and $v_{\rm st}$ in Figs.~\ref{fig:kappa.ecr.vs.weight}-\ref{fig:kappa.ecr.vs.redshift}. Moreover, almost all these effects go in the direction of increasing $L_{\gamma}$ and CR confinement. They also explain why the required $\kappa$ or $v_{\rm st}$ in our simulations are significantly larger than those obtained in ``leaky box'' or flat halo diffusion models models which assume free escape of $\sim$\,GeV protons outside of the thin or thick disk. They demonstrate why the connection between $\kappa$, $L_{\gamma}$, and $e_{\rm cr}$ in Figs.~\ref{fig:kappa.vs.model}-\ref{fig:ecr.vs.model} is not trivially linear as predicted by the toy model in \S~\ref{sec:params}.

\subsubsection{Fast Transport in Neutral Gas is Insufficient}
\label{sec:fast.neutral.not.enough}

In some of our models  $\tilde{\kappa}_{\rm eff}$ can be ``large'' ($\tilde{\kappa}_{29} \gg 1$) in {\em neutral} gas, but relatively small in the ambient warm ionized gas (WIM and inner CGM). This is true by construction in our ``two-$\kappa$'' model in \S~\ref{sec:fast.slow}, or due to ion-neutral damping in self-confinement models. We saw in \S~\ref{sec:results} that this reduces the predicted $L_{\gamma}$ and collisional losses (and therefore the CR ``residence time'' in the disk) by a surprisingly small amount (factor $<2$). There are two reasons for this. First, per \S~\ref{sec:fast.cosmo} above, a neutral cloud or ``slab'' of gas with local $\kappa_{\rm neutral} \rightarrow \infty$ will just converge to constant $e_{\rm cr}$ set by the ``boundary'' condition in the ambient WIM, so if the WIM has low $\kappa_{\rm ion}$ and traps CRs, they will still spend time in the cold clouds inside that WIM. Second, even if we ignore the effect above and assume that the CR residence time in a local ``patch'' simply scales with the local $\sim 1/\tilde{\kappa}_{\rm eff}$ (the ``free escape'' limit), we note that $L_{\gamma}$ and grammage scale with the hadronic losses as $L_{\gamma}\propto \int e_{\rm cr}\,\rho\,d^{3}{\bf x} \propto \int (1/\kappa)\,d M_{\rm gas} \propto M_{\rm ion}/\langle \tilde{\kappa}_{\rm ion} \rangle + M_{\rm neutral} / \langle \tilde{\kappa}_{\rm neutral} \rangle$ (where $M_{\rm ion}$ and $M_{\rm neutral}$ are the total mass of ionized gas and neutrals in the galaxy+CGM). So even if $\kappa_{\rm neutral}\rightarrow\infty$, this can only reduce $L_{\gamma}$ by at most a factor $\sim1 +  M_{\rm neutral}/M_{\rm gas,\,total}$ relative to a model with $\tilde{\kappa}=\tilde{\kappa}_{\rm ion}$ everywhere. In dwarf galaxies, in particular the SMC, LMC, and M33, {\em most} of the gas is ionized, so this is a small correction, and even in the MW or M31, this is  a factor only $\approx 1.5-2$.

\subsubsection{Can Faster Outflows or \Alf\ Speeds Reduce the Required Transport Coefficients?}
\label{sec:faster.outflow.or.B}

\changedtext{It is clear that our ``advection+\Alf{ic} streaming'' ($\kappa_{29}=0$, $v_{\rm st}=v_{A}$) simulations severely over-predict the observed CR grammage, energy density, $\gamma$-ray luminosity, etc. However, large theoretical uncertainties remain in predicted galactic outflow and magnetic field properties \citep[see e.g.][for a review]{2017ARA&A..55...59N}. So although our comparisons between FIRE simulations and observations in previous work (see references in \S~\ref{sec:intro}) suggests plausible agreement, it is possible that real galaxies feature significantly stronger outflows or magnetic fields, reducing the residence time even with $\kappa\rightarrow 0$. But even if we ignore all the complications described above, it seems implausible that this could significantly reduce the values of $\tilde{\kappa}_{\rm eff}$ or $\bar{v}_{\rm st,\,eff}$ required by the observations.}

\changedtext{First consider outflows/pure advection: there are at least three major issues invoking outflows to provide ``most'' of the CR transport. (1) The required outflow speeds must reach at least $\bar{v}_{\rm st,\,eff} \sim 300-3000\,{\rm km\,s^{-1}}$ (Fig.~\ref{fig:kappa.vs.vstream}) in most galaxies at $z\sim0$, including dwarfs -- but these are much larger than outflow speeds observed in all but the most extreme starburst/AGN systems \cite{martin99:outflow.vs.m,martin:2009.outflow.accel.starburst,2018Galax...6..138R}. (2) In the pure-advection limit, the ``residence time'' of CRs is equivalent to the ``residence time'' of gas in the ISM; but the observationally-favored CR residence times $\sim 10^{7}\,$yr are much shorter than even a single Galactic dynamical time $\sim 10^{8}\,$yr. So even gas which accretes falling in the free-fall velocity, mixes in a single dynamical time, and then accelerates outwards to the escape velocity will exceed observed CR residence times. (3) Most of the ISM observed (and simulated), even in dense star-forming regions, is {\em not in outflow} \citep{evans:2009.sf.efficiencies.lifetimes}. Equivalently, most of the CR $\gamma$-ray emission, residence time, and grammage comes not from outflows but from the diffuse ISM; and of course the Galactic constraints on CRs (measured at Earth) come specifically from gas not in outflow. So even infinitely-fast outflows will only reduce the required transport speeds in the non-outflowing ISM by a factor of order the ISM mass fraction in outflows (similar to our argument above regarding cold clouds), which is never larger than tens of percents.}

\changedtext{Next, consider \Alf{ic} streaming. Here the problem is obvious: to approach the required transport speeds and therefore observed grammage/residence times/$\gamma$-ray luminosities without invoking super-\Alf{ic} streaming or diffusion would require $v_{A} \sim v_{\rm st,\,eff} \sim 1000\,{\rm km\,s^{-1}}$ in dwarf and MW-like galaxies, i.e.\ for typical ISM gas densities $n\sim 1\,{\rm cm^{-3}}$ we would require $|{\bf B}| \sim 500\,{\rm \mu G}$, orders-of-magnitude larger than observed. Even if we arbitrarily multiply the magnetic field strengths in our simulations by a factor $\sim 10$ (larger than what is allowed by observations; see \citealt{guszejnov:fire.gmc.props.vs.z}), we would still require a volume-weighted $v_{\rm st,\,eff} \gg v_{A}$ in Fig.~\ref{fig:kappa.vs.vstream}.}

\changedtext{It is therefore difficult if not impossible for these effects to alter our implied constraints on CR transport speeds by more than an order-unity factor.}

\subsection{Extrinsic Turbulence}
\label{sec:discuss.extrinsic}

\subsubsection{\Alf\ Modes}
\label{sec:discuss.extrinsic:alfven}

Consistent with conventional wisdom, we find that most standard extrinsic turbulence models which assume scattering is dominated by resonant \Alf\ waves modes (e.g.\ our ``\Alf-C00'' models and their variants, ``\Alf-YL02,'' ``\Alf-Hi'' and related models) produce negligibly small CR scattering (i.e.\ higher $\kappa$) compared to the observationally-inferred levels at $\sim$\,GeV energies (see Table~\ref{tbl:transport}). Correspondingly, these models alone (i.e.\ including no other scattering sources) under-predict the observed $L_{\gamma}$ and MW grammage, as well as the CR energy density at the solar circle. Even if we neglect anisotropy and its effects on the scattering rate completely, giving $f_{\rm turb}=1$ (our ``\Alf-Max'' model), this is only just barely able to reach the scattering levels observed. 

\subsubsection{Magnetosonic Modes}
\label{sec:discuss.extrinsic:fast}

If we assume a cascade of fast modes down to resonant scales $\sim r_{\rm L}$, assuming such modes are fully-isotropic and ignoring any mode-damping (e.g.\ our ``Iso-K41'' and ``Fast-NoDamp'' models) then we would obtain excessively high scattering rates (low $\kappa$), clearly violating the observational constraints by factors of $\sim 10-100$ (regardless of details of the power spectrum  or whether we assume additional streaming at $\sim v_{A}$). But such models are clearly unphysical: in the warm WIM/CGM discussed above, accounting for just Braginskii viscosity as a damping mechanism and assuming trans-sonic turbulence, the equivalent Kolmogorov scale for fast (or perpendicular slow) modes is a factor $\ell_{\rm Kolm}/r_{\rm L} \sim 10^{5}\,(T/10^{5}\,K)^{2}$ larger than the gyro-resonant scales (in colder gas, ion-neutral damping and atomic/molecular collisional viscosity similarly gives $\ell_{\rm Kolm} \gtrsim 10^{4}\,r_{\rm L}$). Accounting for damping, the power in isotropic magnetosonic modes with wavelengths $\lambda \sim r_{\rm L}$ (hence their contribution to resonant scattering) should be vastly smaller than that in (undamped) \Alf\ waves at similar wavelengths. 

However, \citet{yan.lazarian.04:cr.scattering.fast.modes,yan.lazarian.2008:cr.propagation.with.streaming} argued that non-resonant fast modes with $\lambda \gg r_{\rm L}$ (plus undamped parallel gyro-resonant fast modes) can produce efficient CR scattering: we adopt their proposed scalings in our ``Fast-YL04'' model and show that this could be allowed, and in fact could produce an order-unity fraction of the observed scattering in gas that is both fully-ionized ($f_{\rm neutral} \lesssim 0.001$) and has $\beta \ll 1$. But this represents a small fraction of the ISM and almost none of the CGM, so likely contributes only modestly to observed scattering in total. Only by removing these restrictions (``Fast-Max'') can this model approach the full observed scattering. We also caution that several assumptions in YL04 remain controversial including the degree of resonance-broadening, whether long-wavelength fast modes can efficiently scatter low-energy CRs via transit-time damping, the $k^{-3/2}$ spectrum of the fast-mode power spectrum, and whether parallel fast modes follow the same spectrum below the scales where non-parallel modes are damped. Changing any of these decreases the implied scattering rate from fast modes by a large factor (e.g.\ our ``Fast-Mod'' model).

\subsection{Self-Confinement}
\label{sec:discuss.sc}

Again consistent with conventional wisdom, we find that ``standard'' self-confinement models predict much higher scattering rates and more efficient confinement of low-energy CRs compared to standard extrinsic turbulence models (even the YL04 models). So it is reasonable to expect SC dominates over ET-induced scattering at $\sim $\,GeV. However, we actually find that ``default'' or standard SC models predict {\em excessive} confinement -- higher $\nu$ and lower $\kappa$, resulting in excessively high $\gamma$-ray luminosities, grammage, residence times, and CR energy densities -- compared to observations. For reference, the predicted effective ``residence times'' of CRs in ``SC:Default'' model in MW-like halos are $\gg 10^{8}$\,yr, with CR energy densities $\gtrsim 10\,{\rm eV\,cm^{-3}}$, $\gamma$-ray production near the calorimetric limit, and grammage $X_{s} \gg 100\,{\rm g\,cm^{-2}}$. These characteristics are all in conflict with observations at the factor $\sim 10-1000$ level. 

As we discuss below, many of the model variations considered (see Table~\ref{tbl:transport}) do not resolve this issue: changing the CR energy by a factor $\sim10$, modest changes to the assumed turbulent structure, using equilibrium vs.\ non-equilibrium treatments of CR transport, or adopting $v_{A}^{\rm ideal}$ or $v_{A}^{\rm ion}$ as the relevant \Alf\ speed, all produce order-unity changes that are insufficient to explain these discrepancies. More fundamental changes, either invoking slower gyro-resonant growth rates (or lower scattering rates), or larger resonant-wave damping rates (or new damping mechanisms) by a factor $\sim 100$, are required to reproduce the observations.

It is worth noting that in Table~\ref{tbl:transport} and Figs.~\ref{fig:Lgamma} \&\ \ref{fig:ecr.vs.model}, many of the observable predictions of the SC models appear to be almost ``bimodal.'' Either the models predict excessive confinement near the calorimetric limit (with quite similar observables like those described above; e.g.\ our ``Default,'' ``$\kappa\times6$,'' ``$v_{A}^{\rm ideal}$,'' ``$10\,$GeV,'' ``$f_{\rm turb}$-5/DA,'' ``Non-Eqm,'' models), or they ``jump'' to a new solution with much higher-diffusivity, lower $L_{\gamma}/L_{\rm sf}$ and grammage, and lower $e_{\rm cr}$ at the MW solar circle, all in quite good agreement with the observations (e.g.\ our ``$f_{\rm turb}$-50/500/K41,'' ``NE-$f_{\rm turb}$-100,'' ``$f_{\rm QLT}$-100'' models). This owes to the ``self-confinement runaway'' or ``bottleneck'' effect described in \S~\ref{sec:fast.cosmo}: because SC models limit the absolute CR flux, the transport ``speed'' ($\kappa$ or $v_{\rm st}$) scales inversely with the CR energy density $e_{\rm cr}$ (Eq.~\ref{eqn:kappa.self.confinement}). Thus if there is a rapid injection of  CRs (say from clustered SNe), $e_{\rm cr}$ rises rapidly, lowering $\kappa$, which slows CR escape, increasing $e_{\rm cr}$ and further lowering $\kappa$, in a runaway, until the CRs in that region lose their energy to collisions (hitting the calorimetric limit). To avoid this, the ``pre-factor'' in the diffusive transport speeds, i.e.\ the damping rates $\Gamma_{\rm damp}$ or growth factor $f_{\rm QLT}$ must be large enough that CRs can efficiently escape these ``worst-case'' (most efficiently-trapped) environments. Once they do so, $e_{\rm cr}$ is made smooth by diffusion, and a ``smooth'' or ``average'' diffusivity becomes more reasonable.

\subsubsection{Fast Transport in Neutral Gas \&\ Choice of \Alf\ Speed}
\label{sec:discuss.sc:fast.neutral}

In the {\em neutral} ISM all the self-confinement models here {\em do} predict large $\tilde{\kappa}_{\rm eff} \gg 10^{29}\,{\rm cm^{2}\,s^{-1}}$, regardless of how we treat the \Alf\ speed when $f_{\rm ion} \ll 1$ (\S~\ref{sec:alfven.speed}). If we take $v_{A}=v_{A}^{\rm ion}=f_{\rm ion}^{-1/2}\,v_{A}^{\rm ideal}$ in Eq.~\ref{eqn:kappa.self.confinement}, then this becomes large for $f_{\rm ion} \ll 10^{-6}$ in GMCs, suppressing the ``$\kappa_{\|}$'' term in Eq.~\ref{eqn:kappa.self.confinement}, but giving large $v_{\rm st}=v_{A}$ so $\kappa_{\rm eff} \sim \gamma_{\rm cr}\,v_{\rm st}\,\ell_{\rm cr} \sim 10^{31}\,{\rm cm^{2}\,s^{-1}}\,\ell_{\rm cr,\,kpc}\,B_{\rm 5 \mu G}\,n_{10}^{-1/2}\,(f_{\rm ion}/10^{-8})^{-1/2}$. If, instead, we take $v_{A}=v_{A}^{\rm ideal}$, then (taking $\Gamma\rightarrow \Gamma_{\rm in}$) we have $\kappa_{\rm eff} \sim \kappa_{\|} \sim 0.3\times10^{31}\,{\rm cm^{2}\,s^{-1}}\,\ell_{\rm cr,\,kpc}\,e_{\rm cr,\,eV}^{-1}\,n_{10}^{3/2}\,T_{1000}^{1/2}\,\CRegy$. But for the reasons discussed in \S~\ref{sec:fast.neutral.not.enough} this alone does little to alter $L_{\gamma}$ or the other observational constraints in Table~\ref{tbl:transport} and Fig.~\ref{fig:Lgamma}: the over-confinement from SC models occurs in ionized, not neutral gas. And in the volume-filling WIM/CGM phases $f_{\rm ion}\sim1$ and $v_{A}^{\rm ideal} \approx v_{A}^{\rm ion}$, so the choice of \Alf\ speed does not produce any difference.

\subsubsection{Equilibrium vs.\ Non-Equilibrium Models}
\label{sec:discuss.sc:eqm}

We find that adopting the more detailed non-equilibrium evolution of the coefficients $\kappa_{\|}$, $v_{\rm st}$ as proposed in \citet{thomas.pfrommer.18:alfven.reg.cr.transport} (\S~\ref{sec:non.equilibrium}) makes  little difference to our results, compared to adopting the ``local equilibrium'' description  in Eq.~\ref{eqn:kappa.self.confinement} (using the same damping coefficients). This is not surprising, as the timescale for $\kappa$ to reach the local equilibrium value is short $\sim \Gamma^{-1} \sim 3000\,{\rm yr}\,\Gamma_{-11}^{-1}$. In the non-equilibrium case, CRs do escape the galaxy {\em slightly} more easily, as they can ``free stream'' a bit longer before $e_{A}$ and the scattering rate ``build up.'' However, this is likely at least somewhat artificially enhanced in our simulations  here, because we adopt a ``reduced speed of light'' $\tilde{c} < c$ (which increases the CR ``mean free path'' $\sim \kappa/\tilde{c}$), so we caution against over-interpreting the result.

\subsubsection{Over-Confinement in the WIM \&\ Inner CGM}
\label{sec:discuss.sc:overconfine}

Consider our ``default'' SC models (with $f_{\rm QLT}=f_{\rm cas}=1$), in ionized gas representative of the warm and hot phases of the ISM and CGM. Ion-neutral damping is negligible under these conditions.\footnote{While ion-neutral damping is efficient in dense gas ($n_{1} \gg1$) as $f_{\rm neutral}\rightarrow 1$ (with $f_{\rm ion} \lesssim 10^{-6}$ very small), if $f_{\rm neutral} \lesssim 1$ (so $f_{\rm ion}$ is not $\ll 1$), then achieving an effective isotropic diffusivity $\tilde{\kappa}_{29} \gtrsim 1$ requires $f_{\rm neutral} \gtrsim e_{\rm cr,\,eV} / (\ell_{\rm cr,\,kpc}\,n_{1}^{3/2})$. So at densities $n \lesssim 1\,{\rm cm^{-3}}$, or temperatures $T\gtrsim 2\times10^{4}\,$K (where $f_{\rm neutral} \ll 0.01$ drops exponentially), $\Gamma_{\rm IN}$ is small both compared to other damping mechanisms ($\Gamma_{\rm IN} \ll \Gamma_{\rm turb} + \Gamma_{\rm LL} + \Gamma_{\rm NLL}$) and compared to the observationally-required damping rates.} Non-linear Landau (NLL) damping is also sub-dominant, and in fact {\em cannot} dominate $\Gamma_{\rm eff}$ in the WIM/inner CGM, without violating both the observational constraints on $e_{\rm cr}$ and $\tilde{\kappa}_{\rm eff}$: comparing $\Gamma_{\rm turb}+\Gamma_{\rm LL}$ (Eqs.~\ref{eqn:gamma.turb}-\ref{eqn:gamma.ll}) and $\Gamma_{\rm NLL}$ (Eq.~\ref{eqn:gamma.nll}) in Appendix~\ref{sec:damping}, we see that $\Gamma_{\rm NLL} \gg (\Gamma_{\rm turb}+\Gamma_{\rm LL})$ requires $e_{\rm cr,\,eV} \gg  40\,(1+2.5/\beta^{1/2})^{2}\,\delta v_{10}^{3}\,n_{1}^{2}\,f_{\rm cas}^{2}\,T_{4}^{1/2}\,B_{\mu {\rm G}}^{-2}$. But if this condition were met, inserting these values of $e_{\rm cr}$ and $\Gamma_{\rm eff} \approx \Gamma_{\rm NLL}$ in Eq.~\ref{eqn:kappa.self.confinement} means the diffusivity would have to be {\em less than} $\kappa_{\|} \ll 5\times10^{25}\,{\rm cm^{2}\,s^{-1}}\,\ell_{\rm cr,\,kpc}^{1/2}\,\delta v_{10}^{-3/2}\,n_{1}^{-1/4}\,T_{4}^{1/2}$ (for any $\beta$), because $\kappa_{\|}$ for SC scales inversely with $e_{\rm cr}$. So in these environments $\Gamma_{\rm eff}$ is dominated by turbulent+linear Landau damping, which scale similarly as $\Gamma_{\rm LL} \approx 0.4\,\beta^{1/2}\,\Gamma_{\rm turb}$ and give $\kappa_{\|} \sim 10^{27}\,{\rm cm^{2}\,s^{-1}}\,(1+0.4\beta^{1/2})\,\delta v_{10}^{3/2}\,\ell_{\rm cr,\,kpc}\,\ell_{\rm turb,\,kpc}^{-1/2}\,n_{1}^{3/4}\,\CRegy^{1/2}\,e_{\rm cr,\,eV}^{-1}\,f_{\rm QLT}\,f_{\rm cas}$. 

Although these values of $\kappa$ and the $v_{\rm st}\approx v_{A}$ term\footnote{For $v_{\rm st}=v_{A}$, the corresponding ${\kappa}_{\rm eff} \sim \gamma_{\rm cr}\,v_{\rm st}\,\ell_{\rm cr} \sim 10^{27}\,{\rm cm^{2}\,s^{-1}}\,B_{\mu G}\,\ell_{\rm cr,\,kpc}\,n_{1}^{-1}$.} can become large in the {\em outer} CGM ($\gtrsim 30\,$kpc, where $e_{\rm cr}$ is small, see Fig.~\ref{fig:ecr.vs.model}), for $f_{\rm QLT}\,f_{\rm cas} \sim 1$ these are a factor of $\sim 30-300$ smaller in the WIM/inner CGM than the values needed to explain observations (Table~\ref{tbl:transport}). As discussed above, it is also necessary in these models to overcome the SC runaway or bottleneck effect: this is particularly onerous in regions like super-bubbles, which fill much of the volume around even new SNe (i.e.\ the CR sources, if SNe are clustered). With $n\sim 0.01$ and $e_{\rm cr,\,eV} \sim 10$ in these regions,  the local $\tilde{\kappa}_{\rm eff}$ can be as low as $\sim 10^{24}\,{\rm cm^{2}\,s^{-1}}$ -- equivalently the residence/escape time from a $\sim 100\,$pc-size super-bubble could reach $\sim$\,Gyr!

It is difficult to escape these conclusions: direct observational constraints on e.g.\ the turbulent velocity dispersions, scale-lengths, densities, and CR energy densities in the MW simply do not allow for large enough changes to those parameters to produce the required diffusivity without modifying $f_{\rm QLT}\,f_{\rm cas}$ above. The ISM parameters (e.g.\ $n$, $T$) are uncertain at the order-unity, not factor $\sim 100$ level. The variations across different times in the galaxy history, and different galaxies like {\bf m11f} and {\bf m12i} (as well as other galaxies we have simulated described in Appendix~\ref{sec:alternative.flux.eqns}), fully span the ``allowed'' observational range in these properties, and do not produce anywhere near the required values of $L_{\gamma}$ or grammage with $f_{\rm QLT}\,f_{\rm cas}\sim 1$. And, even if the ``median'' values of the scalings above for a given phase were promising, it is almost impossible to escape the conclusion that there will be substantial regions or local environments in the MW where the particular $\kappa_{\rm eff}$ predicted above would be very low, producing a severe ``bottleneck'' unless, again, $f_{\rm QLT}\,f_{\rm cas}$ or some related factor can be made factor $\sim 100$ larger.

\subsubsection{Possible Resolutions}\label{sec:resolutions.to.sc}

Reconciling self-confinement models with observations fundamentally requires factor $\sim 100$ lower scattering rates $\nu$ (and correspondingly larger $\tilde{\kappa}_{\rm eff}$) in the WIM/inner CGM, compared to the predictions obtained with the most commonly-assumed scalings (our ``default'' model). Qualitatively, there could be several explanations for the discrepancy:
\begin{enumerate}

\item{\bf Inefficient Scattering:} If CR scattering by gyro-resonant waves is much weaker than usually assumed\footnote{Uniformly decreasing the predicted scattering rate $\nu$ by a factor $f_{\rm scatter}$, all else equal, in our models, is equivalent to multiplying $\kappa_{\pm}$ given by the closure-relation in Eq.~\ref{eqn:kappa.closure} by $f_{\rm scatter}$, which in turn multiplies the ``local equilibrium'' $\kappa_{\|}$ in Eq.~\ref{eqn:kappa.self.confinement} by $f_{\rm scatter}$ as well, exactly identical to our ``$f_{\rm QLT}$'' parameter.} (for the same $\dBprl$ or $e_{A}$), this would directly lower $\nu$. Gyro-resonant waves have a reasonably well-understood structure \citep[see e.g.][]{Zirakashvili_2008,Riquelme.2008.bell..simulations,Ohira_2009} and the amplitudes predicted here are generally modest  (for diffusivity $\tilde{\kappa}_{29}$, the gyro-resonant $|\dBprl|/|{\bf B}| \sim 3\times10^{-4}\,(\CRegy/B_{\rm \mu G}\,\tilde{\kappa}_{29})^{1/2}$); however, two recent works studying the 
saturation of the gyro-resonant instability using the PIC method suggest possible ways that the effective $\nu$ might 
be lower than the QLT prediction. First, \citet{bai:2019.cr.pic.streaming} find that the time required for the CR distribution 
to become fully isotropic in the Alfv\'en-wave frame is much longer than predicted by the QLT estimate. This behaviour arises 
because of particularly inefficient scattering across the zero pitch angle ($\mu=0$) barrier, which is both slow and requires scatterers of very short wavelength
compared to $r_{\rm L}$ \citep{Voelk1973}. Second, in the highly anisotropic regime most relevant to 
regions close to sources,  \citet{holcolmb.spitkovsky:saturation.gri.sims} find very inefficient saturation of the gyro-resonant 
instability even when the self-excited Alfv\'en waves reach very large amplitudes, because only a single helicity (handedness) of Alfv\'en 
wave is produced by the CRs. Such an effect may help to limit the self-confinement ``runaway'' (see \S~\ref{sec:fast.cosmo}, \ref{sec:discuss.sc:overconfine}) in regions 
with  high $e_{\rm cr}$.

\item{\bf Lower Gyro-Resonant Growth Rates:} If the growth rate of the gyro-resonant instability is a factor $f^{-1}_{\rm QLT}$ smaller compared to the usual linear-theory expression $\Gamma_{\rm grow}^{\rm linear} \sim \Omega\,(n_{\rm cr}/n_{i})\,(\bar{v}_{\rm st}/v_{A}-1)$, then the quasi-linear saturation amplitude of $\nu \rightarrow \nu/f_{\rm QLT}$ (and $\kappa\rightarrow f_{\rm QLT}\,\kappa$). In the WIM/CGM, we have $\beta \gg 1$, $e_{\rm cr}/e_{\rm B}\gg1$, $\bar{v}_{\rm st}/v_{A} \sim 300-1000 \gg 1$, regimes where the instability is not well-studied and could potentially be strongly modified.\footnote{\changedtext{For the conditions of interest in the WIM/CGM and $\tilde{\kappa}_{29}\sim1$, we expect large $\beta \sim 35\,n_{1}\,T_{4}\,B_{\mu G}^{-2} \gg 1$ (using our standard notation to scale $T$ relative to $10^{4}\,$K, etc.), large ratio of CR to magnetic energy $e_{\rm cr}/e_{\rm B} \sim 40\,e_{\rm cr,\,eV}\,B_{\mu G}^{-2} \gg 1$, small fractional magnetic fluctuations at the gyro scale $|\dBprl|/|{\bf B}| \sim 3\times10^{-4}\,(\CRegy/B_{\rm \mu G}\,\tilde{\kappa}_{29})^{1/2} \ll 1$, small cosmic ray number density relative to ions $n_{\rm cr}/n_{i} \sim 10^{-9}\,e_{\rm cr,\,eV}\,n_{1}^{-1}\,\CRegy^{-1} \ll 1$, and large ratio of ``effective'' streaming speed to \Alf\ speed (corresponding to this diffusivity) $\bar{v}_{\rm st}/v_{A} \sim 300\,\tilde{\kappa}_{29}\,n_{1}^{1/2}\,B_{\mu \rm G}^{-1}\,\ell_{\rm cr,\,kpc}^{-1} \gg 1$.}}  The results of \citet{bai:2019.cr.pic.streaming} may again be of interest, if smaller-scale modes excited by low-$\mu$ and lower-energy particles  are required to fully saturate the gyro-resonant instability. Since such particles are much less numerous, implying the growth rate of the resonant modes is lower, the damping-growth balance that is usually assumed to saturate the instability and determine $\kappa$ (see \S\ref{sec:self.confinement}) might occur at significantly lower Alfv\'en-wave amplitudes than usually assumed. It seems plausible that such an effect could lead to significant enhancements in the self-confinement diffusion rates, although clearly more work is needed.

\item{\bf Larger Damping Rates or Alternative Mechanisms:} Since the saturation amplitude of $|\dBprl|^{2}/|{\bf B}|^{2}$, hence scattering rates, are inversely proportional to the damping rate $\Gamma_{\rm eff}$ in the quasi-linear theory models considered here (giving $\kappa \propto \Gamma_{\rm eff}$), it may instead be that damping rates are under-estimated. We stress that the required damping rates are still very small in absolute terms: $\Gamma_{\rm damp} \gtrsim 10^{-7}\,\Omega$ gives the required $\tilde{\kappa}_{29} \gtrsim 1$. Also, as discussed above, any such damping must operate efficiently in the ionized ISM and inner CGM: ion-neutral damping is efficient where neutral fractions are large but does not resolve the transport bottlenecks that appear in the fully-ionized WIM/HIM and inner CGM. 

One possibility is that the turbulent (or linear Landau) damping rates are larger by a factor $\sim 100$; i.e.\ the turbulent dissipation or cascade time $t_{\rm cascade}$ is shorter by a factor $f_{\rm cas} \sim 100$ at resonant scales. This may appear to be a large factor, but recall that the cascade models used to infer $t_{\rm cascade}$ and $\Gamma_{\rm turb}$ are extrapolated by factors reaching $\sim 10^{8}-10^{10}$ in scale from the ISM/CGM driving scales to $\sim r_{\rm L}$, so even quite small changes to the structure of the cascade could produce such a factor (although at least some of the variations we consider actually change this with the wrong sign, giving lower $\Gamma_{\rm turb}$). If other mechanisms (unresolved here), could directly drive turbulence on small scales (with e.g.\ an isotropic dispersion of $\sim 0.1\,{\rm km\,s^{-1}}$ on scales $\sim r_{\rm L}$) this would also resolve the discrepancy. And even given a particular cascade, we caution that the standard \citet{farmer.goldreich.04} model for how such a cascade  damps resonant \Alf\ waves has a number of uncertainties. Further, it remains untested in  non-linear simulations. 

There could also be additional damping/saturation mechanisms for gyro-resonant instabilities, not considered in our default models: e.g.\ non-linear effects, or self-interactions, or parasitic modes involving other (non-resonant) instabilities. There are many linear instabilities that couple magnetic fields, acoustic modes, gas, and other plasma components on scales $\sim r_{\rm L}$. For example, the acoustic instabilities studied in \citet{1986MNRAS.223..353D,1994ApJ...431..689B,Kempski2020} could be significant precisely in the warm/hot ionized medium when CR pressure gradients are weak. Recently \citet{squire.hopkins:RDI,squire:rdi.ppd,hopkins:2019.mhd.rdi.periodic.box.sims} discovered a class of ``resonant drag instabilities'' (RDIs) between dust and gas or magnetic fields that includes a sub-family of ``\Alf\ RDIs'' and ``cosmic-ray like'' RDIs which directly interact with \Alf\ waves and are unstable at wavelengths $\sim r_{\rm L}$ in the WIM with growth rates (for $\sim 0.1\,\mu{\rm m}$ grains) $\Gamma_{\rm RDI} \gg 10^{-11}\,{\rm s}^{-1}$, making them also potentially interesting here.

\end{enumerate}

\section{Comparison to Other Cosmological Simulations \&\ Previous Work}
\label{sec:comparison.previous}

To our knowledge, there has been no previous work comparing the various ET or SC-motivated CR transport models above in galaxy formation simulations.  Considering ``constant diffusivity'' models, outside of \paperonetwo, only a few other studies have compared galaxy simulations with CR transport to  the any of the observables discussed here. \citet{2016MNRAS.456..582S} considered ``constant diffusivity'' models without MHD or hadronic losses, with isotropic $\tilde{\kappa}_{\rm eff,\,29} \sim 0.03-0.3$ ($v_{\rm st}=0$), arguing that higher diffusivities are needed to match diffuse $\gamma$-ray emission constraints. \citet{2017ApJ...847L..13P} and \citet{buck:2020.cosmic.ray.low.coeff.high.Egamma} considered anisotropic MHD simulations with $v_{\rm st}=0$, and $\kappa_{\|} = 0$ or $\kappa_{\|} = 10^{28}\,{\rm cm^{2}\,s^{-1}}$ (i.e.\ $\kappa_{29}=0.1$). They concluded that with these low $\kappa_{\|}$ values, almost all galaxies produce $L_{\gamma}$ within a factor $\sim 1-3$ of the calorimetric limit, with grammage $X_{s} \gtrsim 100\,{\rm g\,cm^{-2}}$ in MW-like galaxies (see Appendix~\ref{sec:comparison.arepo}), and $e_{\rm cr} \sim 20\,{\rm eV\,cm^{-3}}$ at the ``solar circle.'' All of these results are similar to our constant-diffusivity models with similar $\kappa_{\|}$, supporting  our conclusions  regarding both the transport speeds required and the relatively minor effect from dense gas. However, \citet{buck:2020.cosmic.ray.low.coeff.high.Egamma} argue that their low-$\kappa_{\|}$ models, even their ``advection only'' models ($v_{\rm st}=0$, $\kappa_{\|}=0$), can reproduce the $\gamma$-ray observations (and therefore disagreed with our \paperone\ conclusions). We discuss this in  detail in Appendix~\ref{sec:comparison.arepo}, arguing  that the discrepancy stems not from a theoretical or simulation difference, but from how the $\gamma$-ray observations of the SMC/LMC/M33/MW/M31 are plotted, as well as their neglect of MW grammage and energy-density constraints. 

Within the MW, there is a long history of modeling CR transport in simplified analytic, time-static, smooth ``disk+halo'' models (generally neglecting phase structure or magnetic fields/anisotropy, but see e.g.\ \citealt{2012JCAP...01..011B}), again almost exclusively with ``constant diffusivity'' models (although a few studies have considered models where $\kappa$ varies with e.g.\ galacto-centric radius in some idealized fashion; see \citealt{2018ApJ...869..176L}). As we noted above and in \paperonetwo, our favored values of $\kappa_{\|}$ and the scalings in e.g.\ \S~\ref{sec:params} for our constant-$\kappa_{\|}$ models are broadly consistent with these  studies \citep[compare][]{blasi:cr.propagation.constraints,vladimirov:cr.highegy.diff,gaggero:2015.cr.diffusion.coefficient,2016ApJ...819...54G,2016ApJ...824...16J,2016ApJ...831...18C,2016PhRvD..94l3019K,evoli:dragon2.cr.prop,2018AdSpR..62.2731A}, {\em if} we compare to MW models that include an extended ($\sim 10\,$kpc) gaseous halo, and account for the difference between the isotropically-averaged diffusivity $\tilde{\kappa}_{\rm eff}$ usually measured in those models and the parallel $\kappa_{\rm eff}$ (a factor of $\sim 3$ larger) defined here. These analytic constant-$\kappa$ models generally find $\kappa_{29} \sim 1$ required to reproduce the observations: a factor $\sim 10-100$ larger than the diffusivity implied by older models that ignored any halo and assumed CRs escape outside the thin-disk scale-height ($\sim 200\,$pc).

\section{Conclusions}
\label{sec:conclusions}

We have presented the first numerical simulations that simultaneously follow self-consistent cosmological galaxy formation with CGM and ISM phase structure coupled to explicit  physically-motivated dynamical models of low-energy ($\sim$\,GeV) CR transport, where the relevant transport parameters (effective diffusivity $\kappa$ and/or streaming speed $v_{\rm st}$) are functions of the local plasma properties. We consider a wide range of micro-physical CR transport models, motivated by extrinsic turbulence (ET) and self-confinement (SC) scenarios, and compare the results of these directly to observational constraints in the MW and from nearby galaxies including $\gamma$-ray emission, CR energy densities, grammage, and residence times. We show that this is able to strongly constrain or rule out a large variety of proposed models and scalings for $\kappa$ and $v_{\rm st}$. Our major conclusions include:

\begin{enumerate}

\item{\bf The ``leaky box'' (or ``flat halo diffusion'') is a bad approximation, and the CGM gas is critical}: In {\em all} physically-motivated models we consider, CRs below $\lesssim 10\,$GeV remain confined (mean-free-paths $\lambda_{\rm mfp} \ll r$) at all galacto-centric radii out to well past the virial radius (scales $\lesssim$\,Mpc), even though $\kappa$ tends to increase slowly with radius. This imples that the CR scattering and confinement is strongly influenced by the presence of extended gaseous halos in the CGM (which are ubiquitous and contain {\em most} of the baryons) with scale-lengths $\sim 10-50\,$kpc. ``Toy'' or analytic CR transport models must include such large, {\em continuous} halos when considering $\sim$\,GeV CRs. This, in turn, necessarily implies larger transport speeds, compared to simpler leaky-box or flat-halo diffusion models.

\item{\bf There is no ``single'' diffusivity, and ISM/CGM phase structure is important:} Also in all the physically-motivated models here, CR transport parameters ($\kappa$, $v_{\rm st}$) depend strongly on properties like the local turbulent dissipation rate, magnetic field strength, ionization fraction, and gas density, which vary by {\em orders of magnitude} locally in both time and space along the trajectories of individual CRs owing to, e.g.\ rapidly time-varying ISM phase structure. Because of these variations, even taking spatial-and-time averages within a specific galacto-centric annulus, there is no ``single'' mean $\kappa$ (or $v_{\rm st}$). The volume-weighted and ``residence time'' or ``scattering rate''-weighted $\kappa$ (or $v_{\rm st}/v_{A}$) can differ by factors $\sim 10-100$.

\item{\bf Relatively ``large'' transport speeds are required:} In any models considered which reproduce the observational constraints, the effective {\em scattering-rate-weighted} mean parallel diffusivity $\kappa_{\rm eff,\,\|} \sim 10^{29}-10^{31}\,{\rm cm^{2}\,s^{-1}}$ in the ISM of dwarf and $\sim L_{\ast}$ galaxies within $\lesssim 10\,$kpc. This $\kappa_{\rm eff,\,\|}$ typically rises by factors $\sim 10-100$ in the CGM from $\sim 30-300$\,kpc. It also varies systematically between galaxies (becoming somewhat larger in smaller dwarfs) and with redshift (decreasing, on average, at high-$z$). This corresponds to highly super-\Alf{ic} streaming, with bulk transport speed $\bar{v}_{\rm st} \sim 10-1000\,v_{A}$. If one accounts for large CGM halos, fluctuations in local ISM properties, and isotropic vs.\ anisotropic diffusion, this required diffusivity is consistent with simple analytic and idealized models, but we emphasize that almost all non-linear effects in our simulations tend to enhance CR confinement (increasing the required $\kappa_{\rm eff,\,\|}$). 

\item{\bf Fast CR transport in neutral gas {\em alone} is not enough:} Neutral (molecular or HI) gas clouds in the ISM are embedded in volume-filling WIM and hotter ionized gas, most of which has local neutral fractions $1-f_{\rm ion} \lesssim 0.01$. The entire galaxy is itself embedded in ``inner CGM'' (scales $\lesssim 10-30\,$kpc) gas with densities $n\sim 10^{-3}-0.1\,{\rm cm^{-3}}$, temperatures $\sim 3\times10^{4}-10^{6}\,$K, and $1-f_{\rm ion} \ll 0.01$. So even if $\kappa\rightarrow \infty$ in neutral gas, CRs simply reach a constant energy density inside cold/neutral clouds, with their energy density and transport speed rate-limited by the boundary condition of this ionized ``cocoon.'' 

\item{\bf Extrinsic turbulence (probably) does not dominate:} As widely assumed, most physically-motivated ET models predict lower scattering rates for $\sim\,$GeV CRs, compared to what is observed (indicating that ET does not dominate $\sim$\,GeV CR scattering). However, if we ignore anisotropy and damping (e.g.\ assume an isotropic Kolmogorov turbulent spectrum from the driving scale $\ell_{\rm turb} \sim 0.1\,$kpc down to the gyro scale $r_{\rm L}\sim 0.1\,$au), the scattering rate from ET alone would severely exceed observational limits. Interestingly, one particular version of the proposed model from \citet{yan.lazarian.04:cr.scattering.fast.modes} for scattering by fast modes with wavelengths $\gg r_{\rm L}$ could produce scattering rates similar to SC in gas which is fully-ionized and also has $\beta \ll 1$, but this represents a small fraction of the ISM/CGM and the assumptions made in that model remain highly uncertain. Moreover, once anisotropy and damping are accounted for, {\em all} ET models considered here predict the incorrect qualitative dependence of grammage/residence time on rigidity at energies $\sim$\,GeV-TeV (opposite the observed trend, regardless of the turbulent spectrum). 

\item{\bf Simple quasi-linear expectations for self-confinement produce excessive confinement:} Using the most common quasi-linear estimates for  CR transport governed by SC -- i.e.\ assuming scattering rates are set by resonant \Alf\ wave energy densities  that are themselves set by the competition between  gyro-resonant  streaming instability growth  and damping with standard literature estimates for turbulent, ion-neutral, and Landau damping rates --  we predict galaxy-integrated scattering rates that are a factor $\sim 100$ larger than observationally allowed. This primarily comes from the volume-filling WIM and ``inner CGM'' discussed above, where ion-neutral damping is negligible (transport is fast, in these models, in neutral gas). We discuss possible resolutions in \S~\ref{sec:resolutions.to.sc}. It is plausible that scattering caused by the gyro-resonant instability could be less efficient than naive (quasi-)linear theory expectations by a factor $f_{\rm cas} \sim 100$; for example, due to inefficient isotropization of the CR distribution function across small pitch angles \citep{bai:2019.cr.pic.streaming}, or because  near-source scattering is weaker than expected \citep{holcolmb.spitkovsky:saturation.gri.sims}. Alternatively, damping rates from turbulence or linear-Landau effects could be larger by a factor $f_{\rm cas} \sim 100$, if the turbulence is less-strongly anisotropic (as compared what is implied by  usual critical-balance arguments), or if there are processes which can directly drive turbulence on scales closer to $r_{\rm L}$.  It is also possible that different damping processes, not usually considered, could dominate in the fully-ionized, warm, intermediate density environments that are particularly important for global CR transport.

\item{\bf Models exist which can reproduce CR observations:} We emphasize that if we lower the ``default'' SC scattering rate by a factor $f_{\rm QLT}$ or $f_{\rm cas} \sim 100$, then this model simultaneously reproduces (from fully-cosmological simulations) {\em all} the observational constraints we consider, including $\gamma$-ray measurements from SMC/LMC/M33/MW/M31 through starburst galaxies, the observed CR energy density at the solar circle, MW grammage and residence times {\em and} their dependence on rigidity. That this is possible at all, with just one dimensionless normalization constant ($f_{\rm QLT}\,f_{\rm cas}$) set to a single universal value, is extremely encouraging. We can also reproduce these observations at $\sim1\,$GeV with a constant-$\kappa$ model if we set $\kappa_{29}\sim 3-30$, or with a scaling motivated by ET if we artificially increase the ET scattering rate with e.g.\ our ``\Alf-Max'' or ``Fast-Max'' models, although neither the constant-$\kappa$ model nor these variant ET models predict the observed dependence of grammage/residence time on rigidity (as the SC-motivated models do).

\end{enumerate}

Our goal in this study is primarily to place first observational constraints on various ``a priori'' models which have been proposed in the literature for how the effective CR transport parameters (parallel diffusivity and/or streaming/drift speeds) depend on local plasma properties. We emphasize that our resolution is nowhere near sufficient to {\em predict} these scalings: rather we implement fully-dynamical CR evolution using different scalings derived from analytic models or PIC simulations. The qualitatively important resolution criteria are that we begin to resolve the multi-phase structure within the ISM and CGM (which determines these scalings) and that we at least marginally resolve the deflection length of CRs (so their trajectories through that medium can be followed). Our hope is that the conclusions above motivate some general conclusions for galaxy-scale CR transport, and motivate additional theoretical work exploring CR transport in self-confinement scenarios and/or fast-mode scattering. The simulations are of course an imperfect representation of reality: we discuss a wide range of additional caveats in \S~\ref{sec:alternative.flux.eqns}, including resolution, numerical implementation details, form of the CR flux equation, equilibrium vs.\ non-equilibrium treatments,  statistics (simulating additional galaxies), explicit inclusion of perpendicular diffusivities, and more. The uncertainties owing to some of these choices can be significant for some predictions (for extensive discussion of how resolution influences the ISM structure itself, see e.g.\ \citealt{hopkins:fire2.methods,hopkins:sne.methods}), but for our purposes here they generally produce factor $\lesssim 2$ differences in the predicted $\gamma$-ray luminosity or grammage given a {\em fixed} physical model for CR transport (see \paperonetwo). In contrast, different choices of CR transport models produce factor $\gg 1000$ differences. Given that the most interesting conclusions discussed above are factor $\sim 100$-level effects, it is likely that our conclusions are robust to these and other order-unity effects.

\datastatement{The data supporting the plots within this article are available on reasonable request to the corresponding author. A public version of the GIZMO code is available at \gizmourl. Additional data including simulation snapshots, initial conditions, and derived data products are available at \FIREurl.}

\acknowledgments{We thank the anonymous referee for helpful suggestions. Support for PFH was provided by NSF Collaborative Research Grants 1715847 \&\ 1911233, NSF CAREER grant 1455342, and NASA grants 80NSSC18K0562 and JPL 1589742. CAFG was supported by NSF 1517491, 1715216, and CAREER 1652522; NASA 17-ATP17-0067; and by a Cottrell Scholar Award. DK was supported by NSF grant AST-1715101 and the Cottrell Scholar Award. Numerical calculations were run on the Caltech compute cluster ``Wheeler,'' allocations from XSEDE TG-AST130039 and PRAC NSF.1455342 supported by the NSF, and NASA HEC SMD-16-7592. Data used in this work were hosted on facilities supported by the Scientific Computing Core at the Flatiron Institute, a division of the Simons Foundation.}

\bibliography{ms_extracted}

\begin{appendix}

\section{Default Damping Rates of Gyro-Resonant Alfv\'en Waves}
\label{sec:damping}

In self-confinement models (\S~\ref{sec:self.confinement}), the damping rate $\Gamma$ of gyro-resonant \Alf\ waves ($\dBprl$ or $e_{A}$) plays a central role. In the ISM/CGM, it is generally assumed that $\Gamma$ is dominated by a combination of ion-neutral ($\Gamma_{\rm in}$), turbulent ($\Gamma_{\rm turb}$), linear Landau ($\Gamma_{\rm LL}$) and non-linear Landau ($\Gamma_{\rm NLL}$) damping. \citet{zweibel:cr.feedback.review} and \citet{thomas.pfrommer.18:alfven.reg.cr.transport} summarize literature estimates of these damping rates from quasi-linear theory, which we adopt as our ``default'' set of damping rates, reviewed below.

\begin{enumerate}

\item{\bf Ion-Neutral Damping:} This is well-defined for a partially-neutral, hydrogen-helium plasma, giving:\footnote{In the neutral ISM at the densities we resolve in  our simulations (e.g.\ GMCs), we can just treat  the  hydrogen  and helium terms  here and safely neglect metal  ions and charged dust in Eq.~\ref{eqn:ion.neutral.damping}.}
\begin{align}
\label{eqn:ion.neutral.damping} \Gamma_{\rm in} &= \frac{\alpha_{\rm iH}+\alpha_{\rm iHe}}{2\,\rho_{\rm i}} 
\sim 10^{-9}\,{\rm s}^{-1}\,f_{\rm neutral}\,T_{1000}^{1/2}\,\rho_{-24} 
\end{align}
Here $\rho_{\rm i}$ is the mass density of ions, $\alpha_{\rm i\,X} \equiv (4/3)\,n_{\rm i}\,n_{\rm X}\,\sigma_{\rm i\,X}\,\sqrt{8\,m_{\rm i\,X}\,k_{B}T/\pi}$ where ${\rm X}\in\{ {\rm H},\,{\rm He}\}$, $m_{\rm i\,X}\equiv m_{\rm  i}\,m_{\rm X}/(m_{\rm i}+m_{\rm X})$, $m_{i}$ and $m_{\rm X}$ are the ion and species X masses (and $n_{\rm i}$, $n_{\rm  X}$ their number densities), $\sigma_{\rm i\,H}=10^{-14}\,{\rm cm^{2}}$, and $\sigma_{\rm i\,He}=3\times10^{-15}\,{\rm cm^{2}}$, and the latter  expression assumes an H mass fraction $\approx 0.75$ and defines $T_{1000} \equiv T/1000\,{\rm K}$, $\rho_{-24} \equiv \rho / 10^{-24}\,{\rm g\,cm^{-3}}$, and neutral fraction $f_{\rm neutral}=(1-f_{\rm ion})$.

\item{\bf Turbulent Damping:} Non-resonant motions will interact with and shear gyro-resonant \Alf\ waves: accurately capturing this requires understanding the non-linear behavior of turbulence on scales $\sim r_{\rm L}$, so it remains highly uncertain. Most estimates follow \citet{farmer.goldreich.04}, and assume a  \citet{GS95.turbulence} spectrum for ``strong'' \Alf{ic} turbulence with  an \Alf\ Mach number $\mathcal{M}_{A}[\ell_{A}] \equiv |\delta {\bf v}[\ell_{A}]|/v_{A}^{\rm ideal}=1$ at a scale $\ell_{A}\approx \mathcal{M}_{A}^{-3}\,\ell_{\rm turb}$, giving $\Gamma_{\rm turb} \sim  v_{A}^{\rm  ideal}\,(k_{\rm L}\,k_{\rm turb,\,A})^{1/2}$. Here $k_{\rm L} \sim 1/r_{\rm L}$ and $k_{\rm turb,\,A}\sim 1/\ell_{A}$ represent the resonant and injection wavenumbers, and stand in for appropriate averages over direction and wavenumber (meaning there is order-unity ambiguity here), giving: 
\begin{align}
\label{eqn:gamma.turb} \Gamma_{\rm turb} &
\equiv \frac{v_{A}^{\rm ideal}}{r_{\rm L}^{1/2}\,\ell_{A}^{1/2}}\,f_{\rm cas} \\
\nonumber & \sim 2\times10^{-11}\,{\rm s^{-1}}\,\delta v_{\rm turb,\,10}^{3/2}\,\ell_{\rm turb,\,kpc}^{-1/2}\,\rho_{-24}^{1/4}\,\CRegy^{-1/2}\,f_{\rm cas} 
\end{align}
where $\delta v_{\rm turb,\,10} \equiv  |\delta {\bf v}_{\rm turb}[\ell_{\rm turb}]| / 10\,{\rm km\,s^{-1}}$ and, as in \S~\ref{sec:extrinsic}, we represent our ignorance of the details of turbulence with the parameter $f_{\rm cas}$ (discussed in \S~\ref{sec:define.fcas}). 

\item{\bf Linear Landau Damping:} This is closely related to turbulent damping, and represents damping of oblique waves whose electric fields interact with the gas via Landau resonance when the propagation angle of the \Alf\ waves relative to the local magnetic field is changing owing to turbulent motions  \citep{zweibel:cr.feedback.review}. As a result, $\Gamma_{\rm LL}\approx (\pi^{1/2}/4)\,c_{s}/(r_{\rm L}^{1/2}\,\ell_{A}^{1/2})\,f_{\rm cas}$ scales with the local turbulent cascade time  in exactly the same manner as $\Gamma_{\rm turb}$, but with  a different pre-factor. So following \citet{zweibel:cr.feedback.review}, we can write:
\begin{align}
\label{eqn:gamma.ll} \Gamma_{\rm LL} &\approx \frac{\sqrt{\pi}}{4}\,\frac{c_{s}}{v_{A}^{\rm ideal}}\,\Gamma_{\rm turb} \sim 0.4\,\beta^{1/2}\,\Gamma_{\rm  turb}
\end{align} 

\item{\bf Non-Linear Landau Damping:} This represents wave-wave interactions, scaling non-linearly  with  the \Alf\  wave energy $e_{A\pm}$. For a  given $e_{A\pm}$, $\Gamma_{\rm  NLL,\,\pm} \approx (e_{A\pm}/e_{\rm B})\,\sqrt{\pi}\,c_{s}\,k_{\rm L}/8$ \citep{volk.mckenzie.1981}. As shown  in \S~\ref{sec:deriv}  below, if we assume local quasi-steady state equilibrium of the \Alf\ energy and CR transport coefficients, we do  not need to explicitly evolve the $e_{A\pm}$ terms but obtain the ``effective'' non-linear damping rate $\langle  \Gamma_{\rm NLL} \rangle \approx \Gamma_{\rm NLL}(\langle  e_{A\pm} \rangle)$, which becomes:
\begin{align}
\label{eqn:gamma.nll} \langle \Gamma_{\rm NLL} \rangle &\equiv \left[ 
\frac{(\gamma_{\rm cr}-1)\,\pi^{1/2}}{8}\,\left(\frac{c_{s}\,v_{A}}{r_{\rm L}\,\ell_{\rm cr}} \right)\,\left(\frac{e_{\rm cr}}{e_{\rm B}} \right) 
\right]^{1/2} \\ 
\nonumber &\sim 0.7 \times 10^{-11}\,{\rm s^{-1}}\,\left( \frac{e_{\rm cr,\,eV}}{\CRegy\,\ell_{\rm cr,\,kpc}} \right)^{1/2}\,\left( \frac{T_{10000}}{f_{\rm ion}\,\rho_{-24}} \right)^{1/4}
\end{align}

\end{enumerate}

\section{Non-Equilibrium Model \&\ Derivation of the Local, Quasi-Steady CR Transport Parameters}
\label{sec:deriv}

\subsection{Non-Equilibrium Scattering Rate Expressions}
\label{sec:deriv:non.eqm}

Begin from the non-equilibrium CR flux and gyro-resonant \Alf-wave dynamics equations as derived in \citet{thomas.pfrommer.18:alfven.reg.cr.transport}. Their expression  for $e_{\rm  cr}$ is identical  to ours (see \paperone), with the  definition $\Lambda_{\rm st} \rightarrow {\bf v}_{A}\cdot({\bf g}_{+} - {\bf g}_{-})$, where  the ${\bf g}_{\pm} \equiv (\gamma_{\rm cr}-1)\,({\bf F}\mp {\bf v}_{A}\,h_{\rm cr})/\kappa_{\pm}$ and associated $e_{A\,\pm} \approx |\dBprl|^{2}/4\pi$ represent  the scattering rates and energy in un-resolved \Alf\ waves propagating in the $\pm\bhat$ directions. Their expressions for the CR flux ${\bf F}$ and $e_{A\pm}$ are then:
\begin{align}
\label{eqn:flux.full} \frac{\mathbb{D}_{t}{\bf F}}{{c}^{2}} + \nabla_{\|} P_{\rm cr} &= -({\bf g}_{+} + {\bf g}_{-}) \\ 
\label{eqn:eA.full} \frac{\partial e_{A\pm}}{\partial t} + \nabla\cdot[{\bf u}\,h_{A\pm} \pm {\bf v}_{A}\,e_{A\pm}]
&= {\bf u}\cdot P_{A\pm} \pm {\bf v}_{A}\cdot {\bf g}_{\pm} - \Gamma_{\pm}\,e_{A\pm} 
\end{align}
where $h_{A\pm}\equiv e_{A\pm}+P_{A\pm}$, $P_{A\pm}\equiv e_{A\pm}/2$, and $\Gamma_{\pm}$ includes all the damping terms in \S~\ref{sec:damping}. In the gas momentum equation ($\partial \rho\,{\bf u}/\partial t$), we explicitly add $P_{A+}+P_{A-}$ to the total (magnetic+thermal+CR) pressure, and the additional ``source'' term $\nabla_{\|} P_{\rm cr} + {\bf g}_{+} + {\bf g}_{-} = {c}^{-2}\mathbb{D}_{t}{\bf F}$, to ensure manifest momentum  conservation. The damped \Alf-wave energy $(\Gamma_{+}\,e_{A+} + \Gamma_{-}\,e_{A-}$) is added to the gas thermal energy equation (i.e.\ it is converted from the explicitly-tracked \Alf-energy to thermal energy) instead of directly adding the ``streaming losses'' to the thermal energy. The system is closed by the relation:
\begin{align}
\label{eqn:kappa.closure} \frac{c\,r_{\rm L}}{\kappa_{\pm}} &= \frac{9\pi}{16}\,\left(\frac{e_{A\pm}}{e_{B}} \right)\,\left( 1   + \frac{2\,v_{A}^{2}}{c^{2}}  \right)
\end{align}
With these changes, our equations for the  gas momentum and energy, CR energy and flux, and \Alf-wave energy are exactly identical to the system of equations in \citet{thomas.pfrommer.18:alfven.reg.cr.transport}. 

\subsection{Local Equilibrium Expressions}
\label{sec:deriv:eqm}

Now assume that the CR flux and \Alf\ energy equations have reached local steady-state ($\partial/\partial t \rightarrow 0$, $\mathbb{D}_{t} \rightarrow  0$), and the advection terms (usually smaller by $\sim  \mathcal{O}(|{\bf u}|/c)$ compared to other terms) are negligible. In $e_{A\pm}$, one of the $\pm$ terms --- specifically the one corresponding to waves propagating down  the CR pressure gradient (i.e.\ with the same sign along $\pm\bhat$) to the direction of $-\nabla_{\|} P_{\rm cr}$ --- will have its corresponding $\pm {\bf v}_{A}\cdot {\bf g}_{\pm} $ term be positive-definite, competing against damping,  while  the other is purely-damped. Thus, the anti-parallel  $e_{A\pm} \rightarrow 0$, which implies the corresponding ${\bf g}_{\pm} \propto 1/\kappa_{\pm} \propto e_{A\pm} \rightarrow 0$ as well. Let us denote the ``surviving'' $e_{A\pm}\rightarrow  e_{A}$ and ${\bf g}_{\pm}\rightarrow {\bf g}$. Note that if we write ${\bf g} \equiv  (\gamma_{\rm cr}-1)\,({\bf F} - {\bf v}_{\rm st}\,h_{\rm cr})/\kappa_{\|}$, where $\kappa_{\|}$ corresponds to the appropriate ``surviving'' $\kappa_{\pm}$ and ${\bf v}_{\rm st} \equiv -v_{A}\,\nabla_{\|} P_{\rm cr} / | \nabla_{\|} P_{\rm cr}|$, the correct ``sign'' of the surviving ${\bf g}_{\pm}$ is ensured. So with these definitions in steady-state,  Eq.~\ref{eqn:flux.full} becomes $\nabla_{\|}P_{\rm cr} = -{\bf g}$  and the non-vanishing $e_{A\pm}$ equation (Eq.~\ref{eqn:eA.full}) becomes $0 = \pm{\bf v}_{A}\cdot {\bf g} - \Gamma\,e_{A}$, with $\Lambda_{\rm st} \rightarrow \pm {\bf v}_{A}\cdot {\bf  g}$. Here the $\pm {\bf v}_{A}$ sign corresponds again to the ``surviving'' direction so we can replace $\pm {\bf v}_{A} \rightarrow  {\bf v}_{\rm st}$, giving ${\bf g} = (\gamma_{\rm  cr}-1)\,({\bf  F} - {\bf v}_{\rm st}\,h_{\rm cr})/\kappa_{\|} = -\nabla_{\|}P_{\rm cr}$ and $\Lambda_{\rm st} = {\bf v}_{\rm st}\cdot {\bf g} = -{\bf v}_{\rm st}\cdot \nabla_{\|} P_{\rm cr} = \Gamma\,e_{A}$. 

Note now that $\Lambda_{\rm st}  =-{\bf v}_{\rm st}\cdot \nabla_{\|} P_{\rm cr}$ has exactly the same form as in our ``default'' implementation, and the thermal heating term $\Gamma_{+}\,e_{A+}+\Gamma_{-}\,e_{A-} \rightarrow \Gamma\, e_{A} =  \Lambda_{\rm st}$ from damping the un-resolved \Alf\ waves is exactly the ``streaming loss'' term (i.e.\ the streaming losses can be added directly to the thermal energy, as we do by default). The added term  in the gas momentum equation vanishes: $\nabla_{\|} P_{\rm cr} + {\bf g}_{+} + {\bf g}_{-} \rightarrow \nabla_{\|} P_{\rm cr} + {\bf g} = \mathbf{0}$. From ${\bf g} = -\nabla_{\|} P_{\rm cr}$ we also have ${\bf  F} = \kappa_{\|}\,\nabla_{\|} e_{\rm  cr} + {\bf v}_{\rm st}\,h_{\rm cr}$, i.e.\ our usual  streaming+diffusion approximation with streaming speed $v_{\rm st} = v_{A}$ and diffusivity $\kappa_{\|} = \kappa_{\pm}(e_{A})$. Because $\Gamma\,e_{A} = -{\bf v}_{\rm st}\cdot \nabla_{\|} P_{\rm cr}$, we can solve for $e_{A}$ and therefore $\kappa_{\|}$: but we should note that if the damping  is non-linear, $\Gamma$ is itself a function of $e_{A}$. For the assumptions in \S~\ref{sec:damping}, we can write $\Gamma = \Gamma_{1} +  \Gamma_{2}\,(e_{A}/e_{\rm B})$, where  $\Gamma_{1} = \Gamma_{\rm in}+\Gamma_{\rm turb}+\Gamma_{\rm LL}$ includes the terms independent of $e_{A}$ and $\Gamma_{2}\,(e_{A}/e_{\rm B}) = \Gamma_{\rm NLL}$ gives  the next-order  terms, and we obtain: 
\begin{align}
\label{eqn:eA.eqm} e_{A} &\rightarrow \langle e_{A} \rangle \equiv \frac{v_{A}\,|\nabla_{\|}P_{\rm cr}|}{\Gamma_{\rm eff}} = \frac{(\gamma_{\rm cr}-1)\,v_{A}}{\ell_{\rm cr}\,\Gamma_{\rm eff}}\,e_{\rm cr} \\ 
\label{eqn:kappa.eqm.sc.models} \frac{\kappa_{\|}}{c\,r_{\rm  L}} &\rightarrow \frac{16}{3\pi}\,\left( \frac{\ell_{\rm cr}\,\Gamma_{\rm eff}}{v_{A}} \right)\,\left( \frac{e_{\rm B}}{e_{\rm cr}} \right) \ \ \ \ \ , \ \ \ \ \ v_{\rm st} \rightarrow v_{A} \\ 
\Gamma_{\rm eff} &\equiv \Gamma(e_{A}\rightarrow\langle e_{A} \rangle) = \Gamma_{1}\,\left( \frac{\psi}{2\,\left[\sqrt{1+\psi}  -1 \right]} \right)  \\ 
\nonumber &\approx \Gamma_{1} + \Gamma_{1}\,\frac{\psi^{1/2}}{2} 
\equiv \Gamma_{\rm in} + \Gamma_{\rm turb} +  \Gamma_{\rm LL} + \Gamma_{\rm other} + \langle \Gamma_{\rm NLL} \rangle
\end{align}
where $\psi \equiv  4\,v_{A}\,|\nabla_{\|}P_{\rm cr}|\,\Gamma_{2} / (e_{\rm B}\,\Gamma_{1}^{2})$,  and the $\approx$ expression for $\Gamma_{\rm eff}$ is exact in both small and large-$\psi$ limits with $\langle \Gamma_{\rm NLL} \rangle = \Gamma_{\rm  NLL}(\langle e_{A} \rangle) = \Gamma_{1}\,\psi^{1/2}/2 = (v_{A}\,|\nabla_{\|}P_{\rm cr}|\,\Gamma_{2} / e_{\rm  B})^{1/2}$ (inserting $\Gamma_{2}=\sqrt{\pi}\,c_{s}\,k_{\rm L}/8$ gives $\langle \Gamma_{\rm NLL} \rangle$ in Eq.~\ref{eqn:gamma.nll}).\footnote{From Eq.~\ref{eqn:eA.eqm}, we can also confirm that the contribution  of the gyro-resonant \Alf\ waves to the {\em total} magnetic  pressure is  vanishingly  small, $P_{A}/P_{\rm B} \rightarrow  (8/9\pi)\,(c\,r_{\rm L}/\kappa_{\|}) \sim 3\times10^{-8}\,B_{\rm \mu G}^{-1}\,(10^{30}\,{\rm cm^{2}\,s^{-1}}/\kappa_{\|})$, so whether or not we separately include $P_{A\pm}$ in the total MHD pressure or fold it  into $P_{\rm B} = |{\bf B}|^{2}/8\pi$ as in our ``default'' models makes no difference.}

Finally, using the  fact that we can trivially re-write streaming+diffusion as ``pure diffusion'' or ``pure-streaming'' (\S~\ref{sec:methods:cr.transport}), it is convenient to re-write this in ``pure-streaming'' form, with $\kappa_{\|}\rightarrow 0$ $v_{\rm st}  \rightarrow \bar{v}_{\rm st} = v_{A} + \kappa_{\|}/(\gamma_{\rm cr}\,\ell_{\rm cr})$, i.e.:
\begin{align}
\bar{v}_{\rm st} &\rightarrow v_{A}\,\left[ 1 + \frac{4\,c\,r_{\rm L}\,\Gamma_{\rm eff}}{\pi\,v_{A}^{2}}\,\left( \frac{e_{\rm B}}{e_{\rm cr}} \right)\right] 
\end{align}

Thus, we see that in local steady-state, the full \citet{thomas.pfrommer.18:alfven.reg.cr.transport} expressions reduce to our default expressions with the  appropriate $v_{\rm st}=v_{A}$ and $\kappa_{\|}$. Because, in steady-state, $e_{A} \ll e_{\rm B}$ is miniscule, the non-linear effects of heating and/or pressure changes as the gyro-resonant \Alf\ wave distribution reaches this equilibrium  are negligible. And the timescale to reach this equilibrium is rapid: Eq.~\ref{eqn:eA.full} approaches local equilibrium on the damping timescale $\sim \Gamma^{-1} \sim 3000\,{\rm yr}$ in the warm ISM and $\sim 30\,$yr in the neutral ISM, while Eq.~\ref{eqn:flux.full} should approach steady-state on the scattering timescale $\sim \kappa/c^{2} \sim 10\,{\rm yr}$ (for $\kappa\sim 3\times10^{29}\,{\rm cm^{2}\,s^{-1}}$).

\subsection{Behavior of Solutions: Neither Streaming nor Diffusion}
\label{sec:deriv:behavior}

\begin{figure}
    \plotone{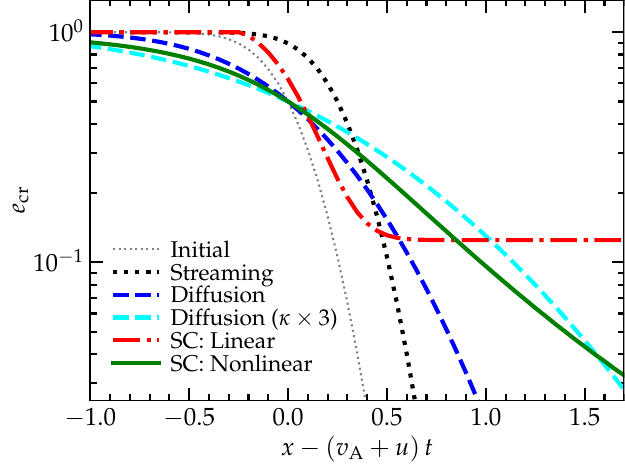}{0.95}
    \vspace{-0.25cm}
    \caption{Illustration of the behavior of the solutions for CR transport with $\kappa_{\|}$ and $v_{\rm st}$ given by self-confinement models (see \S~\ref{sec:deriv:behavior}). We evolve a one-dimensional toy model with parallel fields ($\bhat = \hat{\bf x} = \hat{\nabla}_{\|} e_{\rm cr}$) and constant $v_{A}$, $u$, and other background properties.  Taking $\tilde{c} \rightarrow \infty$ in Eq.~\ref{eqn:flux}, so  ${\bf F}=F\,\hat{\bf x}$ has its local-equilibrium value, and neglecting sources and sinks,  the CR transport equations reduce to $(\partial_{t} + [v_{A}+u]\,\partial_{x})\,e_{\rm cr} = -\partial_{x} F$. We consider an initial step-function-like $e_{\rm cr} = 0.5\,{\rm erfc}(x/0.3)$ evolved to time $t=0.5$ (arbitrary units) assuming: (1) traditional streaming/advection with $F = v_{\rm adv}\,e_{\rm cr}$ where $v_{\rm adv}=1$ is a constant; (2) traditional diffusion with $F = -\kappa\,\partial_{x}\,e_{\rm cr}$ and $\kappa=1$ or $=3$ is constant; (3) the expression for $F=F_{\Gamma} = \kappa_{\|}\,\partial_{x} e_{\rm cr} = (4\,c\,r_{\rm L}\,e_{\rm B}/\pi\,v_{A})\,\Gamma_{\rm eff}$ actually given by SC models (Eq.~\ref{eqn:kappa.eqm.sc.models}) assuming linear-damping terms dominate so $\Gamma_{\rm eff} = \Gamma_{\rm in}+\Gamma_{\rm turb}+\Gamma_{\rm LL}$, giving $\partial_{x} F = -C_{\rm L}\,{\rm SIGN}(\partial_{x} e_{\rm cr})$ with $C_{\rm L}=1/2$; (4) the expression for SC (Eq.~\ref{eqn:kappa.eqm.sc.models}) with non-linear terms dominant ($\Gamma_{\rm eff}=\Gamma_{\rm NLL}$), so $F_{\Gamma} = C_{\rm NL} |\partial_{x}\,e_{\rm cr}|^{1/2}$ with $C_{\rm NL}=1$. These are the simplest expressions that produce non-trivial behavior for each version of the equations, and we choose $v_{\rm adv}$, $\kappa$, $C_{\rm L}$, $C_{\rm NL}$ so that the ``effective'' transport speed $\bar{v}_{\rm st,\,eff}$ is the same around $(x,t)=(0,0)$. Despite the fact that we can write the SC scalings as a ``diffusion'' $\kappa_{\|}$ (Eq.~\ref{eqn:kappa.self.confinement}) or ``super-\Alf{ic} streaming'' $\bar{v}_{\rm st}$ (Eq.~\ref{eqn:vstream.self.confinement}), the behavior of even the simplest solutions is not the same as true diffusion or streaming/advection equations. 
    \label{fig:demo.sc.onedim}}
\end{figure}

Despite the language above, there are three important ways in which the solutions to the CR energy equation (Eq.~\ref{eqn:ecr}) for SC models differ from either a traditional streaming equation (${\bf F} = {\bf v}_{\rm st}\,h_{\rm cr}$, with ${\bf v}_{\rm st}$ constant) or traditional diffusion equation (${\bf F} = -\kappa_{\|}\,\nabla_{\|} e_{\rm cr}$, with $\kappa_{\|}$ constant), as often modeled.

First, and probably most important  as our main focus in this paper (\S~\ref{sec:methods:cr.transport}), $\kappa_{\|}$, ${\bf v}_{\rm st}$, and the ``parallel'' direction $\bhat$ are variable in both space and time. This means an infinite variety of solutions are possible, which need not have any resemblance to the solutions for constant streaming/diffusion models except in an infinitesimally small ``patch'' over an infinitesimally small time.

Second, if the flux is not in equilibrium ($\mathbb{D}_{t}{\bf F} \ne 0$ in Eq.~\ref{eqn:flux}), then obviously Eq.~\ref{eqn:ecr} will not match the expressions for a pure streaming/diffusion equation even if ${\bf v}_{\rm st}$ and $\kappa_{\|}$ are constants. Illustrations of this non-equilibrium behavior for finite $c$ are shown in \citet{jiang.oh:2018.cr.transport.m1.scheme} Figs.~1, 10, 15; \citet{thomas.pfrommer.18:alfven.reg.cr.transport} Figs.~5-6; and \citet{chan:2018.cosmicray.fire.gammaray} Figs.~B1, B4-B5.

Third, even if we assume $\mathbb{D}_{t}{\bf F} = 0$, that ${\bf v}_{\rm st} = {\bf v}_{A}$ has constant magnitude and direction (and $\bhat$ does not change), neglect all collisional losses and source injection, and assume the gas has constant ${\bf u}$, then Eq.~\ref{eqn:ecr} becomes: $d_{t} e_{\rm cr} = \pm \nabla_{\|} F_{\Gamma}$ where $d_{t} e_{\rm cr} = \partial e_{\rm cr}/\partial t + \nabla \cdot [({\bf u}+{\bf v}_{\rm A})\,e_{\rm cr}]$ represents simple advection of the CRs with the \Alf\ speed relative to the gas, $F_{\Gamma} \equiv (4\,c\,r_{\rm L}\,e_{\rm B}/\pi\,v_{A})\,\Gamma_{\rm eff} \approx (10^{5}\,{\rm erg\,s^{-1}\,cm^{-2}})\,n_{1}^{1/2}\,\Gamma_{-11}$ depends only on the gas density and damping rate, and the $\pm$ sign reflects the sign of $\bhat \cdot \nabla_{\|} P_{\rm cr} / |\nabla_{\|} P_{\rm cr}|$. But as others have noted, if $\Gamma\ne 0$, this $F_{\Gamma}$ term behaves {\em neither} as a traditional advection/streaming or as a diffusion term. We illustrate this explicitly with a simplified one-dimensional toy problem in Fig.~\ref{fig:demo.sc.onedim}. If the linear $\Gamma$ terms (e.g.\ ion-neutral, turbulent, linear-Landau) dominate, then $\Gamma$ and $F_{\Gamma}$ are totally independent of the CR properties (though they depend in a complicated manner on gas properties). So $F_{\Gamma}$ behaves as a ``source term'' which ensures the total flux down the CR pressure gradient matches the ``bottleneck'' value set by SC. This behavior is qualitatively distinct from e.g.\ a simple variable or super-\Alf{ic} advection velocity, which would introduce a term $d_{t} e_{\rm cr} = -\nabla \cdot ({\bf v}_{\rm advect}\,e_{\rm cr})$, proportional to the CR energy density. If non-linear Landau damping dominates, $F_{\Gamma} \propto \sqrt{|\nabla_{\|}\,e_{\rm cr}|}$, with a coefficient dependent on gas but not CR properties. This gives  a ``diffusive'' flux proportional to ${|\nabla_{\|} e_{\rm cr}|^{1/2}}$, instead of $\nabla_{\|} e_{\rm cr}$, which again produces qualitatively different behavior  from a standard diffusion equation, with weaker diffusion in the core and super-diffusive ``tails.''

\section{Diffusion Coefficients for Fast-Mode Scattering}
\label{sec:fast}

Here we briefly summarize the scattering rate via fast modes we adopt, directly following the assumptions in \citet{yan.lazarian.04:cr.scattering.fast.modes,yan.lazarian.2008:cr.propagation.with.streaming} [YL04]. Begin with the usual expressions for the $\kappa$ as a function of the pitch-angle diffusion coefficient $D_{\mu\mu}$ for relativistic CRs ($|{\bf v}_{\rm cr}| \approx c$), where $\mu=\cos{\theta_{p}}$ for pitch-angle $\theta_{p}$: $\kappa_{\|} = c\,\lambda_{\rm mfp}/3 = (c^{2}/4)\,\int_{0}^{1}\,d\mu\,(1-\mu^{2})^{2}\,D_{\mu\mu}^{-1}$. Then define the mode angle $\xi \equiv |\cos{\theta_{k}}| = |\hat{\bf k} \cdot \bhat| = k_{\|}/k$ (with $k_{\bot} \equiv (1-\xi^{2})^{1/2}\,k$), driving scale $\ell_{\rm turb}$, dimensionless wavenumber $\tilde{k} \equiv k\,\ell_{\rm turb}$ and $\tilde{r} \equiv r_{L} / \ell_{\rm turb}$, and large-scale $|{\bf B}|=B_{0}$. YL04 then adopt the expression from \citet{1975RvGSP..13..547V} (Eq.~45 therein) for $D_{\mu\mu}$, keeping only the $n=0$ (transit-time damping or TTD) and $n=\pm1$ (gyro-resonant) terms,  and dropping the \Alf{ic} terms. They assume that  fast modes have an isotropic $k^{-3/2}$ power spectrum  with $d^{3}\,{\bf k}\,I^{M}({\bf k}) = \mathcal{M}_{A}^{2}\,(B_{0}^{2}/8\pi)\,\tilde{k}^{-3/2}\,d\tilde{k}\,d\xi$ from the driving scale to some damping scale $k_{\rm damp}(\xi)$ that is angle-dependent, with zero power outside this range, giving:
\begin{align}
D_{\mu\mu}^{(n)} &= \frac{\mathcal{M}_{A}^{2}\,\Omega\,(1-\mu^{2})}{4\pi}\,\int_{0}^{1} d\xi \int_{1}^{\tilde{k}_{\rm damp}(\xi)} \frac{\xi^{2}}{\tilde{k}^{3/2}} \left[ J_{n}^{\prime}(x) \right]^{2} R_{n}\,  d\tilde{k}
\end{align}
where $J_{n}^{\prime}(x) = d J_{n}/dx$ is the derivative of the appropriate Bessel function with $x\equiv k_{\bot}\,v_{\rm cr,\,\bot}/\Omega = \tilde{k}\,\tilde{r}\,(1-\xi^{2})^{1/2}\,(1-\mu^{2})^{1/2}$. YL04 take the ``resonance function'' $R_{n}$ to be $R_{n} = (\pi^{1/2}/\Delta)\,\exp{(-q^{2}/\Delta^{2})}$ where $q = (k_{\|}\,v_{\|}  - \omega_{\rm fast} \pm n\,\Omega)/\Omega \approx k\,r_{L}\,\xi\,\mu - n$ and $\Delta \equiv k_{\|}\Delta v_{\|} / \Omega \approx k\,r_{L}\,\xi\,(1-\mu^{2})^{1/2}\,\mathcal{M}_{A}^{1/2}$, or equivalently $R_{n} = (\pi^{1/2}/\Delta)\,\exp{[-(\mu-n\,\mu_{0})^{2}/\Delta \mu^{2}]}$ with $\Delta \mu^{2} \equiv (1-\mu^{2})\,\mathcal{M}_{A}$ and $\mu_{0}^{-1} \equiv \tilde{k}\,\tilde{r}\,\xi$, as a result of the YL04 assumption that the resonance is broadened with $\Delta \mu \sim \Delta v_{\|} / v_{\bot} \sim \langle (|{\bf B}| - B_{0})^{2} \rangle^{1/4} / B_{0}^{1/2} \sim \mathcal{M}_{A}^{1/2}$. Defining $\tilde{D}_{\mu\mu} = (D_{\mu\mu}^{0} + D_{\mu\mu}^{1})/\Omega$, we have $\kappa_{\|}/(c\,r_{L}) = (1/4)\,\int_{0}^{1}\,d\mu\,(1-\mu^{2})\,\tilde{D}_{\mu\mu}^{-1}$ and the integrals can now be evaluated numerically given $\mathcal{M}_{A}$, $\tilde{r}$, and $\tilde{k}_{\rm damp}(\xi)$. We follow YL04 to calculate $k_{\rm damp}$ by assuming this is where the damping time becomes shorter than the cascade time, assuming  a $k^{-3/2}$ spectrum with $t_{\rm cas}^{-1} \approx (k/\ell_{A})^{1/2}\,v_{A}$, and setting this equal to $\Gamma_{\rm damp}(k,\,\xi,\,...)$ from the sum of collisionless, anisotropic viscous (Braginskii), ion-neutral, and other damping sources (using the expressions in Appendix~A of YL04). 

The simple expressions quoted in the main text are approximate fits to these numerical results over the dynamic range of interest here. They can be approximately derived as follows. When collisionless damping dominates, {\em if} parallel fast modes are undamped ($f_{\rm ion}=1$ and $\beta \ll 1$), then the gyro-resonant term ($n=1$) is sub-dominant in $\kappa$ and depends relatively weakly on plasma properties (see YL04 discussion), implying that the scaling for $\kappa_{\|}$ is dominated by the TTD ($n=0$) term. Ignoring $\mu \rightarrow 1$ (where the $n=1$ term dominates), the broad resonance assumption means $R_{0} \sim 1$, and because the rigidity is small $J_{n}^{\prime}(x) \approx x/2 \sim \tilde{k}\,\tilde{r}$, and $\Gamma_{\rm damp} \sim (\pi\,\beta\,m_{e} / 16\,m_{p})\,k\,v_{A}\,f(\xi)$ where $f(\xi)\sim 1$ for $\xi$ not too close to $0$ or $1$. Combining all of the $\xi$, $\mu$ integrals into a dimensionless function $g(\xi,\,\mu,\,\mathcal{M}_{A}) \sim 1$ we can then extract the dimensional scaling for $\kappa_{\|} \sim (c^{2} / D_{\mu\mu}^{0})\,g(...) \sim c\,\ell_{A}\,(\lambda_{\rm damp}/\ell)^{1/2}$ with $\lambda_{\rm damp}/\ell_{A} \sim (\beta\, m_{e}/m_{p})$. When viscous damping dominates (again assuming $f_{\rm ion}=1$ and $\beta \ll 1$), the resonant $n=1$ term dominates $\kappa_{\|}$ (at $\CRegy \lesssim 100$). Even with $\Delta \mu \sim 1$, the resonant $\mu_{0} \sim 1/k\,r_{\rm L}$ term in $R_{1}$ is large unless $k \gtrsim 1/r_{\rm L}$, which for a $\beta \ll 1$ viscous damping rate   of $\Gamma_{\rm visc}(\beta<1) \approx k^{2}\,\nu_{\rm v}\,(1-\xi^{2}) \sim 2\,k^{2}\,\nu_{v}\,\epsilon_{\xi}$ (defining  $\epsilon_{\xi} = 1-\xi$) requires $|\epsilon_{\xi}| \ll 1$, such that $k_{\rm damp} \gg 1/r_{\rm L}$. Taking these limits and evaluating gives $\kappa_{\|}$ inversely proportional to powers of $\epsilon_{\xi} \sim \tilde{r}^{3/2}\,(\ell_{A}\,v_{A}/\nu_{\rm v})$. 

Finally, regardless of what dominates $\Gamma_{\rm damp}$, if the parallel ($\xi\approx\pm1$) modes are damped on scales $k_{\rm damp} (\xi\rightarrow1) \ll 1/r_{\rm L}$, then $R_{1} \rightarrow 0$ rapidly as $\exp{[-(k_{\rm damp}\,r_{L})^{-2}]}$, and as a result $\kappa_{\|} \rightarrow \infty$ as we integrate to $\mu \rightarrow 1$ ({\em regardless} of the behavior of the TTD terms and broadening $\Delta \mu \sim \mathcal{M}_{A}^{1/2} \sim 1$). This occurs with ion-neutral damping ($\Gamma_{\rm damp} = \Gamma_{\rm in}$, independent of $\xi$), which  gives $k_{\rm damp}\,r_{\rm L} \approx (f_{\rm neutral}/f_{\rm n,\,0})^{-2} \lesssim 1$ where $f_{n,\,0} = 0.001\,(n_{1}\,\beta)^{-3/4}\,T_{4}^{1/4}\,(\ell_{\rm turb,\,kpc}\,\CRegy)^{-1/2}$. It also occurs if $\beta \ge 1$, in which case the viscous damping becomes strong as $\xi\rightarrow1$ with $\Gamma_{\rm visc} \approx k^{2}\,\nu_{\rm v}\,|3\,\xi^{2}-1|$, giving $k_{\rm damp}\,r_{L} \ll 10^{-4}$ for any physically-plausible parameters with Braginskii $\nu_{v}$. These give the damping ``cutoffs'' used in the text (\S~\ref{sec:extrinsic.turb.model}): $f_{\rm cut}=\exp{\{ (f_{\rm neutral}/f_{\rm n,\,0})^{4} + (\beta/0.1)^{1.5}\} }$.

\section{Additional Physical \&\ Numerical Variations Explored}
\label{sec:alternative.flux.eqns}

Here and in \paperonetwo, we have considered a large number of additional tests to confirm that the dominant uncertainty in CR transport is the form of $\kappa_{\ast}$, as opposed to e.g.\ numerical uncertainties or the detailed form of the transport equation. These include:

\begin{enumerate}

\item{\bf Equilibrium vs.\ Non-Equilibrium Transport Expressions:} This is discussed explicitly in the text (and see Appendix~\ref{sec:deriv} above), but we list it here for completeness.

\item{\bf Maximum ``Free Streaming'' Speeds:} $\tilde{c}$ represents the ``effective speed of light'' which determines the maximum free-streaming speed of CRs. In \paperonetwo\ we show this is a ``nuisance parameter,'' because the local steady-state CR flux and energy converge to the same values independent of $\tilde{c}$, so long as it is larger than local advection/diffusion speeds. In addition, we have tested all the models in this paper assuming $\tilde{c}=500\,{\rm km\,s^{-1}}$ or $\tilde{c}=1000\,{\rm km\,s^{-1}}$ as well as $\tilde{c} = {\rm MAX}(1000\,{\rm km\,s^{-1}},\,2\,\kappa_{\ast}/\ell_{\rm cr})$ (our default). So long as $\tilde{c} \gtrsim \kappa_{\ast}/\ell_{\rm cr} \sim 300\,{\rm km\,s^{-1}}\,\tilde{\kappa}_{29}/\ell_{\rm cr,\,kpc}$, then the results are robust to $\tilde{c}$; for the highest-$\kappa_{\ast} \gg 10^{30}\,{\rm cm^{2}\,s^{-1}}$ runs here, this means we require $\tilde{c} \gtrsim 1000\,{\rm km\,s^{-1}}$ to ensure converged results (otherwise $L_{\gamma}$ is artificially large because CRs are ``slowed down''), but even in this limit the qualitative conclusion that CRs escape efficiently is robust. 

\item{\bf Explicit Perpendicular Diffusion:} As shown in \paperonetwo, even assuming pure isotropic diffusion leads only to a factor $\sim 2-3$ lower $\kappa_{\ast}$ required to reproduce the same observed $L_{\gamma}$, grammage, etc. We confirm this in limited tests of our constant-$\kappa$ and ``SC100'' models. Physically, we generally expect the perpendicular diffusivity to be suppressed by a factor $\sim r_{\rm L}/\lambda_{\rm mfp}$: we have experimented with models that explicitly include perpendicular diffusive flux $F_{\bot} = \kappa_{\bot}\,(\nabla - \nabla_{\|}) e_{\rm cr}$ where $\kappa_{\bot} = (r_{\rm L}/\lambda_{\rm mfp})\,\kappa_{\|} \approx r_{\rm L}\,c/3$ and find (as expected) this makes a negligible difference compared to assuming pure parallel diffusion.

\item{\bf Resolution:} \changedtext{We emphasize the importance of resolving the ISM/CGM in the text, yet it is reasonable to worry that the smallest molecular clouds and star-forming regions are under-resolved. Despite this, we have shown in previous papers that GMC properties in these simulations including their size-mass relations (mean densities), linewidth-size relations, mass functions, magnetic field strengths, and lifetimes agree well with observations and appear converged down to clouds with as few as $\sim 10$ resolution elements \citep{hopkins:fire2.methods, grudic:sfe.gmcs.vs.obs, guszejnov:fire.gmc.props.vs.z, guszejnov:2019.imf.variation.vs.galaxy.props.not.variable, orr:ks.law, orr:non.eqm.sf.model, keating:co.h2.conversion.mw.sims, benincasa:2020.gmc.lifetimes.fire}. While this excludes the smallest clouds at our resolution, it includes the complexes that contain $>90\%$ of all galactic star formation \citep{rice:2016.gmc.mw.catalogue}. And as shown in the main text, our key conclusions are not particularly sensitive to the behavior of CRs in the most dense, neutral ISM because of its small volume-filling fraction. Moreover,} \paperonetwo\ consider extensive explicit resolution tests, in both cases varying the mass resolution of the ``constant-$\kappa$'' models by factors of $\sim 100$. In both cases (consistent with further extensive resolution studies in \citealt{hopkins:fire2.methods}) we showed that our predictions for dwarfs were only weakly sensitive to resolution. For MW-mass galaxies some galaxy properties do depend on resolution (for example, the central regions of the galaxies tend to be more dense at lower resolution, owing to less efficient resolution of galactic outflow ``venting''); however the qualitative effects of CRs, and range of allowed transport parameters, were robust to resolution. As $\Sigma_{\rm central}$ changed (weakly) with resolution, the corresponding $L_{\gamma}/L_{\rm sf}$ shifts along the ellipses for a given, single-resolution (i.e.\ systems move along the relations in Fig.~\ref{fig:Lgamma}, for fixed CR transport parameters). We have confirmed this result in our simulations without a constant $\kappa$ by running several of the models here (4 ET models and 4 SC models) for each of ({\bf m11i}, {\bf m11f}, {\bf m12i}) at factor $\sim8$ lower mass resolution (run initially to test and validate our implementation). \changedtext{In more limited tests of {\bf m12i} at $z\sim 0$ we have also confirmed that the exact choices for force softening and star formation criteria have no substantial effects on our conclusions.}

\item{\bf Form of the CR Flux Time Derivative:} The CR flux equation, Eq.~\ref{eqn:flux}, has subtle ambiguities related to the frame in which the CR flux is evaluated, order in $\mathcal{O}(v/c)$, assumptions about the form of the CR distribution function, and extrapolation  of scattering terms from quasi-linear theory. These are  discussed in e.g.\ \citet{zweibel:cr.feedback.review,thomas.pfrommer.18:alfven.reg.cr.transport,chan:2018.cosmicray.fire.gammaray} and references therein, and explored in \paperonetwo, but we briefly discuss them here. The formulations of CR transport in \citet{chan:2018.cosmicray.fire.gammaray}, \citet{jiang.oh:2018.cr.transport.m1.scheme}, and \citet{thomas.pfrommer.18:alfven.reg.cr.transport}, as well as simpler ``pure diffusion/streaming''  models commonly adopted in the literature are -- for a {\em  specific} value of the local $\kappa_{\ast}$  (i.e.\ assuming that $|\delta{\bf B}[r_{\rm L}]|^{2}$ has taken on some local quasi-equilibrium value) --  identical up to the form of the operator $\mathbb{D}_{t}{\bf F}$ in Eq.~\ref{eqn:flux}. In the ``pure diffusion/streaming'' model, $\mathbb{D}_{t}{\bf F}=\mathbf{0}$, so ${\bf F} \equiv  -\kappa_{\ast}\,\nabla_{\|} e_{\rm cr}$ and there is no flux  equation to solve  (simply a single advection+diffusion equation for $e_{\rm cr}$). In  \paperone, $\mathbb{D}_{t}{\bf F} =  \partial {\bf F}/\partial t + \nabla\cdot({\bf u}\otimes{\bf  F})$, and in \citet{jiang.oh:2018.cr.transport.m1.scheme} $\mathbb{D}_{t}{\bf F} =  (\hat{\bf F}\otimes\hat{\bf F}) \cdot [\partial({\bf F} + {\bf u}\,h_{\rm cr})/\partial t]$; neither of these papers attempted to derive the flux equation from first principles, but rather simply adopted a form (inspired by two-moment treatments of radiation hydrodynamics and similar problems) which relaxes to the correct behavior in various limits. \citet{thomas.pfrommer.18:alfven.reg.cr.transport} do attempt such a derivation, and obtain $\mathbb{D}_{t}{\bf F} \equiv \hat{\bf F}\,[\partial |{\bf F}|/\partial t + \nabla\cdot({\bf u}\,|{\bf F}|) +  {\bf F}\cdot\{ (\hat{\bf  F}\cdot \nabla)\,{\bf u}\} ] =  \partial {\bf F}/\partial t + \nabla\cdot({\bf u}\otimes{\bf  F}) + ({\bf F}\cdot \nabla)\,({\bf u}_{\|}-{\bf u}_{\bot})$.\footnote{Note that the \citet{thomas.pfrommer.18:alfven.reg.cr.transport} formulation only differs from  the \paperone\ formulation by the term $({\bf F}\cdot \nabla)\,({\bf u}_{\|}-{\bf u}_{\bot})/\tilde{c}^{2} =[ {\bf F}\{\hat{\bf F} \cdot[ (\hat{\bf F} \cdot \nabla){\bf u}] \}- ({\bf F}\cdot \nabla)\,{\bf u} ]/\tilde{c}^{2}$. This term (1) incorporates a Lorentz term  that manifestly ensures $\hat{\bf F}=\bhat$ is preserved, and (2) includes the ``pseudo-forces'' described by \citet{thomas.pfrommer.18:alfven.reg.cr.transport} which  arise because ${\bf F}$ is defined in the (non-stationary) fluid frame in which the CR distribution function can  be assumed to be gyrotropic.} But all of these are within  the $\mathcal{O}(1/\tilde{c}^{2})$ term  in Eq.~\ref{eqn:flux},  so they vanish when $\tilde{c}\rightarrow\infty$, {\em or} when  the CR flux reaches local quasi-steady state ($\mathbb{D}_{t}{\bf F}\rightarrow 0$), which occurs on the extremely-short CR mean free path/time defined in \S~\ref{sec:deriv}. In fact, the variants with $\mathbb{D}_{t}{\bf F}\ne \mathbf{0}$ above differ {\em only} if $\hat{\bf u}$ and $\bhat$ are non-uniform and time-dependent, on spatial/timescales below the CR mean free path (time) $\sim \kappa/\tilde{c}$ ($\sim  \kappa/\tilde{c}^{2}$), when $\tilde{c}$ is relatively small and the CR flux is out-of-steady-state. But this is exactly the regime where adopting $\tilde{c} < c$ means the CR flux differs from the ``true'' physical solution, so none of these can be exact. To the extent that our results are converged with respect to $\tilde{c}$, as demonstrated  in \paperonetwo, they must also  be independent of the choice of $\mathbb{D}_{t}$ here. Moreover \paperone\ considers  the much  more radical choice $\mathbb{D}_{t}=\mathbf{0}$, and shows the galaxy results are essentially identical. All our constant-$\kappa$ models have been re-run with the different variant $\mathbb{D}_{t}$ forms discussed above in \paperonetwo, where we showed this had a negligible effect on the observables predicted here. We have repeated this with a limited study of models ``Fast-YL04'' and ``SC100'' here, where we find the same result.

\item{\bf Form of the Scattering Terms:} Another ambiguity is whether to represent the scattering term in Eq.~\ref{eqn:flux} as ${\bf F} / \kappa_{\ast}$ with $\kappa_{\ast} \equiv \kappa_{\|} +  \gamma_{\rm cr}\,v_{\rm st}\,\ell_{\rm cr}$ (our default),  or as $({\bf F}-{\bf v}_{\rm st}\,h_{\rm cr})/\kappa_{\|}$, as in \S~\ref{sec:deriv}. Both are consistent with quasi-linear theory, and become exactly identical when $\tilde{c}\rightarrow \infty$ and/or the flux ${\bf F}$ reaches local quasi-steady-state ($\mathbb{D}_{t}{\bf F}$ is small), so again our experiments with different $\mathbb{D}_{t}{\bf F}$ and $\tilde{c}$ indicate our conclusions are robust to this choice. And because our ``favored'' models have a drift velocity $|{\bf F}|/h_{\rm cr} \gg v_{A}$, this is further minimized (generally contributing $<5\%$ corrections, re-running different models for select short periods). Moreover our ``non-equilibrium'' model (\S~\ref{sec:non.equilibrium}) adopts the $({\bf F}-{\bf v}_{\rm st}\,h_{\rm cr})/\kappa_{\|}$ form  and gives similar results to the equilibrium model with $\kappa_{\ast}$.

\item{\bf Form of the ``Streaming Loss'' Term:} The ``streaming loss'' term, $\Lambda_{\rm st}$ in Eq.~\ref{eqn:ecr} is well-motivated in local-steady-state, self-confinement models (where it takes the form $\Lambda_{\rm st} \approx v_{A}\,|\nabla_{\|} P_{\rm cr}|$), as it arises from the damping and thermalization of gyro-resonant \Alf\ waves (well below our simulation resolution limits) excited by CR streaming (see \S~\ref{sec:deriv}). It is less clear how it should behave in our ET models or models with sub-\Alf{ic} streaming. We discuss this and vary the term extensively in our constant-$\kappa$ models in \paperonetwo, considering $\Lambda_{\rm st} = {\rm MIN}(v_{A},\,v_{\rm st})\,|\nabla_{\|} P_{\rm cr}|$ (our default here), or $\Lambda_{\rm st} ={\bf v}_{\rm st}\cdot \nabla P_{\rm cr}$, or $\Lambda_{\rm st} =v_{A}\,|\nabla_{\|}P_{\rm cr}|$, or $\Lambda_{\rm st} =0$. There we showed this had very small ($\sim 10\%$, at high $\kappa$) effects on the observables we predicted. Here we have repeated these comparisons for a subset of our ET models at $z\sim 0$ (restarting them for a short time) to confirm that this produces nearly negligible perturbations to $L_{\gamma}$. We also find that any model where this $\Lambda_{\rm st}$ term is able to produce large CR losses in the ISM or inner CGM (where it might influence our predictions) is already in the well into the regime where collisional losses dominate inside of the galaxy ISM.

\item{\bf Exact Momentum-Conserving Formulation:} In our default formulation, we assume a local strong-coupling approximation so the CRs enter the gas momentum equation via the term $\nabla P_{\rm cr}$. As noted in \S~\ref{sec:deriv}, if we approximate the flux equation in the form described therein or in our second-moment expansion Eq.~\ref{eqn:flux} (both accurate to $\mathcal{O}(v/c)$), then exactly conserving total momentum accounting for the change in inertia of the CRs themselves would require adding a source term $[\nabla_{\|} P_{\rm cr} + {\bf g}_{+} + {\bf g}_{-}] = ({\bf F}-{\bf F}_{\rm eqm})/(3\,\kappa_{\ast}) = \mathbb{D}_{t}\,{\bf F} / \tilde{c}^{2}$ to the gas momentum (where ${\bf F}_{\rm eqm} = -\kappa_{\ast}\nabla_{\|}e_{\rm cr}$ is the local steady-state flux). This obviously vanishes as $\tilde{c}\rightarrow \infty$ or $|\mathbb{D}_{t}{\bf F}|\rightarrow 0$ so our tests of varying $\tilde{c}$, or taking $\mathbb{D}_{t}{\bf F} = \mathbf{0}$ exactly, show that the term should not change our results. We do not include this by default because, as noted in \citet{jiang.oh:2018.cr.transport.m1.scheme} and \paperone, if $\tilde{c} \ll c$, this term is artificially large and the CR contribution to the force will be under-estimated compared to a converged solution with respect to $\tilde{c}$ (because the CR flux deviation from equilibrium is artificially modified by $\tilde{c}$). 

\item{\bf Local Turbulent Velocity Estimator:} Because the local turbulent velocities $\delta v_{\rm turb}$ on a scale (of order our simulation resolution) $\ell_{\rm turb}$ appear in the scalings for both ET and SC (via turbulent damping) CR scattering, we have considered four different local on-the-fly estimators for this quantity. (1) Our default, from \citet{hopkins:virial.sf}, $\delta v_{\rm turb} = \| \nabla \otimes {\bf v} \|\,\ell_{\rm turb} \equiv (\sum_{ij} |\nabla_{j}\,v_{i}\,\ell_{\rm turb}|^{2})^{1/2}$ the Frobenius norm (sum over components) of the velocity difference across a resolution element estimated from the (non-slope-limited) velocity gradient with $\ell_{\rm turb}=\Delta x = (m_{i}/\rho_{i})^{1/3}$ the resolution scale. (2) The ``shear corrected'' norm (norm of the trace-free diagonalized shear tensor of the velocity field, constructed from $\nabla_{j}v_{i}$) times $\Delta x$, as defined and commonly used for \citet{smagorinsky.1963:eddy.approximation.for.diffusion.terms} ``subgrid-scale'' turbulent diffusion models \citep[see e.g.][]{colbrook:passive.scalar.scalings,escala:turbulent.metal.diffusion.fire}. (3) The direct dispersion $|\delta v_{\rm turb}|_{a}^{2} = \sum_{b}|{\bf v}_{b}-{\bf v}_{a}|^{2}$ across neighbors in a sphere of volume $\ell_{\rm turb}^{3}$. (4) The more sophisticated (but computationally expensive) method developed in \citet{rennehan:turb.diff.implementation.fancy}, motivated by detailed turbulence studies, where we smooth the velocity field on multiple scales in multiples of the resolution $\Delta x$, calculate the relative power in velocity fluctuations, and derive the associated turbulent $E(k)$ at $k\rightarrow 1/\Delta x$. On top of these variations, we also note that many of the models which involve $\delta v_{\rm turb}$ really use this as a proxy for $\delta {\bf B}_{\rm turb}$, assuming that at the \Alf\ scale $\ell_{A}$,  $\delta v_{\rm turb} \approx v_{A}$ and $\delta {\bf B}_{\rm turb} \sim |{\bf B}|$. So we have also re-computed {\em all} of the relevant scalings using $\delta {\bf B}_{\rm turb}$ measured directly in the code (with the same four estimators described above), to estimate $\ell_{A}$, and extrapolating the relevant assumed power spectra below this scale. We find that although these eight model variants can produce quite large (order of magnitude, in some cases) differences in the specific value of $\delta v_{\rm turb}({\bf x},\,t)$ estimated at any given point $({\bf x},\,t)$ in the ISM, the {\em statistics} produced by the different estimators are quite similar. A more detailed comparison of these in their own right will be the subject of future work, but relevant for this study, integral quantities like $L_{\gamma}$ are ultimately altered at the factor $\lesssim 2$ level (comparing all these variations), not enough to alter our conclusions.

\item{\bf Additional Statistics (Different Galaxies):} Given the very large number of different CR transport models we survey here, we chose to limit our study to three representative galaxies or ``zoom-in regions'' {\bf m11i}, {\bf m11f}, {\bf m12i} in Table~\ref{tbl:sims}. While this is still an improvement over comparing with a single MW model alone, one might worry that our conclusions could be biased by either limited statistical power or systematic effects owing to e.g.\ the structure or formation history of the particular galaxies. However, we have re-run most of the ``constant-$\kappa$'' models with a much larger number of simulations, presented in detail in \papertwo\ (along with some additional zoom-in regions of local groups following \citealt{garrisonkimmel:local.group.fire.tbtf.missing.satellites}): altogether $>35$ zoom-in regions containing several hundred resolved galaxies ranging in $z=0$ halo mass between $M_{\rm halo} \sim 10^{9}-10^{13}\,M_{\odot}$ (including specifically $10$ ``single'' MW-mass systems and 4 Local Group pairs each containing a MW and Andromeda-like galaxy). We show there that all our conclusions here regarding statistics of e.g.\ comparison with $L_{\gamma}/L_{\rm SF}$ and $e_{\rm cr}$, and the inferred observationally-allowed values of $\kappa$, are robust. We have also run a subset of the non-constant-$\kappa$ models here (``\Alf-C00,'' ``Fast-YL04,'' ``SC:Default,'' and ``SC:100'') on an expanded halo sample including halos ({\bf m10q}, {\bf m11q}, {\bf m11g}, {\bf m12f}) from \papertwo, with halo masses $\log(M_{\rm halo}/M_{\odot}) \sim(10, 11, 11.5, 12)$ and stellar masses $\log{(M_{\ast}/M_{\odot})}\sim(6.3, 9.0, 10, 10.8)$, respectively. Each of halos ({\bf m11q}, {\bf m11g}, {\bf m12f}) behave broadly similarly to our standard ({\bf m11i}, {\bf m11f}, {\bf m12i}), respectively (galaxies with similar mass) for each specific CR transport model. To the extent that they differ in e.g.\ $L_{\gamma}/L_{\rm sf}$ they move (slightly) {\em along}, not with off of, the relation defined by ({\bf m11i}, {\bf m11f}, {\bf m12i}) in Fig.~\ref{fig:Lgamma}. Halo {\bf m10q} (the least massive) is consistent with the extrapolation of these trends, but falls outside the plotted and observed range (with much lower mass/luminosity/density) in our comparisons. All of this is consistent with our larger statistical study in \papertwo.

\item{\bf CR Injection Efficiency:} As discussed in \papertwo, if we add additional sources of CRs (e.g.\ structure formation shocks, AGN) then this will further increase $L_{\gamma}$ without increasing $L_{\rm SF}$, requiring larger diffusivities to reproduce observations, but these are almost certainly sub-dominant for CR production compared to SNe in the galaxies of interest. If we change the assumed efficiency of CR production in SNe ($\epsilon_{\rm cr}$), in the calorimetric limit this changes $L_{\gamma}/L_{\rm SF} \propto \epsilon_{\rm cr}$, so reproducing the observations of the SMC/LMC/M33 with, say, $v_{\rm st} \sim v_{A}$ (so all galaxies are near-calorimetric) while also matching the observed starburst systems would require factor of $\sim 100$ variation in $\epsilon_{\rm cr}$ in SNe {\em as a function of galaxy properties} (which cannot be primarily metallicity, since this is constant for some observed systems with different $L_{\gamma}/L_{\rm SF}$). More importantly, changing $\epsilon_{\rm cr}$ does {\em not} change the median grammage or residence time ``per CR,'' so reproducing the grammage, residence time, and $L_{\gamma}$ observations simultaneously, {\em or} reproducing the $L_{\gamma}$ observations in different galaxies simultaneously with a constant $\epsilon_{\rm cr}$, requires $\epsilon_{\rm cr} \sim 0.1$. We have experimented in \papertwo\ with modest variations $\epsilon_{\rm cr} \sim 0.05-0.2$: the range of observations and simulation spread in predictions make it difficult to rule out factor $\sim 2$ changes in $\epsilon_{\rm cr}$, but at this level these variations have no qualitative effect on our conclusions.

\end{enumerate}

\section{Comparison to Low-Diffusion Models in Other Cosmological Simulations}
\label{sec:comparison.arepo}

\begin{figure}
    \plotone{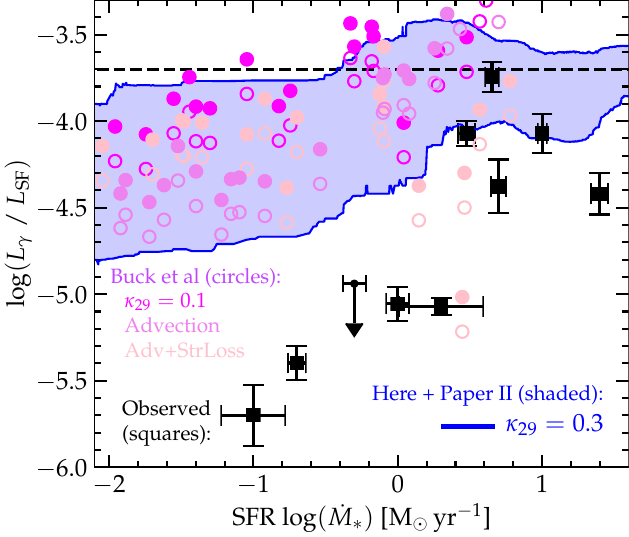}{0.95}
    \vspace{-0.2cm}
    \caption{Comparison of $\gamma$-ray emission $L_{\gamma}/L_{\rm SF}$ versus SFR (as Fig.~\ref{fig:LgammaVsSFRLIR}) in our low-diffusivity CD model $\kappa_{29}=0.3$ (shaded shows $2\sigma$ range) and observed (black points with error bars). We contrast (see \S~\ref{sec:comparison.arepo}) the results from \citet{buck:2020.cosmic.ray.low.coeff.high.Egamma} [B19; circles], who predict $L_{\gamma}$ from independent cosmological simulations without ISM phase structure, considering low-diffusivity models including (1) $\kappa_{29}=0.1$ (with $v_{\rm st}=0$), (2) advection-only ($\kappa_{29}=v_{\rm st}=0$), (3) advection+streaming losses ($\kappa_{29}=v_{\rm st}=0$, but still adding a rapid ``streaming loss'' sink term $=v_{A}\,\nabla P_{\rm cr}$ in the CR energy equation). Open points show the values of $L_{\gamma}$ and $L_{\rm SF}$/SFR taken exactly as given in B19's Fig.~14. Solid points correct these points to adopt the identical stellar and $\gamma$-ray bolometric corrections, $\gamma$-ray bandpass, and assumptions about hadronic loss rates as those adopted in the text here. Their predictions are nearly identical to ours for similar (low) diffusivity, and predict for $\kappa_{29} \ll 1$ that $>90\%$ of galaxies are within a factor $\sim 3$ of the calorimetric limit at any SFR. The B19 models also predict solar-circle grammage $X_{s} \gtrsim 100\,{\rm g\,cm^{-2}}$, CR energy density $e_{\rm cr} \sim 20\,{\rm eV\,cm^{-3}}$, and residence times $\gg 100\,$Myr, similar to our low-$\kappa$ models in Figs.~\ref{fig:ecr.vs.model}-\ref{fig:Xs.vs.model}. The predictions for $\kappa_{29} \lesssim 1$ are consistent between simulations and clearly ruled out by both $\gamma$-ray and MW observations: per \S~\ref{sec:comparison.arepo}, B19's conclusion that low-$\kappa$ models are observationally permitted stems from not considering MW constraints and from plotting the $\gamma$-ray data at the incorrect values of $L_{\rm SF}$.
    \label{fig:Buck}}
\end{figure}

Recently, \citet{2017ApJ...847L..13P,buck:2020.cosmic.ray.low.coeff.high.Egamma} (B19) explored the effects of explicit CR transport models in idealized isolated galaxy and cosmological simulations, similar in spirit to our \paperonetwo. These simulations used a different code and  numerical method  with somewhat lower resolution. They also employ a fundamentally different treatment of the ISM wherein any gas above a density $n>0.1\,{\rm cm^{-3}}$ is assigned a ``stiff'' effective (quasi-adiabatic) equation-of-state, with a SFR set by calibration to observations, and is assumed to launch galactic winds with a mass-loading and velocity set analytically to reproduce the galaxy mass function  following \citet{2017MNRAS.467..179G}. The scheme is designed for large-volume simulations that do not resolve ISM or outflow phase structure, so  we might expect significant differences from our results here.

The authors consider three transport models (1) CR advection only ($\kappa_{29}=0$, $v_{\rm st}=0$, with no ``streaming loss'' term); (2) diffusion-only with $\kappa_{29}=0.1$ ($v_{\rm st}=0$, no ``streaming loss''); and (3) diffusion with ``streaming losses'' but without streaming motion ($\kappa_{29}=0.1$, $v_{\rm st}=0$, but taking the streaming losses to be $v_{A}|\nabla_{\|} P_{\rm cr}|$ with $v_{A} \gtrsim 100\,{\rm km\,s^{-1}}$). These are all akin to a subset of our ``constant diffusivity'' models from \paperonetwo, with low $\kappa$. 

Despite the simulation differences, we find that their conclusions are similar to ours, for similarly low diffusivities: Fig.~\ref{fig:Buck} shows this directly. As the authors state directly in \citet{2017ApJ...847L..13P} (see Fig.~3 therein), in their MW-like halos, all their models predict that almost all of the injected CR energy is lost to collisions, and so produce $L_{\gamma}/L_{\rm SF}$ near the calorimetric limit. Moreover even at LMC and SMC star formation rates their predicted $\dot{E}_{\rm coll} / \dot{E}_{\rm cr} \sim 0.3$ in their favored model (i.e.\ they are always within a factor of $\sim 3$ of calorimetric). The cosmological simulations in B19 give a similar result (Fig.~14 therein): even for the smallest dwarf galaxies (lowest SFRs) plotted, the predicted $L_{\gamma}$ is within a factor $\sim 1.5-3$ of the calorimetric limit. The other diagnostics we consider here also give consistent results. For example, their models (1) and (2) predict a CR energy density at the solar circle in MW-like galaxies of $e_{\rm cr}(r\approx 8\,{\rm kpc}) \sim 15-20\,{\rm eV\,cm^{-3}}$.\footnote{Their model (3) predicts a lower value of $e_{\rm cr}(8\,{\rm kpc})$ only because with streaming losses but no streaming transport (and weak diffusion) and unphysically-large $v_{A} \sim 200\,{\rm km\,s^{-1}}$ in the warm ISM (owing to the artificial ISM ``effective equation of state''), the energy loss timescale from ``streaming'' $\sim 3\,\ell_{\rm cr}/v_{A}$ (see their Figs.~10 \&\ 12) in their simulations at $\sim 8\,$kpc is $\sim 10$ times shorter than the diffusion time ($\sim ({\rm few\ kpc})^{2}/\kappa_{\rm iso}$) for CRs to reach that radius, so most of the CR energy is lost to ``streaming losses'' despite the model not including streaming motion.} Where $\dot{E}_{\rm coll}/\dot{E}_{\rm cr} < 1$, we can use their adopted conversion formulae for their predicted $\gamma$-ray luminosities and injection rates to directly calculate the grammage in their simulations as well.\footnote{If we use the identical adopted parameters from \citet{2017ApJ...847L..13P}, their predicted $\gamma$ ray emission per unit volume in their band $0.1-100$\,GeV is $\dot{e}_{\gamma} = 5.67 \times10^{-17}\,n_{n}\,e_{\rm cr}$, so their $L_{\gamma}^{0.1-100} = \int \dot{e}_{\gamma}\,d^{3}{\bf x}$, while $\dot{E}_{\rm cr} = 3.5\times10^{40}\,{\rm erg\,s^{-1}}\,(\dot{M}_{\ast} / M_{\odot}\,{\rm yr^{-1}})$, and therefore in quasi-steady-state (when $L_{\gamma} \ll L_{\rm calor}$), they must have $X_{s}^{\infty} \approx  380\,{\rm g\,cm^{-2}}\,(L_{\gamma}/10^{40}\,{\rm erg\,s^{-1}})\,(\dot{M}_{\ast} / M_{\odot}\,{\rm yr^{-1}})^{-1}$ (for their quoted values of $L_{\gamma}$ and $\dot{M}_{\ast}$). As $L_{\gamma} \rightarrow L_{\rm calor}$, of course, $X_{s} \rightarrow \infty$.} In all 3 CR transport models, they predict a grammage in MW-mass systems of $X_{s} \sim 80-200\,{\rm g\,cm^{-2}}$, and for all lower-mass/SFR systems (down to $\dot{M}_{\ast} \sim 0.001\,M_{\odot}\,{\rm yr^{-1}}$) they predict $X_{s} \sim 40-130\,{\rm g\,cm^{-2}}$. Finally, although we cannot directly reconstruct their predicted residence times, their predicted $L_{\gamma}/L_{\rm SF}$ or grammage (given their collisional loss rate and mean ISM densities in B19 Fig.~10), or our simple analytic model in \S~\ref{sec:params} all imply similar $\Delta t_{\rm res} \gtrsim 500\,$Myr. 

Each of these conclusions is similar to those from our similar ($\tilde{\kappa} \le 0.3$) simulations in Table~\ref{tbl:transport} and \paperonetwo. Likely the reason we obtain such good agreement, despite considering very different simulations, is simply because the quantities above ``saturate'' once CRs approach the pure-advection/low-diffusion/calorimetric limit. However, B19 claim that their results disagree significantly with ours, arguing that their low-diffusivity models do reproduce the observations. They attribute the difference in predictions primarily to the treatment of dense gas, but as we have shown (1) there is actually very little difference in the predictions, and (2) dense gas has little effect on our predictions. 

The actual differences stem from how the observations are treated. \citet{2017ApJ...847L..13P} and B19 compare only to the $L_{\gamma}-\dot{M}_{\ast}$ correlation: they do not consider grammage or residence time or CR energy density constraints as we do here (all of which clearly rule out these lower-$\kappa$ models). Moreover, for the $L_{\gamma}-\dot{M}_{\ast}$ correlation, the authors estimate $\dot{M}_{\ast}$ of the observed systems (or, equivalently, the far-IR [FIR] $8-1000\,\mu{\rm m}$ luminosity of their simulations) by assuming a universal conversion factor $\dot{M}_{\ast}/(M_{\odot}\,{\rm yr^{-1}}) = 1.34\times10^{-10}\,(L_{\rm FIR}/L_{\odot})$. However, as noted in both \citet{2017ApJ...847L..13P} and B19, it is well-known that this correlation and conversion factor break down quite severely in low-SFR systems including the SMC, LMC, and M33 (and even at factor $\sim2-3$ level in the MW and M31), as the conversion they adopt assumes that {\em all} the light emitted by massive stars is absorbed by cold dust and re-processed into far-infrared (the particular calibration they adopt is derived for luminous infrared galaxies, with typical  extinctions $A_{v} \sim 100$). For the SMC, this means their adopted SFR ($\sim 0.008\,M_{\odot}\,{\rm yr^{-1}}$) is a factor $\sim 10-30$ lower than implied by  high-mass X-ray binary counts \citep{2005MNRAS.362..879S,2016AA...586A..81H}, young stellar object (YSO) counts \citep{2015MNRAS.448.1847H}, long-period variable star counts \citep{2014MNRAS.445.2214R},  simple bolometric ultraviolet continuum \citep{2017MNRAS.466.4540H} or H$\alpha$ emission \citep{2004A&A...414...69W} conversions, or the ``gold standard'' (to which many other methods are calibrated) resolved main-sequence turnoffs (i.e.\ stellar HR or color-magnitude diagram studies; \citealt{2004AJ....127.1531H,2009ApJ...705.1260N,2011A&A...535A.115I,2013MNRAS.431..364W,2015MNRAS.449..639R}). More importantly, this means their assumed SNe rate (which is what $L_{\rm SF}$ is ultimately used for, to estimate $R_{\rm SNe}$ and therefore $\dot{E}_{\rm cr} \approx 10^{50}\,{\rm erg}\,R_{\rm SNe}$) is $\sim 1/15,000\,{\rm yr}$, a factor $\sim 15-30$ lower than inferred from direct observations of SNe remnants in the MCs \citep{2010MNRAS.407.1314M,2017ApJ...837...36L,2019arXiv190811234M}. There are also some differences in the $\gamma$-ray spectral slopes/bolometric corrections assumed, as for example B19 include all emission from $0.1-100\,$GeV (likely including non-negligible pulsar contamination), but these are generally smaller (factor $\sim 2$) effects.

The net result of this is that the SMC is plotted in e.g.\ B19 Fig.~14 as if it has $L_{\gamma} \sim 0.4\,L_{\rm calor}$; this, in turn, means that their theoretical predictions with low $\tilde{\kappa}$ appear consistent -- as would indeed our own low-diffusivity $\kappa_{29}=0.3$ model shown in our Fig.~\ref{fig:Lgamma}. However, observational studies of these systems which carefully account for SNe rates and/or UV luminosities and $\gamma$-ray spectra place the SMC at $L_{\gamma} \sim 0.007\,L_{\rm calor}$ \citep{lacki:2011.cosmic.ray.sub.calorimetric,lopez:2018.smc.below.calorimetric.crs}, a factor of $\sim 50$ lower. If we compare the grammage, residence time, and/or CR energy density constraints in the MW (see values above), this inconsistency is also apparent: all of these numbers are significantly over-predicted (by factors $\sim 10-100$) by the low-$\kappa$ models in B19, so faster transport is clearly required. In short, the difference between our conclusions (here and in \paperonetwo), and those in \citet{2017ApJ...847L..13P} and B19, are driven almost entirely by how those authors compare to the observations, rather than by  theoretical or numerical differences.

\end{appendix}

\end{document}